%% file: PreferredVectorsKdS.tex
\documentclass[11pt]{article}
\usepackage{color}
\usepackage{graphicx}  % standard LaTeX graphics tool
\usepackage[all]{xypic}
\usepackage{psfrag}
\usepackage{pstricks}
\usepackage{tocbibind} %adds contents and bibliography to the table of contents
\usepackage{epsf}
\usepackage{epic}
\usepackage{eepic}
\usepackage{epsfig}
\usepackage{wrapfig}
\usepackage{a4}
\usepackage{amssymb,latexsym}%,amsmath}
\usepackage{amsfonts}

\usepackage[mathscr]{eucal}
\usepackage{citesort}
\usepackage[active]{srcltx}
\usepackage{deleq}
\usepackage{url}
\usepackage{fancybox}
\usepackage{ntheorem}
\usepackage{epstopdf}

\newcommand{\alphaX}{\chi}

\input{LeeNewcommand}
\renewcommand{\rd}{{\rm d}}

\begin{document}
\title{Hamiltonian dynamics in the space of asymptotically Kerr-de Sitter spacetimes\protect\thanks{Preprint UWThPh-2015-14}}

\author{
    Piotr T. Chru\'{s}ciel \\
    Erwin Schr\"odinger Institute and Faculty of Physics\\
    Universit\"at Wien
\and
    Jacek Jezierski\\
    Uniwersytet Warszawski
%    \\ KMMF, Ho\.za 69, 00-682 Warszawa
\and
Jerzy Kijowski
\\
CFT PAN, Warszawa
}

\maketitle

\begin{abstract}
We determine all spacetime flows which lead to a Hamiltonian dynamics in the space of general relativistic initial data sets with asymptotically Kerr-de Sitter ends. The corresponding Hamiltonians are calculated. Some implications for black-hole thermodynamics are pointed out.
\end{abstract}

\tableofcontents

\section{Introduction}
% \ptcr{ rewrites 17 VII 14}

There is  growing astrophysical evidence that spacetimes with positive cosmological constant provide physically correct models for cosmology.
Large families of such non-compact, vacuum, general-relativistic models have been constructed (see~\cite{ChPollack,CPP,CortierKdS,CMP,CM,BMW,GabachClement} and references therein). In particular we have now large classes of initial data sets with one or more ends of cylindrical type, in which the metric becomes periodic when one recedes to infinity along  half-cylinders. This then raises the question,
how to define the total mass, or energy, of such configurations. We have answered this question for asymptotically Schwarzschild-de Sitter metrics in~\cite{CJK2} using a Hamiltonian formalism, and the aim of this paper is to generalize the analysis there to metrics with asymptotically Kerr-de Sitter ends.

Consider, thus, the variational identity associated with the motion of a hypersurface in a vacuum space-time  in the direction of a vector field $X$, Equation~\eq{homogeneous} below.
The requirement of convergence of the volume integral there, when the boundary recedes to infinity on an asymptotically periodic end, leads to the requirement that the Lie derivative of the metric in the direction of $X$ approaches zero asymptotically, at least in an integral sense. Barring the case where $X$ is everywhere spacelike at large distances, one expects that this will only happen for metrics which are asymptotic to Kerr-de Sitter metrics. Hence our interest in asymptotically Kerr-de Sitter metrics. We note that a large family of non-trivial vacuum spacetimes which are exactly Kerr-de Sitter outside the domain of dependence of a compact set has been constructed in~\cite{CortierKdS}.

Now, in the coordinate system of \eq{kds.metric} below, all Killing vectors of Kerr-de Sitter metrics with rotation parameter $a\ne 0$ take the form
\bel{18VII14.1}
 X= \alphaX \partial_t + \beta \partial_\varphi
 \, ,
\ee
where $\alphaX$ and $\beta$ do not depend upon the spacetime coordinates, but might depend upon the solution at hand.

The vector field $\partial_\varphi$ is singled-out by the requirement that all its orbits are $2\pi$-periodic. Not unexpectedly, we check that the spacetime flow of $\partial_\varphi$ generates a Hamiltonian flow on the space of metrics. The corresponding Hamiltonian is usually interpreted as the total angular momentum of the solution.

In the case where the rotation parameter $a$ of the metric vanishes,  the vector $\partial_t$ is singled-out as the unique, up to a multiplicative constant, Killing vector field orthogonal everywhere to $\partial_\varphi$.  We  showed in~\cite{CJK2} that the spacetime flow of $\partial_t$ generates a Hamiltonian dynamical system in the space of gravitational initial data with asymptotically Schwarzschild-de Sitter ends, with the associated Hamiltonian equal to the mass parameter $m$.

When $a\ne 0$, there does not seem an obvious choice for a second preferred Killing vector. A natural question arises, which of the Killing vector fields of the form \eq{18VII14.1} generate a Hamiltonian dynamical system on the space of asymptotically Kerr-de Sitter vacuum metrics.   We give an exhaustive description of such vector fields in Theorem~\ref{T18VII14.1} below. This is the main result of this work.
The result requires a detailed analysis of the variational identities arising in this context, to be found in Section~\ref{s16IV13.1} which occupies a significant part of  this paper. We believe this analysis has interest on its own.
 %\ptcr{rewordings 4 VII 15}

Much to our surprise, the vector field $\partial_t$ of the coordinates of \eq{kds.metric} is \emph{not} associated with a Hamiltonian flow. However, we show that the vector field $\sqrt{1 + a^2\Lambda/3}\partial_t$ is, the corresponding Hamiltonian being equal to $m (1+a^2 \Lambda/3)^{-3/2}$. Some further explicit examples of interest of Hamiltonian flows are given.

As such, the simplest asymptotically Kerr-de Sitter spacetime is the Kerr-de Sitter spacetime itself. The variational identities that result from a Hamiltonian analysis are then known in the literature under the name of ``black hole thermodynamics".
The ambiguity in the choice of the energy, equivalently, in the choice of the Hamiltonian vector field, leads to ambiguities in the definition of thermodynamical variables. Here we point out that in the Kerr-de Sitter spacetimes one can define preferred Killing vector fields $X$ by requiring that $X$ is tangent to some bifurcate Killing horizon, and has surface gravity {equal to} one there. (This enables one to define e.g. the rotation velocities of the remaining horizons with respect to the selected one.) It turns out that these vector fields are Hamiltonian, with Hamiltonian equal the area of the selected horizon. The choice of ``total energy''  as the value of the associated Hamiltonian, leads to a set of geometrically unique thermodynamical identities.

We note the recent analysis of~\cite{HollandsWaldStability}, where thermodynamical instabilities are tied to instabilities of associated black string solutions. We believe that our analysis provides a good starting point for similar considerations for solutions with a positive cosmological constant, and hope to be able to address this issue in a near future.
%\ptcr{comment added 2 VI 15}

 %\ptc{previous footnote move here, reference to  the appendix with CYK, remark on Komar added 4 VII 15}
We refer the reader to~\cite{AbbottDeser,AnninosMusings,BalaBD,TodSzabados,AshtekarBK} and references therein for alternative approaches to a definition of mass in the presence of a positive cosmological constant. In Appendix~\ref{A4VII15.1} we shortly discuss a definition using conformal Killing-Yano tensors. We note that in space-times with a non-zero cosmological constant the Komar integral does not provide a surface-independent integrand except for the angular-momentum of hypersurfaces asymptotically invariant under rotations.

\section{The variational formula}
 \label{s16IV13.1}

We wish to define a Hamiltonian dynamical system on a set of Lorentzian metrics on an $(3+1)$-dimensional manifold $\mcM$, assuming the existence of a spacelike hypersurface $\hyp\subset \mcM$ on which the metric asymptotes to a Kerr-de Sitter metric as one recedes to infinity along an end of $\hyp$. This is a special case of a more general construction which can be done in all space-dimensions $n\ge 2$ and proceeds as follows:  Given a Lorentzian metric $g$ on $\mcM$ we  choose an $n$-dimensional manifold $\Sigma$ and a one-parameter family of embeddings
\bel{10IV15.1}
 \R\times \Sigma \ni (t,\mathbf{x}) \mapsto \psi(t,\mathbf{x}) \in \mcM
 \,.
\ee
Note that while for all $t\in \R$  the maps $\Sigma \ni   \mathbf{x}  \mapsto \psi(t,\mathbf{x}) \in \mcM$ are assumed to be embeddings of $\Sigma$, we do not assume that the whole map \eq{10IV15.1} is an embedding. Thus, the hypersurfaces
$$
 \hyp_t:=\{ \psi(t, \mathbf{x})\,,\ \mathbf{x}\in \Sigma\}
$$
do not need to form a foliation,  they are allowed to cross each other, etc.

We emphasise that the maps $\psi$ can depend upon the metric $g$ under consideration. Indeed, this will be the case in our main application here, namely the description of a family of Hamiltonian dynamical systems in the space of asymptotically Kerr-de Sitter metrics.

To continue, we choose a compact volume $V \subset \Sigma$ with a smooth boundary $\partial V$.
The equations will be analysed on $V$ before taking an exhaustion of $\Sigma$  and passing to the limit. We set
$$
 \mcV=\psi(0,V)
\,.
$$

Along each hypersurface $\hyp_t$ we define a spacetime vector field $X$ by the formula
\bel{10IV15.1+z}
  X := \psi_* \partial_t
 \,.
\ee
Note that this defines a vector field on a neighbourhood of $\hyp:=\hyp_0$ if $\psi$ is a diffeomorphism near $\{0\}\times \Sigma$  but,  in general, we will only have a spacetime vector field defined along $\hyp_t$ for each $t$.

\subsection{Adapted coordinates}
 \label{ss5VI15.1}

In~\cite{KijowskiGRG} it was assumed that $X$ is actually a vector field on $\mcM$, which is moreover supposed to be timelike and non-vanishing. Then, the spacetime domain %
$$
 \Omega := \cup_{t }\{{\cal G}^X_t (\mcV)\}\subset \mcM
$$
was considered, where by $\{{\cal G}^X_t \}$ we denote the (local) group of diffeomorphisms generated by the field $X$. The boundary  defined as  $\partial \Omega :=  \cup_{t }\{{\cal G}^X_t (\partial V)\}$ was, therefore, a smooth timelike submanifold, with  $\mcV:=\Omega\cap \hyp$.
Let $(x^k)$, $k=1,\dots,n$, be a coordinate system on  $\Sigma$. In~\cite{KijowskiGRG} local coordinates $y^\mu$ on $\mcM$ were chosen so that
$$
 \mbox{$\hyp_t=\{y^0 = t\} $, and $y^i(\psi(t,\mathbf{x}))=x^ i(\mathbf{x})$,}
$$
whence $X= \partial_0$. Using this particular coordinate system, the following variational formula has been proved in~\cite{KijowskiGRG} in dimension $3+1$, for a Lorentzian metric interacting with a matter field $\varphi$:
 %\ptc{tu teraz calki po $V$ a nie po $\mcV$}
%
\begin{eqnarray}
0 & = & \frac 1{{ 2 \gamma}} \intvolV   \left( {\dot P}^{kl}  \delta g_{kl} -
{\dot g}_{kl} \delta P^{kl}\right)  +
\intvolV  \left({\dot p}\delta \varphi - {\dot \varphi}\delta p\right) +
\frac 1{{  \gamma}} \intdvolV  ( {\dot \lambda} \delta \alpha  -
{\dot \alpha} \delta \lambda )
\nonumber \\
 & + & \frac 1{{ 2 \gamma}} \intdvolV  (   2 \nthree\delta {\bf Q}
- 2\nthree^A \delta {\bf Q}_A
+ { \nu}{\bf Q}^{AB}
 \delta g_{AB}) + \intdvolV  p \delta \varphi \;  \label{homogeneous}
\end{eqnarray}
with $\gamma=8\pi$ when $n=3$ (see, e.g.,~\cite[Appendix~D]{CJK2} for other dimensions).
Here, $g_{kl}$ and $P^{kl}$ describe gravitational Cauchy data on $V$, i.e.~the pull-back to $V$ of the metric induced by the space-time metric on $
\mcV$ and of the ADM~momentum, respectively, whereas $\varphi$ and $p$ represent symbolically the corresponding Cauchy data for the matter fields, if any. ``Dot'' stands for the time derivative $\partial_t$.  As explained in detail below or in~\cite{KijowskiGRG}, the
${\bf Q}$'s describe various components of the extrinsic curvature of the world tube $\partial \Omega$, whereas $\nu$, $\nu^A$, and $ g_{AB}$ encode the metric induced on $\partial \Omega$ using an $((n-1)+1)$ decomposition. Finally, $\lambda = \sqrt{\det g_{AB}}$ is the volume form on $\partial V$ whereas $\alpha$ denotes the hyperbolic angle between $\hyp$ and $\partial \Omega$. For a Hamiltonian flow the second line of \eq{homogeneous} equals the variation of the Hamiltonian, which is at the origin of the occurrence of the $\gamma$ factors in the formula.

As such, in~\cite{KG-JK}
the formula was generalized to the case when $X$ is spacelike. Next, in~\cite{CJKbhthermo}
the field was allowed to be null-like.
Finally, it has been pointed out in~\cite{CJK2} that the formula remains true for vacuum metrics, possibly with a cosmological constant, in any space-dimension $n\ge 2$,
with a constant $\gamma $ which depends upon dimension.

Our first aim is to remove the coordinate condition $X=\partial_0$, and the hypothesis that $X$ is a vector field on $\mcM$, and allow   for  non vanishing variation $\delta X^\mu$ of the field $X$. The latter is motivated by the fact, that imposing coordinate conditions leads typically to vector fields $X$ which depend upon the configuration of the field, so varying the latter necessarily requires varying the former. This issue will indeed have to be addressed in our analysis below of asymptotically Kerr-de Sitter spacetimes.

Now, on each $V_t := \{t\}\times V$ we have a non-degenerate
 $n$-dimensional metric $g$ and the ADM momentum $P$, defined {\em via} the $\psi$-pull-back of the corresponding objects from $\psi(\{t\}\times V) \in \mcM$. Given a coordinate system $(x^k)$ on $V$, we denote components of $g$ and $P$, together with their derivatives with respect to the parameter $t$, as $g_{kl}$, $P^{kl}$,  ${\dot g}_{kl}$ and ${\dot P}^{kl}$, respectively. On each boundary $\partial V_t$ we have the induced $(n-1)$-metric, which we describe by its components $g_{AB}$. For computational purposes it is useful to choose the last coordinate $x^n$ in such a way that it is constant on $\partial V$, and then the collection $(x^A)$, $A=1,\dots , n-1$, provides a coordinate chart on the boundary.
%However, this choice is not necessary and, in principle, any other coordinate system on %$\partial V$ can be used to encode the components $g_{AB}$.
By $\lambda := \sqrt{\det g_{AB}}$ we denote the $(n-1)$-volume density on the boundary.

At each point of $  \partial \mcV$ we decompose the field $X$ into
the part tangent to $\partial \mcV$, which we denote by
$X^\parallel=\nu^A\partial_A$, and the part $X^{\perp}$, orthogonal
to $\partial \mcV$. We have, therefore, $X=X^{\perp}+X^{\parallel}$ and
we define
$$
 \nu:=\pm\sqrt{|(X^\perp|X^\perp)|}
 \,,
$$
%,
where the $+$ sign is taken
if $X$ is timelike and ``$-$'' if $X$ is spacelike. At points at which $X^\perp$ is non-zero and not null we define the unit vector
$\text{\bf N}:= \frac 1\nu X^\perp$, so that there we have:
\begin{equation}\label{pole-X}
    X=\nu \text{\bf N} + \nu^A\partial_A \, .
\end{equation}
We will use the same symbols $\nu^A$ and $\nu$ for the corresponding pull-backs to $\partial V$.

\begin{figure}
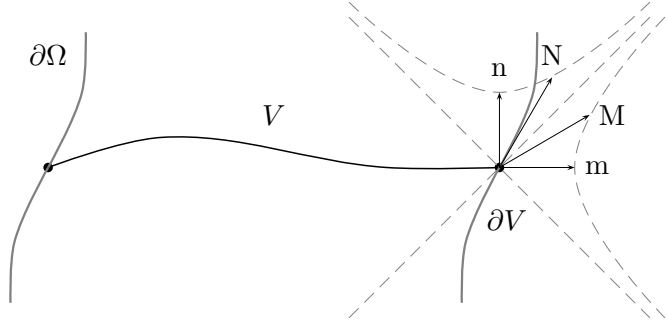

  \begin{center}
$$
\pscurve[linewidth=.02](-3,0)(-1.5,.4)(1.5,0)(3,0)
\psdot(-3,0)
\psdot(3,0)
\psline[linewidth=.01, linestyle=dashed, linecolor=gray](1,2)(5,-2)
\psline[linewidth=.01, linestyle=dashed, linecolor=gray](1,-2)(5,2)
\pscurve[linewidth=.01, linestyle=dashed, linecolor=gray](1,2.2)(3,1)(5,2.2)
\pscurve[linewidth=.01, linestyle=dashed, linecolor=gray](5.2,2)(4,0)(5.2,-2)
\rput(-3,1.5){\partial\Omega}
\pscurve[linewidth=.03,linecolor=gray](-3,0)(-2.55,1)(-2.5,1.8)
\pscurve[linewidth=.03,linecolor=gray](-3,0)(-3.45,-1)(-3.5,-1.8)
\pscurve[linewidth=.03,linecolor=gray](3,0)(3.45,1)(3.5,1.8)
\pscurve[linewidth=.03,linecolor=gray](3,0)(2.55,-1)(2.5,-1.8)
\psline[linewidth=.01]{->}(3,0)(3,1)
\psline[linewidth=.01]{->}(3,0)(3.7,1.2)
\psline[linewidth=.01]{->}(3,0)(4,0)
\psline[linewidth=.01]{->}(3,0)(4.2,.7)
\rput(3.1,-.7){\partial V}
\rput(0,.7){V}
\rput(3,1.3){\mathrm{n}}
\rput(3.7,1.5){\mathrm{N}}
\rput(4.3,0){\mathrm{m}}
\rput(4.5,0.7){\mathrm{M}}
$$
  \end{center}
\vspace{1.2cm}
 \caption{ The tangent space at $\partial V$.} \label{F7IV15.1}
 \end{figure}

To define the remaining objects appearing in (\ref{homogeneous}), on $\partial \mcV$ consider the two-dimensional tangent plane in $T \mcM$ orthogonal to $T {\boundaryimage }$ (the plane may be identified with the two-dimensional Minkowski space) and use the following four normalized vectors: $\text{\bf N}$, $\text{\bf
M}$ -- orthogonal to $\text{\bf N}$, $\text{\bf m}$ -- tangent
to $\hyp$, directed outwards and, finally, $\text{\bf n}$ -- orthogonal to $\text{\bf m}$, directed in the future. See Figure~\ref{F7IV15.1}.

%\ptc{figure commented out, to be reinstated at the end}
%\vspace{2cm}

By ``$\alpha$'' we denote the ``hyperbolic angle''
between $\text{\bf N}$ and $\text{\bf n}$, defined as follows:
\begin{equation}
\alpha=\left\{\begin{array}{ll} \text{arsinh}(\text{\bf
N}\,|\,\text{\bf m}) & \quad \text{for }X^\perp\text{ timelike,}
\\ \text{sgn}(\text{\bf N}\,|\,\text{\bf m})\text{arcosh}(\text{\bf
N}\,|\,\text{\bf m}) & \quad \text{for }X^\perp\text{
spacelike.}\end{array}\right. \label{alfa}
\end{equation}
Of course, the definition is meaningful only for non-vanishing $\nu$, which is the case if we assume that $X^\perp$ is everywhere  timelike, or everywhere spacelike non-vanishing everywhere. The hypothesis that $
\nu$ has no zeros is needed for the derivation of the elegant formula \eq{homogeneous}. However, we stress that this hypothesis will \emph{not} be needed in our main results below.

In order to define the remaining objects ${\bf Q}$, ${\bf Q}_A$ and ${\bf Q}^{AB}$ we consider the hypersurface $\partial \Omega = \{{\cal G}^X_t (\partial V)\} \subset \mcM$, parameterized by coordinates $(x^a) = (x^0, x^A)=(t, x^A)$, and the ADM version $Q^{ab}$
 of its extrinsic curvature:
\begin{eqnarray}
Q^{ab} & := & \sqrt{|\det g_{cd}|} \ (L {\hat g{}}^{ab} - L^{ab} ) \, ,
\label{ku}  \\
 \label{2VII15.1}
L_{ab} & := & - \frac 1{\sqrt{g^{nn}}} {\Gamma}^n_{ab} =
- \frac 1{\sqrt{g^{nn}}} A^n_{ab} \, ,
\end{eqnarray}
where ${\hat g{}}^{ab}$ is the $n$-dimensional inverse with respect to the
induced metric $g_{ab}$ on $\partial \Omega$, with $L=L^{ab}g_{ab}$.

Assuming again that $\nu$ has no zeros, we set:
\begin{eqnarray*}
  {\bf Q} &=& \nu Q^{00} \, , \\
  {\bf Q}_A &=& Q^0_{\ A} = Q^{0a}g_{aA} \, , \\
  {\bf Q}^{AB} &=& { \frac 1\nu}\ttg^{AC}Q_{CD} \ttg^{DB}
  = { \frac 1\nu}\left(
  Q^{AB} + Q^{0A}\nu^B + Q^{0B} \nu^A + Q^{00} \nu^A \nu^B \right)
  \, ,
\end{eqnarray*}
where we use $(n-1)$-dimensional inverse $\ttg^{AB}$ of the metric
$g_{AB}$ on $\partial V$.

We can also reformulate these definitions in terms of the extrinsic curvature of $\partial \mcV$, namely: the torsion covector
\begin{equation}
\ell_A:=(\nabla_A\text{\bf N} \,|\,\text{\bf
M}) \, ,
\end{equation}
and the extrinsic curvature tensor in direction of $\text{\bf M}$
\begin{equation}\label{ka}
  k_{AB}\equiv k_{AB}( \text{\bf M}):= (\nabla_A \partial_B
  | \text{\bf M}) \, ,
\end{equation}
and its trace
\begin{equation}\label{k-tr}
    k=\widetilde{\tg}^{AB} k_{AB} \, .
\end{equation}
Equivalently, for any pair $(Y,Z)$ of vector fields tangent to
$\partial \mcV$ it holds that $k(Y,Z) = (\nabla_Y Z|\text{\bf M})$. We have %It holds that
\begin{eqnarray}
  {\bf Q} &=& \lambda k \, , \label{Qu}
\\
  {\bf Q}_A &=& \lambda  \ell_A \, ,
   \label{QuA}
\\
  {\bf Q}^{AB} &=& \lambda\left(k^{AB} - \ttg^{AB} k + \ttg^{AB} s \right)
 \, ,
  \label{QuAB}
\end{eqnarray}
where by $s$ we denote the following ``acceleration scalar''
\begin{equation}\label{es}
  s=(\nabla_{\text{\bf N}} \text{\bf N}\,|\,\text{\bf M})= -
  (\text{\bf M}\,|\,\nabla_{\text{\bf N}} \text{\bf M}) \, .
\end{equation}
This completes the list of geometric objects used in \eq{homogeneous}.

In Appendix~\ref{A5VI15.1} we prove
the following:%
\footnote{More precisely, in Appendix~\ref{A5VI15.1} we prove the second part of our theorem, namely that the vector field $X$ is allowed
to depend upon the field configuration without changing \eq{homogeneous}. The first part of the theorem is implicit in the considerations in~\cite{KijowskiGRG}, and follows explicitly from the remaining analysis in the current work in any case.}

\begin{Theorem}
 \label{T25III15.1}
Formula \eq{homogeneous} is valid for $V$ which are either spacelike everywhere or timelike everywhere, for any vector field $X$ defined along $V$ such that $X$ is nowhere vanishing and everywhere transverse to  the image $\mcV$ of $V$ in spacetime.
The vector field $X$ is allowed to depend upon the metric and its derivatives (hence
 with non-vanishing variation $\delta X^\mu$) provided that   $\mcV$ is kept fixed in spacetime.
\end{Theorem}

\input{KijowskiALaWald}

\input{tangentX}
\section{The Kerr-de Sitter metrics}
 \label{s3III13.1}

We wish to analyse \eq{homogeneous} for metrics defined on a half-cylinder $ [0,\infty)\times S^2 $ which  asymptote to a periodic spacelike hypersurface in a maximal extension of the Kerr-de Sitter metric. (The reader is referred  to~\cite{CarterlesHouches}, or~\cite{COS} and references therein, for a discussion of maximal extensions.) Locally, in    Boyer-Lindquist coordinates~\cite[p.~102]{CarterlesHouches}, the metric takes the form
\begin{eqnarray}
  \label{kds.metric}
  \nonumber
    g &=& \rho^2\left(\frac{1}{\Delta_r} dr^2+\frac{1}{\Delta_\theta} d\theta^2\right)
      +\frac{\sin^2(\theta)\Delta_\theta}{\rho^{2}\Xi^{2}} \left(a dt-(r^2+a^2)d\varphi\right)^2  \\
    &\quad& -  \frac{\Delta_r}{\rho^{2}\Xi^{2}}\left(dt-a \sin^2(\theta) \, d\varphi\right)^2 \, ,
\end{eqnarray}
where
\begin{eqnarray}
  \label{rho.2}
    \rho^2 &=& r^2+a^2\cos^2(\theta) \vphantom{\frac11} \, ,\\
  \label{delta.r}
    \Delta_r  &=& (r^2+a^2)\left(1-\frac{\Lambda}3 r^2\right)-2 m r \, ,\\
    %&=& -\frac{\Lambda }{3} r^4 + \left(1-\frac{a^2 \Lambda }{3}\right)r^2 -2 m r +a^2 \\
  \label{delta.theta}
    \Delta_\theta &=& 1+\frac{a^2 \Lambda}3  \cos^2(\theta) \, , \\
  \label{xi}
    \Xi &=& 1+\frac{a^2\Lambda}3 \, ,
\end{eqnarray}
with $t \in \mathbb{R}$, $r \in \mathbb{R}$, and $\theta$, $\varphi$ being the standard coordinates parameterizing the sphere.

 %\ptc{paragraph added 2 VII}
Above we have used the standard notation $(t,r,\theta,\varphi)$ for the Boyer-Lindquist coordinates. This leads to a conflict of notation,
as in all our considerations so far $t$ was a parameter along the Hamiltonian flow. The Boyer-Lindquist coordinate $t$ would correspond to the coordinate $y^0$ in our previous considerations in cases in which the initial data surface is given by $\{t=0\}$. Note, however, that such coordinates can  work at most in regions where $\partial_t$ is spacelike, so the identification of the Boyer-Lindquist coordinate $t$ with $y^0$ is also to be avoided. Our explicit calculations will be done for a boundary surface $\partial \mcV$ taken to be one of the coordinate-surfaces $\{t=\const\,\ r=\const\}$, before passing with the boundary to infinity.

Throughout we will assume $\Lambda>0$ and $a\ne 0$; the case $a=0$ has been covered in \cite{CJK2}, while the case $\Lambda<0$ is briefly discussed in Appendix~\ref{A5VI15.11}.

\begin{figure}
\begin{center}
{\includegraphics[scale=.7,angle=90]{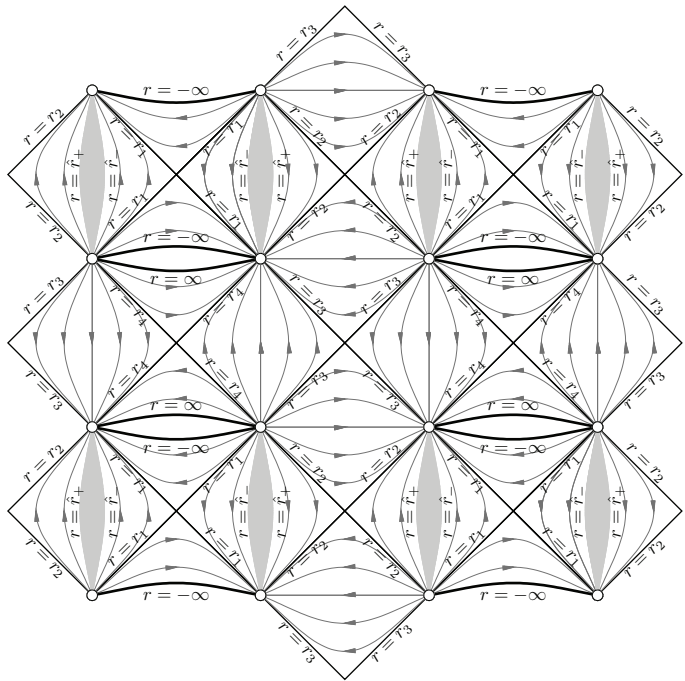}}
\caption{A projection diagram for the Kerr-Newman - de Sitter metric with four distinct zeros of $\Delta_r$, from~\cite{COS}. The pattern continues indefinitely in all directions.}
\label{F2VII15.1}
\end{center}
\end{figure}

We will keep away from zeros of $\rho$, where the geometry is singular, and ignore the trivial coordinate singularities $\sin(\theta)=0$. Recall that the metric \eq{kds.metric} can be smoothly extended across the zeros of $\Delta_r$, which become Killing horizons in the extended spacetime. Note that under our assumptions $\Delta_r$ has at least two and up to four distinct real zeros. A projection diagram for a natural maximal analytic extension in the case of four distinct real zeros is shown in Figure~\ref{F2VII15.1}, see~\cite{COS} for the remaining cases. The figure illustrates clearly that KdS space-times with the corresponding range of parameters contain complete periodic spacelike hypersurfaces. Inspection of the remaining diagrams in~\cite{COS} shows that complete periodic spacelike hypersurfaces meeting an infinite number of Killing horizons exist only when $\Delta_r$ has four distinct real zeros. In the remaining cases there exist asymptotically cylindrical complete spacelike hypersurfaces which interect only one or two horizons.}

When $m=0$,  an explicit coordinate transformation bringing the metric to the usual de Sitter form can be found in~\cite[p.~102]{CarterlesHouches}, see \eq{AkcayMatzner} below, compare~\cite{AMatzner,WarnickUnpublished,HT}.

In what follows the following formula will be useful:
\bel{4V13.1}
 \lambda := \sqrt{\det g_{AB}} =    \frac{ \sin (\theta)}{ \sqrt{\Delta_\theta}\Xi }\sqrt{ \Delta_\theta(r^2+a^2) ^2
       -  {a^2 \Delta_r}   \sin^2(\theta)     }
       \, .
\ee

\section{Hamiltonian flows for asymptotically Kerr-de Sitter metrics}
 \label{s3III13.1x+}

Recall \eq{finalresultbisbis}, which we rewrite as
\bean
 \lefteqn{
  \intvol   \left({\cal L}_X g_{kl} \delta P^{kl} - {\cal L}_X P^{kl}\delta g_{kl}\right)
 + 2 \intdvol  ({\cal L}_{X^0\partial_0} \alpha\delta\lambda - {\cal L}_{X^0\partial_0} \lambda \delta \alpha )
 }
 &&
\\
  &&
 +   16 \pi  \intvol
    X^0  {\cal E}^ {\mu\nu}  \delta g_{\mu\nu}
  -2X^\mu\delta {\cal E}^0{}_{\mu}
   \nn
\\
 & = &
  \intdvol X^0 (   2 \nthree\delta {\bf Q}
    - 2\nthree^A \delta {\bf Q}_A + { \nu}{\bf Q}^{AB}\delta g_{AB})
% \nn
%\\
%    &&
+  \int_{\partial \mcV}
    \left(
  2 Y^k  \delta{P }^{n}{}_{  k} -  Y^n P ^{kl}  \delta g_{kl}
   \right)
 \nn
\\
 & =: &
  \theta_b(S,X)
   \, .
  \label{5VI15.21}
\eea
The subscript "$b$" stands for ``boundary''.

Given a compact spacelike surface $S$ and a spacetime vector field $X$ defined along $S$,  $\theta_b(S,X)$
is thus the ``non-dynamical''
boundary term  from our variational identity.
$\theta_b$ can be thought of as a one-form
on the space of vacuum initial data.

As such, we consider the collection of vacuum initial data which approach  Kerr-de Sitter initial data in an asymptotically periodic end, together with first derivatives. (A large class of such initial data has been constructed in \cite{CortierKdS,ChPollack,CM}.) Our aim is to calculate $
\theta_b$ for such metrics.

Now, the value of the
integral defining $\theta_b(S,X)$ approaches the value of the same integral calculated in an exact Kerr-de Sitter metric when $S$ is moved to infinity along
the end. So, the problem of calculating $\theta_b$ for asymptotically Kerr-de Sitter metrics is reduced to one of calculating this
integral for exact Kerr-de Sitter metrics.

In spite of various attempts, we have not been able to carry out a direct calculation of $\theta_b(S,X)$ for  the metric (\ref{kds.metric})  on a general surface $t=\const$, $r=\const'$. However, one can proceed as follows:

\Eq{5VI15.21} shows that, \emph{for all metric variations within the family of
exact Kerr-de Sitter or Kerr-anti de Sitter metrics}, and for all Killing vectors $X$ of those metrics we have
\bel{27IX13.6}
 \theta_b(S,X)=  \theta_b(S',X)
 ,
\ee
whenever there exists a \emph{spacelike or timelike}
 hypersurface $\mcV$ so that $\partial \mcV = S\cup S'$.

In particular, if $S$ is a sphere
$$
 S_{\tau,\rho}:=\{t=\tau, \ r=\rho \}
$$
in a region where the slices $\{t=\const\}$ are timelike or spacelike,
we obtain
\bel{27IX13.7}
 \partial_r\big(\theta_b(S_{t,r} ,X)\big)= 0
 .
\ee
Since the Kerr-de Sitter metric is $t$-independent, it obviously holds that
\bel{27IX13.8}
 \partial_t\big(\theta_b(S_{t,r} ,X)\big)= 0
 .
\ee
We conclude that we can choose any convenient value of $t$ and $r$ to calculate $\theta_b(S_{t,r} ,X)$ within each of the regions, where $(t,r,\theta,\varphi)$ form a coordinate system.
Although this is not immediately apparent, it is the case that the integrand depends smoothly upon the metric when approaching the Killing horizons, hence surfaces lying on the boundaries of the relevant regions can also be used to calculate $\theta_b(S_{t,r} ,X)$. So, in fact, to determine $\theta_b(S_{t,r} ,X)$ it suffices to calculate the  integral at a bifurcation surface $\Delta_r=0$. This simplifies the calculations enough to make them tractable.

The reader should keep in mind that
{when seeking a primitive for $\theta_b$ on bifurcation spheres},    variations  $\delta m$ and $\delta a$ have to be accompanied with a variation $\delta r$ of $r$ so that $\delta \Delta_r=0$.

\subsection{``Energy''}
As such, the condition $\Delta_r=0$ implies that $\nu=0$ and $ {\bf Q}=0$; recall that  $\nu^\theta=0$ everywhere in our coordinates. Denoting by $\int_{\Delta_r=0}$ the limit $\lim_{\epsilon \to 0} \int_{\Delta_r=\epsilon}$,
 %\ptcr{comment added 5 VII 15}
we find
\[ \int_{\Delta_r=0} - 2 \nu^A \delta {\bf Q}_A = \int_{\Delta_r=0} -2\nu^\varphi\delta{\bf Q}_\varphi
= \frac{2a}{r^2+a^2} \int_{\Delta_r=0} \delta{\bf Q}_\varphi = \frac{16\pi a}{r^2+a^2}
 \delta \left(\frac{ma}{\Xi^2}
 \right)
 \, ,
\]
\[ \int_{\Delta_r=0} \perpQ{}{^{AB}} \delta g_{AB} = \frac{\partial_r\Delta_r}{\Xi(r^2+a^2)}
 \delta
  \left(\frac{4\pi (r^2+a^2)}{\Xi}
   \right)
 \, .
\]
To obtain the above, the following formulae are useful:
\begin{equation}
 \nthree^\varphi %= \frac{ g_{0\varphi}}{g_{\varphi \varphi }}
  =-a\frac{\left(a^2+r^2\right)\Delta_{\theta}-\Delta_r}{{\left(a^2+r^2\right)^2 \Delta_{\theta} - a^2  \sin^2(\theta)\Delta_r}}
 \;\; =_{\Delta_r=0} -\frac{a}{{a^2+r^2}}
 \, , \label{nuphi30VIII13}
\end{equation}

\begin{equation}
{\bf Q}_\varphi = \frac{ma \sin^3 (\theta)}{\Xi^2\rho^4}\left[ (r^2-a^2)\rho^2+ 2r^2(r^2+a^2)\right]
 \, ,
\end{equation}
\begin{equation}
\nthree = \frac 1{\sqrt{|{\hat g{}}^{00}|}}
 = \frac{\sqrt{\Delta_r \Delta_{\theta}} \rho } {\Xi \sqrt{\left(a^2+r^2\right)^2 \Delta_{\theta} - a^2  \sin^2(\theta)\Delta_r }}
 \, ,
\end{equation}
\begin{eqnarray}
 \nonumber
 \perpQ{}{^{AB}} \delta g_{AB} &=&  Q_{CD} \, {\tilde{\tilde g}}{^{CA}}
 \, {\tilde{\tilde g}}{^{DB}} \delta g_{AB} = - Q_{CD} \delta {\tilde{\tilde g}}{^{CD}}
\\
\nonumber
 & = &
 -\nu \lambda \left( L g_{AB} - L_{AB}
 \right) \delta {\tilde{\tilde g}}{^{AB}}
\\
 & = &
 \nu \lambda \left( L \, {\tilde{\tilde g}}  ^{AB}\delta g  {_{AB}} + L_{AB}
 \delta {\tilde{\tilde g}}{^{AB}}\right)
\nonumber
  \\
  &=&
    \nu \lambda \left(  \frac{1}{   \rho ^2 }  \partial_r\big(   { \sqrt{\Delta_r  } \rho  } \big)
     \times \frac 1 {\lambda^2} \delta (\lambda ^2) +   \frac { \sqrt{\Delta_r}}{2\rho} \partial_r g_{AB}
    \delta {\tilde{\tilde g}}{^{AB}}\right)
\nonumber
  \\
  &=&
    \nu \left(  \frac{2}{   \rho ^2 }  \partial_r\big(   { \sqrt{\Delta_r  } \rho  } \big)
      \delta  \lambda  +   \frac {\lambda  \sqrt{\Delta_r}}{2\rho} \partial_r g_{AB}
    \delta {\tilde{\tilde g}}{^{AB}}\right)
% \, .
  \label{1III13.2x}\\
&=& \hspace*{-2ex} \strut_{\Delta_r=0} \; \frac{2\sqrt{\Delta_r } } {\Xi \left(a^2+r^2\right)  }
      \partial_r\big(   { \sqrt{\Delta_r  }  } \big)
      \delta  \lambda
      \, ,
\end{eqnarray}
\[ \sqrt{\Delta_r} \partial_r\big(   { \sqrt{\Delta_r  }  } \big)
 =
  \frac12 \partial_r \Delta_r = r-m -\frac{r\Lambda}3(2r^2+a^2) \, ,
\]
\bel{4V13.1+}
 \lambda %= \sqrt{\det g_{AB}}
 =    \frac{ \sin (\theta)}{ \sqrt{\Delta_\theta}\Xi }\sqrt{ \Delta_\theta(r^2+a^2) ^2
       -  {a^2 \Delta_r}   \sin^2(\theta)     }
 \;\; =_{\Delta_r=0} \frac{ \sin (\theta)}{ \Xi } (r^2+a^2)
       \, .
\ee
We also note
\bel{18V13.21}
 J:= -\frac 1 {8\pi} \int_{S^2} { P}^r{}_\varphi   =  \frac 1 {8\pi} \int_{S^2} {\bf Q}_\varphi =  \frac {ma}{\Xi^2}
 \, ,
\ee
where the last equality
can be used to simplify the calculation of the term $\displaystyle \int_{\Delta_r=0} \delta{\bf Q}_\varphi$.

{}From $\delta \Delta_r=0$ we get $\frac12\partial_r\Delta_r \delta r = r\delta m -(1-\frac\Lambda{3}r^2)a\delta a$.
Finally, we are led to
\begin{eqnarray} \nonumber
 \lefteqn{
 \int_{\Delta_r=0} 2 \nu \delta{\bf Q} - 2 \nu^A \delta {\bf Q}_A + \perpQ{}{^{AB}} \delta g_{AB}
 }& &
\\
  \nonumber
  & & = 16\pi\frac{a}{r^2+a^2}\delta\frac{ma}{\Xi^2} +4\pi\frac{\partial_r\Delta_r}{\Xi(r^2+a^2)}\delta\left( \frac{r^2+a^2}{\Xi}\right) \\
 & &= \frac{16\pi}{\sqrt{\Xi}} \delta \left(\frac{m}{\Xi^{3/2}}\right)
 \label{24V13.hor}
 \, .
\end{eqnarray}
The last equality is not completely obvious, and its proof proceeds as follows:
%\begin{eqnarray} \nonumber
%  \nonumber
%  & & \frac{a}{r^2+a^2}\delta\frac{ma}{\Xi^2} +\frac14 \frac{\partial_r\Delta_r}{\Xi(r^2+a^2)}\delta\left( \frac{r^2+a^2}{\Xi}\right) \\
%& = & \frac{a}{r^2+a^2}\delta\frac{ma}{\Xi^2} +\frac12 \frac{\partial_r\Delta_r}{\Xi^2(r^2+a^2)}\left(r\delta r +a\delta a\right) + \frac14 \frac{\partial_r\Delta_r}{\Xi}\delta\frac{1}{\Xi} \\ \nonumber
%& = & \frac{1}{\Xi^2}\delta{m} - \frac{r\Lambda}{3\Xi^2}a\delta a
%+ \frac1\Xi\left[ \frac{2ma^2}{r^2+a^2}+\frac12(r-m) -\frac{r\Lambda}{6}(2r^2+a^2)\right]\delta\frac{1}{\Xi} \\
%& = & \frac{\delta{m}}{\Xi^2} - \frac{\Lambda}{3\Xi^3}\left[ r\Xi+\frac{4ma^2}{r^2+a^2}+ r-m -\frac{r\Lambda}{3}(2r^2+a^2)\right]a\delta a \label{22VIII.bis}\\
%& = & \frac{\delta{m}}{\Xi^2} - \frac{2\Lambda}{3\Xi^3}\left[ r -\frac{\Lambda}{3}r^3+\frac{2ma^2}{r^2+a^2}-\frac12 m \right]a\delta a \\
%& = & \frac{\delta{m}}{\Xi^2} - \frac{\Lambda}{\Xi^3}ma\delta a \\
% &= & \frac{1}{\sqrt{\Xi}} \delta \left(\frac{m}{\Xi^{3/2}}\right) =
%  \delta  {\left(\frac{m}{\Xi^{2}}\right)}  -\left(\frac{m}{\Xi^{3/2}}\right) \delta \frac{1}{\sqrt{\Xi}}   \label{bis24V13.hor}
% \, .
%\end{eqnarray}
%We used the identity:
%\[ \Delta_r=0 \Rightarrow r -\frac{\Lambda}{3}r^3+\frac{2ma^2}{r^2+a^2} = 2m \]
%
% \ptcr{z tych dwoch rachunkow wpisz ten rachunek ktory wolisz, a to co prawdziwe ale na razie niepotrzebne wrzuc do pliku leftovers.tex; i tekst w pliku pisz prosze po angielsku; JJ: wpisuje jeszcze raz ten dowod}
%{\bf Alternative proof:}
\begin{eqnarray} \nonumber
  \nonumber
  & & \frac{a}{r^2+a^2}\delta\frac{ma}{\Xi^2} +\frac14 \frac{\partial_r\Delta_r}{\Xi(r^2+a^2)}\delta\left( \frac{r^2+a^2}{\Xi}\right) \\
   \nonumber
& = & \frac{a}{r^2+a^2}\delta\frac{ma}{\Xi^2} +\frac12 \frac{\partial_r\Delta_r}{\Xi^2(r^2+a^2)}\left(r\delta r +a\delta a\right) + \frac14 \frac{\partial_r\Delta_r}{\Xi}\delta\frac{1}{\Xi} \\ \nonumber
& = & \frac{1}{\Xi^2}\delta{m} - \frac{r\Lambda}{3\Xi^2}a\delta a
+ \frac1\Xi\left[ \frac{2ma^2}{r^2+a^2}+\frac12(r-m) -\frac{r\Lambda}{6}(2r^2+a^2)\right]\delta\frac{1}{\Xi} \\
% & = & \frac{\delta{m}}{\Xi^2} - \frac{\Lambda}{3\Xi^3}\left[ r\Xi+\frac{4ma^2}{r^2+a^2}+ r-m
%  -\frac{r\Lambda}{3}(2r^2+a^2)\right]a\delta a \label{22VIII.bis}\\
& = &   \delta {\left(\frac{m}{\Xi^{2}}\right)} +
\bigg(\underbrace{r-\frac\Lambda{3}r^3- \frac{2mr^2}{r^2+a^2}}_{\frac{r\Delta_r}{r^2+a^2}}-\frac12 m\bigg) \frac{1}{\Xi}\delta\frac{1}{\Xi} \label{altbis24V13.hor}
 \, ,
\end{eqnarray}
where the following identity is useful:
\[ -\frac\Lambda{3\Xi^2}a\delta a=\frac12\delta \frac{1}{\Xi} \, .\]

\Eq{24V13.hor} is equivalent to the statement:

\begin{Proposition}
 \label{P18VII14.1}
The flow generated by the vector field
$$
  X=\sqrt{\Xi}\partial_t
$$
is Hamiltonian, with Hamiltonian $H_X$ given by
\bel{18VII14.2}
 E:= H_{ \sqrt{\Xi}\partial_t} =
 \frac{m}{\Xi^{3/2}}
 \, .
\ee
\end{Proposition}

%\reject
%\input{massKdSvJJ18VII14}
%
%
%\reject

\subsection{Angular momentum}
 \label{ss26IV15.1}

There is a standard calculation which shows that $\theta_b(S_{t,r} ,\partial_\varphi)$ is, up to a multiplicative factor, the variation of total angular momentum, which we reproduce here for completeness; compare Proposition~\ref{P26IV15.2}. For this, let $X$ be any vector field tangent to $V$. The divergence theorem and  the vacuum vector constraint equation $D_k P^{ki}=0$ give%
\footnote{The minus sign in \eq{27IX13.7+} has been intriguing to us. In this context it is helpful to realize that in a field theory where the energy density is given by $T_{00}$, energy-conservation in its usual form requires the momentum density to be defined as $-T_{0i}$. A simple consistency check is provided by a massless scalar field of the form $\phi=f(x-vt)$, where the minus sign is needed to have the field momentum positively directed along the velocity vector. Compare~\cite{BogShi}.}
\bel{27IX13.7+}
 -J:= \frac 1 { \gamma}\intdvol   P^i{}_j X^j dS_i =   \frac 1 { \gamma}\intvol  P^{ij} D_{i}X_j =   \frac 1 {2\gamma}\intvol  P^{ij}\mymcL_X g_{ij}
 \, .
\ee
Hence, if $X$ does not depend upon the metric and is further tangent to $\partial \mcV$,
\bean
 -\delta J &= &   \frac 1 {2\gamma}\intvol  \delta P^{ij}\mymcL_X g_{ij}
  +  P^{ij}\mymcL_X \delta g_{ij}
  % { + P^{ij}\mymcL_{\delta X} \delta g_{ij}})
\\
 &= &     \frac 1 {2\gamma}\intvol  \delta P^{ij}\mymcL_X g_{ij}
  + \underbrace{\mymcL_X( P^{ij} \delta g_{ij})}_{D_k(X^k  P^{ij} \delta g_{ij})} -  \mymcL_X  P^{ij}\delta g_{ij}
%  { + P^{ij}\mymcL_{\delta X} \delta g_{ij}} \big)
   \nonumber
\\
 & = &     \frac 1 {2\gamma}\intvol   \delta P^{ij}\mymcL_X g_{ij}
  -  \mymcL_X  P^{ij}\delta g_{ij}
   \big)
%{ +\frac1{\gamma}\intdvol   P^i{_j} \delta X^j dS_i}
    \nonumber
\\
    &&
    +\frac 1 {2\gamma} \intdvol   P^{ij} \delta g_{ij}   \underbrace{X^k dS_k}_0
 \, ,
  \phantom{xxx}
\eeal{27IX13.8+}
providing the usual Hamiltonian formula for angular momentum:
\bean
 -\delta J
 & = &     \frac 1 {2\gamma}\intvol  \frac{\partial g_{ij}}{\partial \varphi}\delta P^{ij}
  -  \frac{\partial P^{ij}}{\partial \varphi}\delta g_{ij}
 \, .
  \phantom{xxx}
\eeal{27IX13.8++}
%
%
%{
%\[
% -\delta J -\frac1{\gamma}\intdvol   P^i{_j} \delta X^j dS_i
% = \frac1{\gamma}\intdvol ( \delta P^i{_j}) X^j dS_i
%\]
%}
%
Comparing with \eq{5VI15.21},  we obtain an alternative proof of Proposition~\ref{P26IV15.2} for vector fields tangent to $S$:
\bel{28III15.1}
\theta_b(S_{t,r},\partial_\varphi) = - 2 \gamma \delta J
 \,.
\ee

When $X$ remains tangent to $\mcV$ but is allowed to depend on the metric,
there arises a supplementary term
\bel{27IX13.13}
   \frac 1 {2\gamma}\intvol  P^{ij}\mymcL_{\delta X} g_{ij} =  \frac 1 { \gamma}\intdvol   P^i{}_j \delta X^j dS_i
 \, .
\ee
We see that whenever $\delta X$ is a Killing vector (which is the case in our considerations here), or when $\delta X$ vanishes at
$\boundaryimage $, the integral \eq{27IX13.13} vanishes.

{}From \eq{18V13.21} we obtain
\bel{27IX13.11}
 \theta_b(S_{t,r},\partial_\varphi) =   -2 \gamma \delta J
  = -16 \pi\delta\left(
  \frac{m a}{\Xi^2}
   \right)
  \, .
\ee

\subsection{The Henneaux-Teitelboim vector field}
 \label{ss27III15.1}

In~\cite{HT}, in the closely related case of negative cosmological constant, the authors consider the vector field
\bel{27IX13.2}
 X_{HT}:= \partial_t + \frac {\Lambda a}{3\Xi} \partial_\varphi
 \, .
\ee
The volume part of the variational formula \eq{homogeneous} is linear in $X$, which gives
\bel{27IX13.5}
 \theta_b(S_{t,r} ,X_{HT})= \theta_b(S_{t,r} ,\partial_t)+ \frac {\Lambda a}{3\Xi}\theta_b(S_{t,r} ,\partial_\varphi)
 \, .
\ee
%
%
% \ptc{rachunek jacka w pliku, commented out}
%
%dla pola $d/d\phi$ mozna wziac wiaz wektorowy:
%$P^k_\phi,k -\frac12 P^{kl}g_{kl,\phi}=0$
%i dla formy hamiltonowskiej otrzymujemy:
%$$
%\intvol  P^{kl,\phi}\delta g_{kl} - g_{kl,\phi}\delta P^{kl} =
%- \intvol  P^{kl}\delta g_{kl,\phi} + g_{kl,\phi}\delta P^{kl} =
%- \delta \intvol  P^{kl} g_{kl,\phi}= -2 \delta \intvol  P^k_\phi,k
%= -2 \delta \intdvol  P^r_\phi
%$$
%
Hence,
\bean
  \theta_b(S_{t,r} ,X_{HT}) &=& 16 \pi \left[
  \frac {1} {\Xi^{1/2}} \delta \left(\frac {m }{\Xi^{3/2}}\right)
  - \frac { \Lambda a}{3\Xi}\delta\left(
  \frac{ m a}{\Xi^2}
   \right)
   \right]
\\
 & = &
  \frac { 16 \pi} {\Xi } \delta \left(\frac {m }{\Xi^{2}}\right)
 \, .
\eeal{27IX13.12}
We see that $X_{HT}$ is \emph{not} Hamiltonian.

On the other hand, \eq{27IX13.12} and our remaining considerations so far show that:

\begin{Proposition}
   \label{P18VII14.2}
The vector field
\bel{28III15.2}
 X=\Xi X_{HT} \equiv \Xi \partial_t + \frac {\Lambda a}{3} \partial_\varphi
\ee
generates a Hamiltonian flow on the space of metrics, with Hamiltonian $H_X$ given by
\bel{18VII14.6}
 H_{ \Xi X_{HT} }
  =
  \frac m { \Xi^2}
  \, .
\ee
 \end{Proposition}

%\ptc{the file leftover.tex contains various leftovers,  for recycling purposes, including some details of the asymptotic expansion calculations}
%\input{leftovers}

\subsection{Kerr-Schild coordinates}
 \label{ss20VII14.1}

In~\cite[p.102]{CarterlesHouches} it is emphasized that the Kerr-de Sitter metric is \emph{asymptotically de Sitter} as it tends to the de Sitter metric in the limit
as $r$ goes to infinity. One way of making this precise is to invoke the Kerr-Schild coordinates, with a de Sitter background metric.
Following~\cite{GibbonsGen,AMatzner} we use the coordinate transformation
%X
\bean
    d\tau &=& \frac{1}{\Xi}dt+\frac{2 m r}{\left(1-\frac{r^2 \Lambda}{3}\right)\Delta_r} dr\, , \\
    d\phi &=& d\varphi- \frac{a\Lambda}{3 \Xi}dt+\frac{2 m r \;  a}{(r^2+a^2)\Delta_r}dr
\eea
to obtain
\be
  \label{kerrschild}
  g = g_{dS}+ \frac{2 m r}{\rho^2}(k_\mu dx^{\mu})^2 \, ,
\ee
with
\bean
    g_{dS} &=& -\frac{\left(1-\frac{r^2 \Lambda}{3}\right)\Delta_\theta}{\Xi}d\tau^2
    +\frac{\rho^2}{\left(1-\frac{r^2\Lambda}{3}\right)(r^2+a^2)}dr^2
    +\frac{\rho^2}{\Delta_\theta}d\theta^2 \\
      &\;&  +\frac{(r^2+a^2)\sin^2(\theta)}{\Xi}d\phi^2\, , \\
    k_\mu dx^{\mu} &=& \frac{\Delta_\theta}{\Xi}d\tau+\frac{\rho^2}{\left(1-\frac{r^2\Lambda}{3}\right)(r^2+a^2)}dr
    -\frac{a \sin^2(\theta)}{\Xi} d\phi \, .
  \label{ds.schild}
\eea
The metric
$g_{dS}$ is the de Sitter metric in unusual coordinates, which can be verified by using~\cite{CarterlesHouches,GibbonsGen} to find the transformation
\bea
  \nonumber
    R^2 &=& \frac{r^2\Delta_\theta+a^2 \sin^2(\theta)}{\Xi} \, , \\
  \nonumber
    R^2 \sin^2(\Theta)&=&\frac{r^2+a^2}{\Xi}\sin^2(\theta) \, , \\
  \nonumber
    T &=& \tau  \, , \phantom{\frac1{3}} \\
  \nonumber
    \Phi &=& \phi \phantom{\frac1{3}}
\eea
between \eq{ds.schild} and the de Sitter metric in static coordinates:
\begin{eqnarray}
  \label{ads.static}
  g_{dS} = - (1-\frac{\Lambda R^2}{3}) dT^2 + \frac1{1-\frac{\Lambda R^2}{3}} dR^2 + R^2\left(d\Theta^2+\sin^2(\Theta) d\Phi^2\right) \, .
\end{eqnarray}
The vector $k^\mu\partial_\mu$ is  null  both for $g$ and $g_{dS}$.

According to the above, the natural ``time-evolution'' Killing vector $X_{\mathrm{KS}}$ in the Kerr-Schild setting is
\bel{29IX13.1}
 X_{\mathrm{KS}} \equiv \partial_\tau =
 \Xi \partial_t + \frac{a\Lambda}3 \partial_\varphi
 \, .
\ee
This coincides with the vector field \eq{28III15.2}, with the associated Hamiltonian given by \eq{18VII14.6}. We can also calculate directly:
% \ptcr{rachunek ptc in mathematica 28 III 15}
%
\bean
 \theta_b(S_{t,r} ,X_{\mathrm{KS}}) &= &  \Xi \theta_b(S_{t,r} ,\partial_t)+\frac {\Lambda a}{3}\theta_b(S_{t,r} ,\partial_\varphi)
\\
 & = &
    {16 \pi}\left[ {\Xi^{1/2}} \delta \left(\frac {m }{\Xi^{3/2}}\right)
  - \frac { \Lambda a}{3}\delta\left(
  \frac{ m a}{\Xi^2}
   \right)\right]
   \nonumber
\\
 & = &
    {16 \pi}  \delta \left(\frac {m }{\Xi^{2}}
   \right)
 \, .
\eeal{29IX13.3}

\section{All Killing vector fields generating Hamiltonian flows}
 \label{s28III15.1}

We have seen that the ``energy functional''
$$
 E:= \frac{m}{\Xi^{3/2}}
$$
generates the flow of the vector field   $\sqrt{\Xi}\partial_ t$, while the  ``angular momentum functional''
$$
 J :=  \frac{ma}{\Xi^2}
$$
generates the flow of $-\partial_\varphi$. This allows us to prove the following:

\begin{theorem}
   \label{T18VII14.1}
A vector field $X$ defined along an initial data hypersurface $\hyp$ which asymptotes to a Kerr-de Sitter Killing vector field
as one recedes to infinity along an asymptotically KdS end,
\bel{17IV14.2}
 X \to \alphaX  \sqrt{\Xi}\partial_t +\beta \partial_\varphi
 \, ,
\ee
where $\alphaX $ and $\beta$ are allowed to depend upon $J$ and $E$, but not upon coordinates, generates a Hamiltonian flow on the space of initial data sets with one asymptotically Kerr-de Sitter end if and only if
\bel{17IV14/1}
 \frac{\partial\alphaX }{\partial J}+\frac{\partial\beta}{\partial E}=0 \, .
\ee
If \eq{17IV14/1} holds, the Hamiltonian $H_X= H_X(E,J)$ generating the flow of initial data associated to $X$ is the unique, up to a constant, solution of the equations
\bel{19VII14.1}
 \partial_E H_X = \chi
  \, ,
   \quad
   \partial_J H_X = -\beta
   \, .
\ee
\end{theorem}
%jest polem hamiltonowskim jeÅ›li

\proof
%
 %\ptcr{proof rewritten}
Let us denote by $\mcV_R\subset \mcS$ a family of domains with smooth boundaries $S_R$ such that $S_R$ is homologous to spheres of constant $t$ and $r$ in the asymptotically KdS  region of $\mcS$, with $\cup_{R} \mcV_R =\hyp$.

Let $X_1$ denote any vector field which asymptotes to $\sqrt{\Xi}\partial_t$, and let $X_2$ denote any vector field which asymptotes to $\partial_\varphi$.
Then the vector field
$$
 X_3:=X-\chi X_1 - \beta X_2
$$
asymptotes to zero.

We have seen that, for all vacuum variations of asymptotically KdS initial data it holds that
\bean
 \lefteqn{
\lim_{R\to\infty} \chi  \intvolR   \left({\cal L}_{X_1} g_{kl} \delta P^{kl} - {\cal L}_{X_1} P^{kl}\delta g_{kl}\right)
 + 2
\underbrace{\lim_{R\to\infty}\chi  \intdvolR  ({\cal L}_{X^0_1\partial_0} \alpha\delta\lambda - {\cal L}_{X^0_1\partial_0} \lambda \delta \alpha )
 }_0
 }
 &&
\\
 &  &
 \phantom{xxxxxxxx}
  =\lim_{R\to\infty} \chi \theta_b(S_R,X_1) =\chi  dE
   \, ,
 \phantom{xxxxxxxxxxxxxxxxxxxxxxxxxxxxxx}
  \label{2VII15.2}
\\
 \lefteqn{
\lim_{R\to\infty} \beta \intvolR   \left({\cal L}_{X_2} g_{kl} \delta P^{kl} - {\cal L}_{X_2} P^{kl}\delta g_{kl}\right)
 + 2
\underbrace{\lim_{R\to\infty}\beta \intdvolR  ({\cal L}_{X^0_2\partial_0} \alpha\delta\lambda - {\cal L}_{X^0_2\partial_0} \lambda \delta \alpha )
 }_0
 }
 &&
  \nonumber
\\
 &  &
 \phantom{xxxxxxxx}
  =\lim_{R\to\infty} \beta \theta_b(S_R,X_2) =
   - \beta  dJ
   \, .
 \phantom{xxxxxxxxxxxxxxxxxxxxxxxxxxxxxx}
  \label{2VII15.3}
\\
 \lefteqn{
\lim_{R\to\infty} \intvolR   \left({\cal L}_{X_3} g_{kl} \delta P^{kl} - {\cal L}_{X_3} P^{kl}\delta g_{kl}\right)
 + 2
\underbrace{\lim_{R\to\infty} \intdvolR  ({\cal L}_{X^0_3\partial_0} \alpha\delta\lambda - {\cal L}_{X^0_3\partial_0} \lambda \delta \alpha )
 }_0
 }
 &&
  \nonumber
\\
 &  &
 \phantom{xxxxxxxx}
  =\lim_{R\to\infty} \theta_b(S_R,X_3) = 0
   \, .
 \phantom{xxxxxxxxxxxxxxxxxxxxxxxxxxxxxx}
  \label{2VII15.4}
\eea
(In the first two equations, the boundary integrals involving $\alpha$ and $\lambda$ vanish because $\partial_t$ and $\partial_\varphi$ are asymptotic Killing vectors; in the last equation, all boundary integrals vanish because $X_3$ asymptotes to zero.)
It easily follows that
\begin{eqnarray}
 \alphaX  \delta E - \beta \delta J
 & = &
{
\lim_{R\to\infty}  \intvolR   \left({\cal L}_{X} g_{kl} \delta P^{kl} - {\cal L}_{X} P^{kl}\delta g_{kl}\right)
 }
  \label{homogeneousc}
\end{eqnarray}
The left-hand side  will be equal to $\delta H_X$, for some function $H_X=H_X(E,J)$, if and only if \eq{17IV14/1} holds.
\qedskip

The theorem implies that neither the field $\partial_t$, nor the Henneaux-Teitelboim field $X_{HT}$, generate Hamiltonian flows
on the space of metrics.

A corollary of Theorem~\ref{T18VII14.1} is that every Killing vector field of the Kerr-de Sitter metric can be rescaled by an $m$-- and $a$--dependent factor so that the rescaled field is Hamiltonian. Indeed, consider a vector field of the form \eq{17IV14.2} which does not satisfy \eq{17IV14/1}. Then the vector field
\bel{17IV14.3}
 \gamma(\alphaX \sqrt{\Xi}\partial/\partial t +\beta \partial/\partial\varphi)
\ee
%]
will be Hamiltonian if and only if $\gamma$ is a solution of the ODE
\bel{17IV14.4}
 \bigg( \alphaX\frac{\partial}{\partial J}+\beta\frac{\partial}{\partial E}
   +
  \frac{\partial\alphaX}{\partial J}
 +
 \frac{\partial\beta}{\partial E}
  \bigg)\gamma
 = 0
   \, .
\ee
This equation can always be solved locally  using the method of characteristics.

Note that many such rescalings exist: Suppose that $X$ has been rescaled so that $\gamma X$ has a Hamiltonian $H_{\gamma X}$. Then for any function $f$ the vector field $f'(H_{\gamma X}) \gamma X$ is Hamiltonian, with Hamiltonian
$$
 H_{f' \gamma X} = f(H_{\gamma X})
 \, .
$$

As an example, we note that the vector field
$$
 \mbox{$\displaystyle \partial_t  - \frac {\Lambda a}{3\Xi} \partial_\varphi $=``$
 X_{\mathrm{KS}}$ with $a$ replaced by its negative''
  }
$$
is not Hamiltonian. But the Killing vector field
$\displaystyle\frac1{\sqrt{2\Xi-1}}\left(\partial_t  - \frac {\Lambda a}{3\Xi} \partial_\varphi\right)$ is, with Hamiltonian
$\displaystyle\frac{m}{\Xi^2}\sqrt{2\Xi-1}$. Similarly
$\displaystyle\frac m{\Xi^2}\left(\partial_t  - \frac {\Lambda a}{3\Xi} \partial_\varphi\right)$ is, with Hamiltonian
$\displaystyle\frac{m^2}{2 \Xi^2} ({2\Xi-1})$.

Further examples of Hamiltonian vector fields include
\bel{5VII15.1}
 \frac{\partial a}{\partial M} X_{\mathrm{KS}} -
  \frac{\partial a}{\partial J} \partial_\varphi
   =
-\frac{J}{M^2}X_{\mathrm{KS}}
- \frac 1 M \partial_\varphi
\,,
\ee
generated by $H=a\equiv J/M$, and
\bel{5VII15.2}
 \frac{\partial m}{\partial M} X_{\mathrm{KS}} -
  \frac{\partial m}{\partial J} \partial_\varphi
   =
    \Xi\left[
 \left(1 - \frac{\Lambda J^2}{M^2}\right)
    X_{\mathrm{KS}}
    - \frac {2\Lambda J}{3 M} \partial_\varphi
    \right]
\,,
\ee
generated by $H=m\equiv M [1+ \Lambda J^2/(3M^2)]^2$.

\section{Black hole thermodynamics}
 \label{ss18VII14.5}

``Black hole thermodynamics'' can be thought of as considerations based on variational identities involving  global invariants such as mass and angular momentum and local invariants associated to event horizons. In stationary spacetimes with a preferred ``stationary'' Killing vector, say $\mathring X$, {and} a Killing horizon, one  proceeds as follows: One defines the total energy $E$ as the Hamiltonian associated to the flow generated by $\mathring X$. If $\mathring X$ is tangent to the generators of a horizon with compact cross-sections, a rather general calculation shows that
\bel{28III15.4}
 \delta E =  \frac{\kappa}{8 \pi} \delta A
 \,,
\ee
where $A$ is the area of the section of the horizon, and $\kappa$ is the surface gravity.
If, however, $\mathring X$ is not tangent to the generators, then there must exist another Killing vector $X$ tangent to the generators. In spacetime dimension four, mild supplementary conditions guarantee that one can find a linear combination, say $\eta=\partial_\varphi$, of $\mathring X$ and $X$ which has $2\pi$ periodic orbits:
\bel{28III15.5}
 \mathring X = X - \Omega \eta
 \,.
\ee
Note that a specific normalisation of $X$, determined by $\mathring X$, has been chosen in \eq{28III15.5}.
\Eq{28III15.4} is now replaced by
\bel{28III15.6}
 \delta E = \frac{\kappa}{8 \pi}  \delta A + \Omega \delta J
 \,.
\ee

As already discussed in e.g.~\cite{DolanKastor},
 %\ptcr{ref added 5 VII}
a basic problem with this program for Kerr-de Sitter spacetimes is the choice of the Killing vector $\mathring X$ above. Let us illustrate this with some examples.

\subsection{$\partial_t$}
 \label{ss28III15.1}

Suppose, first, that we decide to use the $\partial_t$-vector field of the Boyer-Lindquist coordinates as the preferred Killing vector field $\mathring X$. As we have seen that this Killing vector is not Hamiltonian,  the prescription cannot be applied.
This last problem can be fixed by taking instead $\mathring X= \sqrt{\Xi} \partial_t$, which is associated with the Hamiltonian $  m \Xi^{-3/2}$. We will return to this choice in Section~\ref{ss28III15.3}.

\subsection{$\partial_T$}
 \label{ss28III15.2}

Yet another choice,  which is directly Hamiltonian  without the need to do any rescalings, is the Kerr-Schild  vector field $X_{\mathrm{KS}}$ of \eq{28III15.2}, and let us therefore investigate this choice in detail.

First, some general considerations are in order. In maximally extended Kerr-de Sitter spacetimes, the Killing horizons are located at
$$\Delta_r\equiv (r^2+a^2)\left(1-\frac{\Lambda}3 r^2\right)-2 m r=0
 \, .
$$
%.
Let $r_*$ be a root of this equation, then it is easily seen from \eq{kds.metric} that the Killing vector field
\bean
 X_* &=& \Xi \left(\partial_t + \frac{a}{a^2+r_*^2} \partial_\varphi\right) =
 X_{\mathrm{KS}} + \left( \frac{a \Xi}{a^2+r_*^2} - a \frac \Lambda 3\right)
 \partial_\varphi
 = \partial_T +\underbrace{\frac{a \left(3- \Lambda
   r^2_* \right)}{3
   \left(a^2+r^2_*\right)} }_{=:\Omega_*}\partial_\varphi
\\
 & =&
 X_{\mathrm{KS}} +\underbrace{\frac{( 2 m a\, r_*)}{
   \left(a^2+r^2_*\right)^2} }_{ \Omega_*}\partial_\varphi
% \equiv  X_{\mathrm{KS}} + \frac{a}{
%   \left(a^2+r^2_*\right) } \left (1 - \frac{\Lambda r_*^2}{2}\right)\partial_\varphi
\eeal{3X13.2}
is tangent to the generators of the corresponding Killing horizon. (In the last equality above the equation satisfied by $r_*$ has been used.)

We have seen that the  Hamiltonian  mass $M_H$ associated with $X_{\mathrm{KS}}$ is
\bel{3X13.3}
 M_H = \frac {m}{\Xi^2}
 \, ,
\ee
and that the total angular momentum $J$ is
\bel{3X13.3J}
 J  =  \frac {ma}{\Xi^2} =   M_H a
 \, .
\ee

Using \eq{4V13.1}, the area $A_*$ of the horizon is
\bel{3X13.4}
 A_*= 2 \pi \int_0^\pi \lambda|_{r=r_*} \,d\theta =  \frac{2 \pi(r^2_*+a^2)} \Xi  \int_0^\pi     \sin (\theta)
        \,d\theta
       = \frac{4 \pi(r^2_*+a^2)} \Xi
       \, .
\ee

As such,   the surface gravity $\kappa_*$ of a Killing horizon $\mcH_*$ can be defined through the formula
\bel{5X13.1}
\partial_\mu(X^\alpha X_\alpha)|_{\mcH_*} = -2 \kappa_* X_\mu
 \, .
\ee
A convenient procedure to calculate $\kappa_*$ proceeds as follows: let $b$ be any one-form which extends smoothly across the horizon
and such that $b(X)=1$. Then $\kappa_*$ can be obtained as the value at the horizon of minus one half of the left-hand side of \eq{5X13.1}:
\bel{28III15.11}
 \kappa_*= - \frac 12 b\big(\nabla (X^\alpha X_\alpha)\big)|_{\mcH_*}
  \,.
\ee
Note that the left-hand side is independent of the choice of $b$, and so is therefore the right-hand side.

Recall~\cite{CarterlesHouches}
that an extension of the Kerr-de Sitter metric \eq{kds.metric} across a Killing horizon  can be obtained by introducing new coordinates
\bea
    d\hat{v} &=& dt + \Xi \frac{r^2+a^2}{\Delta_r} dr \, , \label{1XII12.1} \\
    d\hat{\varphi}&=&d\varphi + \Xi \frac{a}{\Delta_r} dr \, . \label{1XII.12.2}
\eea
Hence the one form
$$
 b = \frac 1 \Xi dt +   \frac{r^2+a^2}{\Delta_r} dr
$$
extends smoothly across the Killing horizon $\mcH_*$ and satisfies $b(X_*)=1$. Thus
\bean
 \kappa_* &= &-  \lim_{r\to r_*} \frac{(r^2+a^2)}{2\Delta_r} g^{r\mu}\partial_\mu(g(X_*,X_*)\big)
\\
 &= &
 -  \lim_{r\to r_*} \frac{(r^2+a^2)}{2\Delta_r} g^{rr}\partial_r(g(X_*,X_*)\big)
  \nonumber
\\
 &= &
   \frac{\partial_r \Delta_r}{2(r^2+a^2)} \bigg|_{r=r_*}
  \, .
\eeal{5X13.2}
Here we have used
% \ptcr{this is wrong, because \eq{5X13.4} should depend upon $r_*$ but does not; the calculation appears to come from a Christa file thermodnamics Kds.nb}
%
\bean%
% g(X_*,X_*) & = &
% -\frac{\text{$\Delta $r}(r) \left(a^2
%   \cos (2 \theta )+a^2+2 r^2\right)}{2
%   \left(a^2+r^2\right)^2}
%   \, ,
%\\
%  \partial_r(g(X_*,X_*)\big) &
% = & \frac{4 r \text{$\Delta $r}(r) \left(a^2
%   \cos (2 \theta
%   )+r^2\right)-\left(a^2+r^2\right)
%   \text{$\Delta $r}'(r) \left(a^2 \cos
%   (2 \theta )+a^2+2 r^2\right)}{2
%   \left(a^2+r^2\right)^3}
%   \, , \nonumber \\
% & &
%  %\displaystyle
   g^{rr} & = & \frac{\Delta_r}{r^2+a^2\cos^2\theta}
 \, ,
 \\
 g(X_*,X_*)  &= &
-\frac{\Delta_r \left(r_*^2+a^2\cos^2\theta\right)}{\left(a^2+r_*^2\right)^2} +O((r-r_*)^2)
   \, .
\eeal{5X13.4}

We claim that we have, consistently with what has been said above,
\be\label{tbh1}
 \underbrace{ \left(1-{ \frac\Lambda{3}r^2}\right)\frac{a}{r^2+a^2}
  }_{\Omega_*}
\underbrace{ \delta \frac{ma}{\Xi^2}}_{\delta J}  + \underbrace{\frac14 \frac{\partial_r\Delta_r}{ (r^2+a^2)}}_{\kappa_*/2}\delta
    \underbrace{\left( \frac{r^2+a^2}{\Xi}\right)}_{A_*/ 4 \pi}
  =
    \underbrace{\delta \left(\frac{m}{\Xi^{2}}\right)}_{\delta M_H}
\,,
  \ee
where $r=r_*$. Indeed, our variational identities lead to (compare~\cite{LarranagaMojica,GibbonsHawkingCEH,SudarskyWald93,Sudarsky:Wald,BoussodeSitter})
\bel{28III15.4+}
 \delta M_H = \frac 1 {8\pi} \kappa_* \delta A + \Omega_* \delta J
 \,.
\ee
To prove \eq{tbh1},
we rewrite \eq{24V13.hor}  in the form
\bel{28III15.15}
 \frac{a\sqrt{\Xi}}{r^2+a^2}\delta\frac{ma}{\Xi^2} +\frac14\frac{\partial_r\Delta_r}{\sqrt{\Xi}(r^2+a^2)}\delta\left( \frac{r^2+a^2}{\Xi}\right)
  =  \delta \left(\frac{m}{\Xi^{3/2}}\right)
 \,.
\ee
Dividing by $\sqrt\Xi$, we are lead to
\be \frac{a}{r^2+a^2}\delta\frac{ma}{\Xi^2} +\frac14 \frac{\partial_r\Delta_r}{\Xi(r^2+a^2)}\delta\left( \frac{r^2+a^2}{\Xi}\right) = \delta {\left(\frac{m}{\Xi^{2}}\right)} +\frac{m}{2\Xi^3}\delta {\Xi}
 \,.
  \label{28III15.12}
\ee
Inserting the identity
\be \frac{m}{2\Xi^3}\delta {\Xi} = \frac\Lambda{3}\frac{m}{\Xi^3}a\delta a= \frac\Lambda{3}
\left[ r^2\frac{a}{r^2+a^2}\delta\frac{ma}{\Xi^2} - a^2\frac{\partial_r\Delta_r}{4\Xi(r^2+a^2)}\delta\left( \frac{r^2+a^2}{\Xi}\right)\right]
% \,
\ee
into \eq{28III15.12} one obtains \eq{tbh1}.

\subsection{$\sqrt \Xi\partial_t$}
 \label{ss28III15.3}

The dynamics associated with the vector field $\sqrt \Xi \partial _t$ of Proposition~\ref{P18VII14.1} is generated by the Hamiltonian
$$
 \hat M_H:= m \Xi^{-3/2}
 \,.
$$
%.
We have the decomposition
\bel{28III15.14}
 \sqrt\Xi\partial_t
  =
   \sqrt  \Xi \left(\partial_t + \frac{a}{a^2+r_*^2} \partial_\varphi\right)
   -   \underbrace{\frac{a\sqrt  \Xi }{a^2+r_*^2}}_{=:\hat \Omega_*} \partial_\varphi
   = \underbrace{\Xi^{-1/2} X_*}_{=:\hat X _*} - \hat \Omega_* \partial_\varphi
   \,,
\ee
where $\hat X_*$ is tangent to the generators of $\mcH_*$. The surface gravity of $\hat X_*$ is $\hat \kappa_* = \Xi^{-1/2} \kappa_*$, with $\kappa_*$ given by \eq{5X13.2}. By inspection of \eq{28III15.15} one obtains:
\bel{28III15.15+}
 \delta \hat M_H = \frac 1 {8\pi} \hat \kappa_* \delta A + \hat \Omega_* \delta J
 \,.
\ee

\input{ShortAreaHamiltonian}

\appendix

\input{Notations}
\input{Liederivatives}

\input{invarianceJK}

\input{SpacelikeVectorsJK}

\input{NegativeLambda}
\input{CYK}

%\input{SpaceTimeFormulaeForMass}

%\ptc{see recycling.tex for recycling material}
%\input{recycling}

\bigskip

{\noindent \sc Acknowledgements}
We are grateful to Christa Raphaela \"Olz for providing a figure,
and to Bobby Beig for useful discussions. Supported in part by Narodowe Centrum Nauki under the grant DEC-2011/03/B/ST1/02625
and the Austrian Science Fund (FWF) under project P 23719-N16.  JJ and JK wish to thank the Erwin Schr\"odinger Institute, Vienna, for hospitality and support during part of work on this paper.

\bibliographystyle{amsplain}
\bibliography{%
../references/reffile,%
../references/newbiblio,%
../references/hip_bib,%
../references/newbiblio2,%
../references/bibl,%
../references/howard,%
../references/bartnik,%
../references/myGR,%
../references/newbib,%
../references/Energy,%
../references/netbiblio,%
../references/PDE}

\end{document}

%% file: LeeNewcommand.tex
\input{switches}

\input{fixmes}

\input{renewcommands}

\newcommand{\partialstot}{\partial_t}

\newcommand{\eqref}[1]{(\ref{#1})}

\newcommand{\intdvolV}{\int_{\partial V}}
\newcommand{\intvolV}{\int_V}

\newcommand{\perpQ}{\nu {\bf Q}}

\newcommand{\mymcL}{{\color{red}{\mcL}}}

\newcommand{\ttg}{\tilde{\widetilde g}{}}

\newcommand{\boundaryimage}{{\color{red} {\partial \mcV}}}% was \psi_\lambda(\{t\}\times \partial V)
\newcommand{\intvol}{{\color{red}\int_{\mcV}}}
\newcommand{\intdvol}{{\color{red}\int_{\partial \mcV}}}
\newcommand{\intvolR}{{\color{red}\int_{\mcV_R}}}
\newcommand{\intdvolR}{{\color{red}\int_{S_R}}}

\newcommand{\nthree}{\nu}

\newtheorem{theorem}{\sc Theorem}[section]

\newtheorem{Proposition}[theorem]{\sc Proposition}

\newtheorem{Remarks}[theorem]{\sc Remarks}
\newtheorem{Theorem}[theorem]{\sc Theorem}

\newcommand{\hs}{\cH_{\mbox{\scriptsize sing}}}

\newcommand{\beadl}[1]{\begin{deqarr}\label{#1}}
\newcommand{\eeadl}[1]{\arrlabel{#1}\end{deqarr}}%
\def\nz{\ifmmode {I\hskip -3pt N} \else {\hbox {$I\hskip -3pt N$}}\fi}
\def\zz{\ifmmode {Z\hskip -4.8pt Z} \else
       {\hbox {$Z\hskip -4.8pt Z$}}\fi}
\def\qz{\ifmmode {Q\hskip -5.0pt\vrule height6.0pt depth 0pt
       \hskip 6pt} \else {\hbox
       {$Q\hskip -5.0pt\vrule height6.0pt depth 0pt\hskip 6pt$}}\fi}
\def\rz{\ifmmode {I\hskip -3pt R} \else {\hbox {$I\hskip -3pt R$}}\fi}
\def\cz{\ifmmode {C\hskip -4.8pt\vrule height5.8pt\hskip 6.3pt} \else
       {\hbox {$C\hskip -4.8pt\vrule height5.8pt\hskip 6.3pt$}}\fi}
\def\au{{\setbox0=\hbox{\lower1.36775ex\hbox{''}\kern-.05em}\dp0=.36775ex\hs
kip0pt\box0}}
\def\ao{{}\kern-.10em\hbox{``}}

\newcommand\Gregbeq{\begin{eqnarray}}
\newcommand\Gregeeq{\end{eqnarray}}

\def\cH{{\cal H}}

\def\h1{{\hat 1}}
\def\h2{{\hat 2}}

\def\3f{\frac{3}{2}}

\newcommand{\roscoff}[1]{}

{\catcode `\@=11 \global\let\AddToReset=\@addtoreset}
\AddToReset{figure}{section}

\DeclareFontFamily{OT1}{rsfs}{}
\DeclareFontShape{OT1}{rsfs}{m}{n}{ <-7> rsfs5 <7-10> rsfs7 <10-> rsfs10}{}
\DeclareMathAlphabet{\mycal}{OT1}{rsfs}{m}{n}

{\catcode `\@=11 \global\let\AddToReset=\@addtoreset}
\AddToReset{equation}{section}
\renewcommand{\theequation}{\thesection.\arabic{equation}}
\newcounter{mnotecount}[section]

\renewcommand{\themnotecount}{\thesection.\arabic{mnotecount}}

\def\pa{\partial}
\newcommand{\jlcax}[1]{}
\newcommand{\eean}{\nonumber\end{eqnarray}}

\newcommand{\kk}[1]{}%{\mnote{{\bf If we consider the KK case:} #1}}

\newcommand{\mcH}{{\mycal H}}

\newcommand{\beq}{\begin{equation}}
\newcommand{\sgn}{\ \mbox{sgn}}

\newcommand{\rd}{\,{ d}} % exterior differential

\newcommand{\FS}       %{F_1} %
                  {F}
\newcommand{\HS} %{F_2}
       {H_{\mbox{\scriptsize volume}}}

{\ptc{this should be removed in the oberwolfach version}}%

\newcommand{\mcE}{{\mycal E}}%

\newcommand{\eeal}[1]{\label{#1}\end{eqnarray}}
\newcommand{\bed}{\begin{deqarr}}
\newcommand{\eed}{\end{deqarr}}
\newcommand{\bedl}[1]{\begin{deqarr}\label{#1}}
\newcommand{\eedl}[2]{\arrlabel{#1}\label{#2}\end{deqarr}}

\newcommand{\tg}{{\widetilde{g}}}
\newcommand{\bel}[1]{\begin{equation}\label{#1}}
\newcommand{\bea}{\begin{eqnarray}}
\newcommand{\bean}{\begin{eqnarray}\nonumber}
\newcommand{\beal}[1]{\begin{eqnarray}\label{#1}}
\newcommand{\eea}{\end{eqnarray}}

\newcommand{\nn}{\nonumber}
\newcommand{\Eq}[1]{Equation~\eq{#1}}

\newcommand{\qed}{\hfill $\Box$}
\newcommand{\qedskip}{\hfill $\Box$\medskip}
\newcommand{\proof}{\noindent {\sc Proof:\ }}
\newcommand{\be}{\begin{equation}}
\newcommand{\eeq}{\end{equation}}
\newcommand{\ee}{\end{equation}}
\newcommand{\beqa}{\begin{eqnarray}}
\newcommand{\eeqa}{\end{eqnarray}}
\newcommand{\beqan}{\begin{eqnarray*}}
\newcommand{\eeqan}{\end{eqnarray*}}
\newcommand{\ba}{\begin{array}}
\newcommand{\ea}{\end{array}}

\newcommand{\const}{\mbox{\rm const}} %constants

\newcommand{\hyp}{\mycal S}

\newcommand{\mcM}{{\mycal M}}

\newcommand{\mcV}{{\mycal V}}

\newcommand{\mnote}[1]%{}
{\protect{\stepcounter{mnotecount}}$^{\mbox{\footnotesize
$%\!\!\!\!\!\!\,
\bullet$\themnotecount}}$ \marginpar{%\color{red}%
\raggedright\tiny\em
$\!\!\!\!\!\!\,\bullet$\themnotecount: #1} }

\newcommand{\warn}[1]%{}%{}
{\protect{\stepcounter{mnotecount}}$^{\mbox{\footnotesize
$%\!\!\!\!\!\!\,
\bullet$\themnotecount}}$ \marginpar{%\color{red}%
\raggedright\tiny\em
$\!\!\!\!\!\!\,\bullet$\themnotecount: {\bf Warning:} #1} }

\newcommand{\R}{\mathbb R}

\newcommand{\eq}[1]{(\ref{#1})}

\newcommand{\ptc}[1]{\mnote{{\bf ptc:}#1}}

\newcommand{\mcS}{{\mycal S}}
\newcommand{\mcL}{{\mycal L}}

\newcommand{\beqar}{\begin{deqarr}}
\newcommand{\eeqar}{\end{deqarr}}

\newcommand{\beaa}{\begin{eqnarray*}}
\newcommand{\eeaa}{\end{eqnarray*}}

\newcommand{\bethm}{\begin{Theorem}}
\newcommand{\et}{\end{Theorem}}
\newcommand{\bl}{\begin{Lemma}}

%% file: switches.tex
\newcommand{\mnotex}[1]%{}
{\protect{\stepcounter{mnotecount}}$^{\mbox{\footnotesize
$%\!\!\!\!\!\!\,
\bullet$\themnotecount}}$ \marginpar{%%
\raggedright\tiny\em
$\!\!\!\!\!\!\,\bullet$\themnotecount: #1} }

\newcommand{\jamesx}[1]{}%{\james{ {#1}}}
\renewcommand{\jamesx}[1]{{\mnote{{{\bf jg:}
#1} }}}

%% file: fixmes.tex
\newcommand{\cref}[1]{\mbox{{FIXME; what is the 4 for?}}4\emph{\ref{#1})}}

\newcommand{\h}[2]{#1\dotfill\ #2\\\ptc{fixme}}

%% file: renewcommands.tex
\renewcommand{\color}[1]{}

%% file: KijowskiALaWald.tex
\subsection{General case}

As made clear in Theorem~\ref{T25III15.1}, the analysis of \eq{homogeneous} so far assumed that the field $X$ is a) a spacetime vector field b) which is metric-independent and c) is transverse to $\mcV$ with $X^\perp$ spacelike or timelike. However, in our analysis of asymptotically Kerr-de Sitter metrics we need to allow vector fields $X$ which depend upon the metric, which might vanish and/or change type, and in fact hypersurfaces $\mcV$ which might be either spacelike everywhere or timelike everywhere. We will remove all these unwanted conditions. This forces us to revisit the calculations leading to \eq{homogeneous} for general vectors $X$.

The essential difference between our final formula below and that in~\cite{KijowskiGRG} is, that here the geometric quantities appearing on $\partial V$ are \emph{not} necessarily those associated with a hypersurface obtained by flowing $\partial \mcV$ along $X$, as this set will not form a hypersurface in general, but those associated with a hypersurface $ \{y^n=0\}$ obtained by choosing some coordinate system near $\mcV$ in which $ \mcV = \{y^0=0\}$, $ \partial\mcV \subset \{y^n=0\}$. When $X$ is everywhere transverse to $\mcV$, we can choose coordinates so that $X=\partial/\partial y^0$, in which case all objects appearing in this section coincide with those of~\cite{KijowskiGRG}, as described in Section~\ref{ss5VI15.1}. However, transversality will \emph{not} be assumed in this section.

We emphasise that even in the transverse case we do not need to make the choice $X=\partial/\partial y^0$  (or $X$ proportional to $\partial/\partial y^0$, which leads to the same geometric objects). There is indeed a freedom here, which is equivalent to the choice of the vector field $\partial/\partial y^0$ at the boundary $\partial \mcV$. Every such choice will lead to the variational formula \eq{finalresultbisbis} below, with different  geometric quantities appearing there at $\partial \mcV$ for two not-everywhere-colinear-at-$\partial\mcV$ choices of $\partial/\partial y^0$.
A natural possibility is to choose $\partial/\partial y^0$ to be normal to $\mcV$, but other choices might be more convenient depending upon the problem at hand.

In any case, we stress once again that the index $0$ does \emph{not} indicate the variable $t$ associated with Hamiltonian flow, but an auxiliary coordinate $y^0$.

To continue,
consider the Lagrangian for vacuum Einstein's equations with a cosmological constant $\Lambda$:
\begin{equation}
L =   \frac 1{16 \pi} \sqrt{|g|} ( R+2\Lambda) = {\pi}^{\mu\nu}
R_{\mu\nu} + \frac \Lambda{8\pi} \sqrt{|g|} \, , \label{Hilbert}
\end{equation}
where $R_{\mu\nu} = R^\lambda{}_{\mu\lambda\nu}$ is expressed in terms of a metric and a symmetric connection ${\Gamma}^{\lambda}_{\mu\nu}$. The multiplicative coefficient $1/16 \pi$ is physically motivated only in dimension $3+1$, using units where $G=1$.
Following~\cite{KijowskiGRG} we define
\beaa
%  &
%  {\cal G}^{\mu\nu} := \sqrt{|g|} \ (R^{\mu\nu} - \frac 12 g^{\mu\nu} R)
% \, ,
% &
%\\
 &
 {\pi}^{\mu\nu} := \frac 1{16 \pi} \sqrt{|g|} \  g^{\mu\nu}
 \, .
 &
\eeaa

Consider a one-parameter family of field configurations $\lambda \mapsto  \left( g_{\mu\nu}(\lambda), {\Gamma}^{\lambda}_{\mu\nu}(\lambda)\right)$ and define a ``variation''  $\delta $ as the operation
$$\delta :=\frac{d}{d\lambda}\bigg|_{\lambda=0} \, ,$$
thus $\delta g = \frac{dg}{d\lambda}|_{\lambda=0}$, etc.
The following identity can be easily verified:
\begin{eqnarray}
\delta L &=& R_{\mu\nu} \delta {\pi}^{\mu\nu} + \frac \Lambda{8\pi} \delta \sqrt{|g|}
+ {\pi}^{\mu\nu} \delta R_{\mu\nu}
\nonumber \\
 &=& \underbrace{-\frac 1{16 \pi}\big( {\cal G}^{\mu\nu} - \sqrt{|g|}\Lambda g^{\mu\nu}\big)}_{=:{\mcE}^{\mu\nu}} \delta g_{\mu\nu}\underbrace{ -
 \left( \nabla_\kappa {P}_{\lambda}^{\ \mu\nu\kappa} \right)}_{=:{\mcE}^{\mu\nu}_\lambda}
  \delta
{\Gamma}^{\lambda}_{\mu\nu}  +
\partial_\kappa \left( {P}_{\lambda}^{\ \mu\nu\kappa} \delta
{\Gamma}^{\lambda}_{\mu\nu} \right)  \nonumber \\
&=& {\mcE}^{\mu\nu} \delta g_{\mu\nu}
-
 \left( \nabla_\kappa {P}_{\lambda}^{\ \mu\nu\kappa} \right)
  \delta
{\Gamma}^{\lambda}_{\mu\nu}  +
\partial_\kappa \left( {P}_{\lambda}^{\ \mu\nu\kappa} \delta
{\Gamma}^{\lambda}_{\mu\nu} \right)
 \label{dL=pidgamma-1}
\, ,
\end{eqnarray}
where
\begin{equation}\label{big-P}
    {P}_{\lambda}^{\ \mu\nu\kappa} = \delta^\kappa_\lambda \pi^{\mu\nu} -
    \delta^{(\mu}_\lambda \pi^{\nu)\kappa} \, .
\end{equation}
%%
%We see that the metricity condition for the connection: $\nabla_\lambda g_{\mu\nu} = 0$,
%can either be assumed {\em a priori} or derived in an equivalent form as the Palatini variational principle:
%\begin{equation}\label{metricity}
%    0={\mcE}^{\mu\nu}_\lambda\equiv \frac{\delta L}{\delta {\Gamma}^{\lambda}_{\mu\nu}}=-\nabla_\kappa {P}_{\lambda}^{\ \mu\nu\kappa}
%    \, .
%\end{equation}
Due to the identity (\ref{dL=pidgamma-1}), the (vacuum) Euler-Lagrange equations,
\begin{equation}\label{E-L-Lambda}
    0=\mcE^{\mu\nu}:=\frac{\delta L}{\delta g_{\mu\nu}} =
    -\frac 1{16 \pi}\left({\cal G}^{\mu\nu} - \sqrt{|g|}\Lambda g^{\mu\nu}\right)
    %\, .
\end{equation}
are equivalent to
\begin{equation}
\delta L =
\partial_\kappa \left( {P}_{\lambda}^{\ \mu\nu\kappa} \delta
{\Gamma}^{\lambda}_{\mu\nu} \right)   \label{dL=pidgamma}
\, .
\end{equation}
To simplify notation we set
\begin{equation}
A^{\lambda}_{\mu\nu} := {\Gamma}^{\lambda}_{\mu\nu} -
{\delta}^{\lambda}_{(\mu} {\Gamma}^{\kappa}_{\nu ) \kappa} \, ,
\end{equation}
which enables us to rewrite identity \eq{dL=pidgamma-1} in yet another, equivalent form.
\begin{equation}
\delta L -{\mcE}^{\mu\nu} \delta g_{\mu\nu} - {\mcE}^{\mu\nu}_\lambda
\delta
{\Gamma}^{\lambda}_{\mu\nu}=
\partial_\lambda \left( {\pi}^{\mu\nu} \delta
A^{\lambda}_{\mu\nu} \right) \, .
\label{dL=pidA}
\end{equation}
This equation holds regardless of whether or not the metric $g$ satisfies itself the vacuum field equations or the connection fulfills the metricity condition.

Now, each field configuration $\lambda\mapsto g(\lambda)$ comes with its own family of maps
$$
 (\lambda,t)\mapsto \psi_\lambda(t,\cdot)
% \,,
$$
as at the beginning of Section~\ref{s16IV13.1},
where each $\psi_\lambda(t,\cdot):\Sigma\mapsto \mcM$ is an embedding of $\Sigma\supset V$ into the associated spacetime $\mcM$.
Let
$$\lambda \mapsto X(\lambda):=(\psi_\lambda )_*\partial_t
$$
be the one-parameter family of spacetime vector fields defined along $\psi_\lambda(\{t\}\times \Sigma)$ associated with $\psi_\lambda$. At given  $t$, the field $X(\lambda)$  can be  extended to a smooth vector field defined in a spacetime neighborhood of $\psi_\lambda(\{t\}\times V)$ in many ways. If the $\psi_\lambda$'s  are local diffeormorphisms near  $\{t\}\times\Sigma$, then $\psi_\lambda^*\partial_t$ defines directly such a vector field, and we will make this choice. Otherwise we choose any smooth extension; one of the important outcomes of our calculations below is to show that the result does not depend upon this choice.

To derive the ``Hamilton-type'' variational identity, \eq{dL-Waldbis} below, we rewrite the Lie derivative of the connection,
\begin{equation}
 {\mycal L}_X \Gamma^{\lambda}_{\mu\nu} =
 \nabla_\mu \nabla_\nu X^\lambda - X^\sigma R^\lambda{_{\nu\mu\sigma}}
  \, ,
   \label{2IV15.5}
\end{equation}
in terms of $A$ (compare~\cite{KijowskiGRG,KijowskiTulczyjew})
\begin{eqnarray}
{\pi}^{\mu\nu}  {\mycal L}_X A^{\alpha}_{\mu\nu} & = & \nonumber
 {P}_{\lambda}^{\ \mu\nu\kappa} {\mycal L}_X
{\Gamma}^{\lambda}_{\mu\nu} \\
&=& (\delta^\alpha_\lambda {\pi}^{\mu\nu} -
\delta^\mu_\lambda {\pi}^{\alpha\nu} ) (\nabla_\mu \nabla_\nu X^\lambda
- X^\sigma R^\lambda_{\ \ \nu\mu\sigma}) \nonumber
\\
    & = & \frac {\sqrt{|g|}}{16 \pi}
\bigg\{ \nabla_\mu (\nabla^\mu X^\alpha - \nabla^\alpha X^\mu )
+ 2 R^{\alpha}_{\ \sigma} X^\sigma \bigg\}
  \nonumber
\\
 & = & \frac {1}{16 \pi}  \bigg\{
\partial_\mu \left[ \sqrt{|g|} (\nabla^\mu X^\alpha - \nabla^\alpha
X^\mu ) \right] + 2 \sqrt{|g|} R^{\alpha}{_{\sigma}} X^\sigma \bigg\} \label{piXA}
\, .
 \phantom{xx}
\end{eqnarray}
This leads to the identity
\begin{eqnarray} \nonumber
-\delta \left({\pi}^{\mu\nu}  {\mycal L}_X A^{\lambda}_{\mu\nu} - X^\lambda L \right) & = &
({\mycal L}_X{\pi}^{\mu\nu} )\delta  A^{\lambda}_{\mu\nu} -
  ({\mycal L}_X A^{\lambda}_{\mu\nu}) \delta {\pi}^{\mu\nu} \\ & &
+ \partial_\sigma \left[ X^\lambda {\pi}^{\mu\nu} \delta A^{\sigma}_{\mu\nu}
 -X^\sigma {\pi}^{\mu\nu} \delta A^{\lambda}_{\mu\nu} \right]
   \label{dL-Wald} \\ \nonumber & &
  {  + L\delta X^\lambda + \pi^{\mu\nu} [{\mycal L}_X,\delta] A^{\lambda}_{\mu\nu}
  +X^\lambda \mcE^{\mu\nu}\delta g_{\mu\nu}}
\, ,
\end{eqnarray}
keeping in mind that we allow $X$ to depend upon the field configuration, and where we do not assume that the vacuum Einstein equations are satisfied. However, we have assumed that the connection $\Gamma$ is metric; equivalently,  the field $\mcE^\alpha_{\beta\gamma}$ defined in \eq{dL=pidgamma-1} vanishes. Using (\ref{piXA}), we find
\begin{eqnarray} \nonumber
 \lefteqn{
{\pi}^{\mu\nu}  {\mycal L}_X A^{\lambda}_{\mu\nu} - X^\lambda L +X^\lambda \mcE^{\mu\nu}\delta g_{\mu\nu}
= X^\lambda \mcE^{\mu\nu}\delta g_{\mu\nu}+}
&&
\\ &&
 \frac {1}{16 \pi}  \bigg\{
 \partial_\mu \left[ \sqrt{|g|} (\nabla^\mu X^\lambda - \nabla^\lambda
 X^\mu ) \right]
  +
   (\underbrace{2 \sqrt{|g|} R^{\lambda}{_{\sigma}}X^\sigma  -16\pi X^\lambda L}_{=-32 \pi X^\sigma
   {\cal E}^\lambda{_{\sigma}}})
\bigg\}
 \, .
 \phantom{xx}
\end{eqnarray}

Recall that
$$
 \delta X= \frac{dX}{d\lambda}|_{\lambda=0}
  \, .
$$
The identity $[{\mycal L}_X,\delta]=-{\mycal L}_{\delta X}$ enables one to rewrite the last line in (\ref{dL-Wald}) as follows:
\begin{eqnarray} \nonumber
 \lefteqn{
  L \delta X^\lambda - {\pi}^{\mu\nu}  {\mycal L}_{\delta X} A^{\lambda}_{\mu\nu} =
  }
  &  &
\\
 &&
-\frac {1}{16 \pi}  \bigg\{
\partial_\mu \left[ \sqrt{|g|} (\nabla^\mu \delta X^\lambda - \nabla^\lambda
\delta X^\mu ) \right] - 32 \pi {\cal E}^\lambda{_{\sigma}   \delta X^\sigma}
\bigg\}
 \, .
 \phantom{xxx}
\end{eqnarray}
We conclude that for solutions of the field equations the term $\displaystyle {\mycal L}_X{\pi}^{\mu\nu} \delta  A^{\lambda}_{\mu\nu} -
  {\mycal L}_X A^{\lambda}_{\mu\nu} \delta {\pi}^{\mu\nu}$ in \eq{dL-Wald} is
the divergence of a bivector density:
\begin{eqnarray}
 \nonumber
 \lefteqn{ ({\mycal L}_X A^{\lambda}_{\mu\nu}) \delta {\pi}^{\mu\nu} - ({\mycal L}_X{\pi}^{\mu\nu}) \delta  A^{\lambda}_{\mu\nu}
   =  -2  X^\sigma \delta {\cal E}^\lambda{}_{\sigma}
   + X^\lambda {\cal E}^ {\mu\nu}  \delta g_{\mu\nu}
   }
   &&
\\
 \nonumber
 &&  +\frac {1}{16 \pi} \partial_\mu \bigg\{
 \delta\big[ \sqrt{|g|} (\nabla^\mu X^\lambda - \nabla^\lambda
X^\mu ) \big]
\bigg\}
% \\ & &
+ \partial_\sigma \left[ X^\lambda {\pi}^{\mu\nu} \delta A^{\sigma}_{\mu\nu}
 -X^\sigma {\pi}^{\mu\nu} \delta A^{\lambda}_{\mu\nu} \right]
   \\
    & & {
   -\frac {1}{16 \pi}  %\bigg\{
\partial_\mu \left[ \sqrt{|g|} (\nabla^\mu \delta X^\lambda - \nabla^\lambda
\delta X^\mu ) \right]
%\bigg\}
}
\, .
   \label{dL-Waldbis}
\end{eqnarray}

Let  $(y^0,y^i)$ be local coordinates near $\mycal V=\psi_0(\{0\}\times V)\equiv \psi (\{0\}\times V)$ so that $y^0$ is constant on $\mycal V$. The $y^\mu$'s here will always be local coordinates on $\mcM$, with $\partial_\mu \equiv \partial/\partial y^\mu$, while $(t, x^i)$ will always denote local coordinates on $\R\times \Sigma$. In particular we emphasise that $\partial_0=\partial/\partial y^0$ will \emph{not} be equal to $\partial/\partial t$ in general.

{From now on we assume that  $\mycal V $ is spacelike everywhere or timelike everywhere; equivalently, ${\pi}^{00}$ is nowhere vanishing on $\mcV$.
 As such, studies of Hamiltonian dynamics assume that  $\mcV$ is spacelike. However, for the analysis in Section~\ref{s3III13.1x+} the variational identities that we are about to derive will also  be needed for $\mcV$'s which are timelike.

In coordinates such that $\mcV\subset \{y^0=\const\}$, the integral of the left-hand side of (\ref{dL-Waldbis})
over the $n$-dimensional manifold  $\mycal V $ reads
\bean
 \lefteqn{
    \intvol  \left[({\mycal L}_X A^{\lambda}_{\mu\nu}) \delta {\pi}^{\mu\nu} - ({\mycal L}_X{\pi}^{\mu\nu}) \delta  A^{\lambda}_{\mu\nu}\right]\rd\Sigma_\lambda
    }
    &&
\\
 &&     =
\intvol  \left[({\mycal L}_X A^{0}_{\mu\nu}) \delta {\pi}^{\mu\nu} - ({\mycal L}_X{\pi}^{\mu\nu}) \delta  A^{0}_{\mu\nu}\right]  \rd^n y
  \, . \label{hamformXApi}
\eea
We wish to rewrite this formula in terms of the usual initial data on $\mcV$.
%
\input{logikawzorow.tex}

\input{Christoffel}

\input{ExtensionIndependence}
Our considerations so far have taken care of all the terms involving $X^i$ and $\partial_0 X^\mu$. We continue with the analysis of the remaining terms, with the aim of proving \eq{finalresultbis}. For all practical purposes, the calculations are the same as if we wanted to analyze the variational identity \eq{dL-Waldbis} for vector fields such as $X^i \equiv 0 \equiv \partial_0 X^\mu$, and so we proceed accordingly. \Eq{dL-Waldbis} specialised to this case reads
\begin{eqnarray} \nonumber
\lefteqn{ ({\mycal L}_X A^{0}_{\mu\nu}) \delta {\pi}^{\mu\nu} - ({\mycal L}_X{\pi}^{\mu\nu}) \delta  A^{0}_{\mu\nu} = } & &
\\
\nonumber
& &
 X^0\left[( \partial_0 A^{0}_{\mu\nu}) \delta {\pi}^{\mu\nu} - (\partial_0 {\pi}^{\mu\nu}) \delta  A^{0}_{\mu\nu}\right]
 + { \partial_l(\partial_\mu X^0 \delta\pi^{\mu l})} \\
 & & +\partial_k X^0 \left[ \pi^{\mu\nu}\delta A^{k}_{\mu\nu} +\delta(\pi^{k\mu}\Gamma^{0}_{\mu 0} -\pi^{0\mu}\Gamma^{k}_{\mu 0})
 \right]
  \,.
 \label{4VI15.11}
\end{eqnarray}

In order to analyze the first term on the right-hand side above we use
 \eq{boundtermpidpi+k}. Integrating over $\mcV$ gives
\begin{eqnarray}
 \nonumber
\intvol  X^0 \left(\partial_0 A^{0}_{\mu\nu}\delta{\pi}^{\mu\nu} -\partial_0{\pi}^{\mu\nu} \delta A^{0}_{\mu\nu} \right)
 & = &  \frac 1{16 \pi}  \intvol   X^0 \left(\partial_0 g_{kl} \delta P^{kl} - \partial_0 P^{kl}\delta g_{kl}\right)
  \\
   \nonumber
 &  & \hspace*{-4cm}
   +
  \intvol \partial_k X^0\left[ \partial_0{\pi}^{00} \delta \left( \frac {{\pi}^{k0}}{{\pi}^{00}} \right)
 - \partial_0\left( \frac {{\pi}^{k0}}{{\pi}^{00}} \right) \delta {\pi}^{00}
\right]
  \\
\label{boundtermpidpi+}
 &  & \hspace*{-4cm}
  -
  \intdvol X^0 \left[ \partial_0{\pi}^{00} \delta \left( \frac {{\pi}^{n0}}{{\pi}^{00}} \right)
 - \partial_0\left( \frac {{\pi}^{n0}}{{\pi}^{00}} \right) \delta {\pi}^{00}
\right] \, .
\end{eqnarray}

To analyze the boundary term in (\ref{boundtermpidpi+}), we start with the identity
\begin{equation}
{\pi}^{00} \delta \left( \frac {{\pi}^{n0}}{{\pi}^{00}} \right) +
{\pi}^{nn} \delta \left( \frac {{\pi}^{n0}}{{\pi}^{nn}} \right)
= 2 \sqrt{|{\pi}^{00}{\pi}^{nn}|} \ \delta
\frac {{\pi}^{n0}}{\sqrt{|{\pi}^{00}{\pi}^{nn}|}}
\, .
\end{equation}
Writing
\begin{equation}\label{defq}
 q  :=  \frac {{\pi}^{n0}}{\sqrt{|{\pi}^{00}{\pi}^{nn}|}} =
 \frac {g^{n0}}{\sqrt{|g^{00}g^{nn}|}}
 \, ,
\end{equation}
and assuming  that $\sgn\, {\pi}^{00} \sgn\, {\pi}^{nn}=-1$,
one finds
\begin{equation} \label{qbis}
 2 \sqrt{|{\pi}^{00}{\pi}^{nn}|} = \frac 2{16 \pi} \sqrt{|g|}
 \sqrt{|g^{00}g^{nn}|} = \frac 1{8 \pi}
 \frac {\sqrt{\det g_{AB}}}{\sqrt{1 + q^2}}
 \, .
\end{equation}
We will write
\begin{equation}
 \lambda:=\sqrt{\det g_{AB}}
\end{equation}
for the pull-back to $\partial V$ of the $(n-1)$-dimensional volume density on the boundary $\boundaryimage $.
In this notation we have
\begin{equation}
{\pi}^{00} \delta \left( \frac {{\pi}^{n0}}{{\pi}^{00}} \right) +
{\pi}^{nn} \delta \left( \frac {{\pi}^{n0}}{{\pi}^{nn}} \right) =
\frac {\lambda}{8 \pi} \frac {\delta q}{\sqrt{1 + q^2}}
= \frac {\lambda}{8 \pi}  \delta \alpha
\, ,  \label{dalpha}
\end{equation}
where $\alpha := {\rm arcsinh} \, (q) $ is the hyperbolic angle between
the vector orthogonal to the hypersurface ${\psi (\{t\}\times V)}$
and the world-tube $\{ y^n = {\rm const} \}$ (e.g., $\alpha = 0$ corresponds to the situation where the vector $\psi_* \partial _t$
is {\em tangent} to the tube).
{ When $\sgn\, {\pi}^{00} \sgn\, {\pi}^{nn}=1$,   instead of
\eq{qbis} and \eq{dalpha} we get
\begin{equation} \label{qminus}
 2 \sqrt{|{\pi}^{00}{\pi}^{nn}|} = \frac 2{16 \pi} \sqrt{|g|}
 \sqrt{|g^{00}g^{nn}|} = \frac 1{8 \pi}
 \frac {\sqrt{\det g_{AB}}}{\sqrt{1 - q^2}}
 \, ,
\end{equation}   and
\begin{equation}
{\pi}^{00} \delta \left( \frac {{\pi}^{n0}}{{\pi}^{00}} \right) +
{\pi}^{nn} \delta \left( \frac {{\pi}^{n0}}{{\pi}^{nn}} \right) =
\frac {\lambda}{8 \pi} \frac {\delta q}{\sqrt{1 - q^2}}
= \frac {\lambda}{8 \pi}  \delta \alpha
\, ,  \label{dalphaminus}
\end{equation}
where now $q=\sin\alpha$.}

This, and a calculation identical to the one leading to \eq{8IV15.1} imply that the boundary integral in (\ref{boundtermpidpi+}) takes the form:
\[
  \intdvol X^0\left[ \partial_0{\pi}^{nn} \delta \left( \frac {{\pi}^{n0}}{{\pi}^{nn}} \right)
  - \partial_0\left( \frac {{\pi}^{n0}}{{\pi}^{nn}} \right) \delta {\pi}^{nn}
  + \frac {1}{8\pi} (\partial_0\alpha\delta\lambda - \partial_0\lambda \delta \alpha )
 \right]
 \, .
\]
Finally, we obtain
\begin{eqnarray}
 \lefteqn{
 \intvol    X^0 \left(\partial_0 A^{0}_{\mu\nu}\delta{\pi}^{\mu\nu} -\partial_0{\pi}^{\mu\nu} \delta A^{0}_{\mu\nu} \right)
 }
  &&
   \nonumber
\\
 \nonumber
 &=&
   \frac 1{16 \pi}  \intvol   X^0 \left(\partial_0 g_{kl} \delta P^{kl} - \partial_0 P^{kl}\delta g_{kl}\right)
+ \frac 1{8 \pi} \intdvol  X^0 (\partial_0\alpha\delta\lambda - \partial_0\lambda \delta \alpha )
\\ \nonumber &  &   %\hspace*{-4cm}
   +
  \intvol \partial_k X^0\left[ \partial_0{\pi}^{00} \delta \left( \frac {{\pi}^{k0}}{{\pi}^{00}} \right)
 - \partial_0\left( \frac {{\pi}^{k0}}{{\pi}^{00}} \right) \delta {\pi}^{00}
\right]
\\ & & \label{X0hamform}
+ { \intdvol X^0 \left[ \partial_0{\pi}^{nn} \delta \left( \frac {{\pi}^{n0}}{{\pi}^{nn}} \right)
 - \partial_0\left( \frac {{\pi}^{n0}}{{\pi}^{nn}} \right) \delta {\pi}^{nn} \right] }
  \, .
\end{eqnarray}

%
 %\ptcr{certainly repetitive with Kijowski, presumably not needed}
%As already pointed out in~\cite{KijowskiGRG}, we see that for a finite region $V$ with
%non-empty boundary, the standard ADM canonical structure has
%to be modified by a boundary form $\delta \lambda \wedge \delta
%\alpha$ in such a way, that the symplectic form becomes
%\begin{equation}
%\omega_{\psi_\lambda(\{t\}\times V)} =
%- \frac 1{16 \pi}  \intvol   \left(
%\delta g_{kl} \wedge \delta P^{kl} \right)
%+ \frac 1{8 \pi} \intdvol  (\delta \lambda \wedge \delta
%\alpha) \, ,
%\end{equation}
%%
%It is  proved in~\cite{Kij1} that the boundary
%term is the necessary correction which has to be added to the volume
%term to make the symplectic structure invariant under all diffeomorphisms which
%are the identity on
%the boundary $\partial V$.

We continue with an analysis of the boundary term with $X^i $ set to zero:
\begin{equation}\label{bdtermzX0}   X^0 {\pi}^{\mu\nu} \delta A^{n}_{\mu\nu}
 -X^{n} {\pi}^{\mu\nu} \delta A^{0}_{\mu\nu} = X^0\pi^{\mu\nu}\delta A^{n}_{\mu\nu}
  \, .
\end{equation}
%
%We will use the fact that the identities (\ref{A000})
%and (\ref{A00k}), with $x^n$ replaced by $x^0$, can be viewed as constraints
%on the boundary of the world-tube $\R \times \partial V$, where $\R$ is
%the time-axis:
%%
%\begin{eqnarray}
%A^{n}_{nn} & = & \frac 1{{\pi}^{nn}} \left(
%\partial_a {\pi}^{na} + A^{n}_{ab} {\pi}^{ab} \right) \, , \\
%A^{n}_{3a} & = & - \frac 1{2 {\pi}^{nn}} \left(
%\partial_a {\pi}^{nn} + 2 A^{n}_{ab} {\pi}^{nb} \right) \, ,
%\end{eqnarray}
%%
%and $a,b=0,1,2$. Analogously to \eq{pi-dA0} it holds that
%%
%\begin{eqnarray}
%{\pi}^{\mu\nu} \delta  A^{n}_{\mu\nu} & = &
%{\pi}^{ab} \delta  A^{n}_{ab} +
%2 {\pi}^{na} \delta  A^{n}_{3a} +
%{\pi}^{nn} \delta  A^{n}_{nn}
% \nonumber
%\\
% & = & - \frac 1{16 \pi} g_{ab} \delta Q^{ab} +
%\partial_a \left[
%{\pi}^{nn} \delta \left( \frac {{\pi}^{na}}{{\pi}^{nn}}
%\right) \right]  \, ,
% \label{17I15.5}
%\end{eqnarray}

Let us exchange the role of $y^n$ and $y^0$. Identities (\ref{A000})
and (\ref{A00k}) become constraints
on the boundary of the world-tube $T \times \partial V$, where $T$ is
the time-axis:
\begin{eqnarray}
A^n_{nn} & = & \frac 1{{\pi}^{nn}} \left(
\partial_a {\pi}^{na} + A^n_{ab} {\pi}^{ab} \right) \, , \\
A^n_{na} & = & - \frac 1{2 {\pi}^{nn}} \left(
\partial_a {\pi}^{nn} + 2 A^n_{ab} {\pi}^{nb} \right) \, ,
\end{eqnarray}
and $a,b=0,1,2,\ldots, n-1$. They imply
\begin{eqnarray}
 \nonumber
    {\pi}^{\mu\nu} \delta  A^{n}_{\mu\nu}
     & = &
    {\pi}^{ab} \delta  A^{n}_{ab} +
    2 {\pi}^{na} \delta  A^{n}_{na} +
    {\pi}^{nn} \delta  A^{n}_{nn}
\\
 & = &
  {\frac{(n-3)}{16 \pi} \delta (\sqrt{|\det g_{cd}|}L)} - \frac 1{16 \pi} g_{ab} \delta Q^{ab} +
\partial_a \left[
{\pi}^{nn} \delta \left( \frac {{\pi}^{na}}{{\pi}^{nn}}
\right) \right]  \, ,
 \phantom{xxxxx}
 \label{17I15.5}
\end{eqnarray}
where $Q^{ab}$ and $L$ have been defined in \eq{ku}.
%
%where we have set
%%
%\begin{eqnarray}
%Q^{ab} & := & \sqrt{|\det g_{cd}|} \ (L {\hat g}^{ab} - L^{ab} ) \, ,
%\label{ku}  \\
%\nonumber
%L_{ab} & := & - \frac 1{\sqrt{g^{nn}}} {\Gamma}^{n}_{ab} =
%- \frac 1{\sqrt{g^{nn}}} A^{n}_{ab} \, ,
%\end{eqnarray}
%and where ${\hat g}^{ab}$ is the 3-dimensional inverse with respect to the
%induced metric $g_{ab}$ on the world-tube.
\Eq{17I15.5} gives
\begin{eqnarray}
\lefteqn{
\intdvol X^0 {\pi}^{\mu\nu} \delta  A^{n}_{\mu\nu}}  \nonumber &  &\\
& = & %\hspace*{-3cm}
\nonumber
{\frac{(n-3)}{16 \pi} \intdvol X^0 \delta (\sqrt{|\det g_{cd}|}L)}
 - \frac 1{16 \pi} \intdvol X^0 g_{ab} \delta Q^{ab} + \intdvol X^0
\partial_a \left[
{\pi}^{nn} \delta \left( \frac {{\pi}^{na}}{{\pi}^{nn}}
\right) \right] \\ &= & %\hspace*{-2cm}
\nonumber  {\frac{(n-3)}{16 \pi} \intdvol X^0 \delta (|\sqrt{\det g_{cd}|}L)}
 - \frac 1{16 \pi} \intdvol X^0 g_{ab} \delta Q^{ab}
 + \intdvol X^0 \partial_A \left[ {\pi}^{nn} \delta \left( \frac {{\pi}^{nA}}{{\pi}^{nn}} \right) \right] \\ \label{boundtermcanc}
 & & + {
\intdvol X^0\left[ \partial_0{\pi}^{nn} \delta \left( \frac {{\pi}^{n0}}{{\pi}^{nn}} \right)
 - \partial_0\left( \frac {{\pi}^{n0}}{{\pi}^{nn}} \right) \delta {\pi}^{nn} \right] }
 + \intdvol X^0 \delta \left[ {\pi}^{nn} \partial_0 \left( \frac {{\pi}^{n0}}{{\pi}^{nn}} \right)\right]
 %+ \intdvol  X^0 {\pi}^{nn} [\partial_0,\delta ] \left( \frac {{\pi}^{n0}}{{\pi}^{nn}} \right)
\, .
\end{eqnarray}

We turn attention now to the Komar-type boundary terms in \eq{dL-Waldbis}:
\[ \hspace*{-1cm} \frac 1{16 \pi}  \delta \left[\sqrt{|g|} (\nabla^{n} X^0 - \nabla^0 X^{n} )\right]
 - { \frac 1{16 \pi}  \sqrt{|g|} (\nabla^{n} \delta X^0 - \nabla^0 \delta X^{n} ) } \]
\[ = X^0\delta \left(\pi^{n\mu}\Gamma^0_{\mu 0} -\pi^{0\mu}\Gamma^{n}_{\mu 0} \right)
 +  \partial_{\mu} X^0 \delta \pi^{n\mu}
 \, .
\]
It holds that
\bel{4IV15.41}
  \pi^{n\mu}\Gamma^0_{\mu 0} -\pi^{0\mu}\Gamma^{n}_{\mu 0}
   =
    \frac1{8\pi}  \sqrt{|\det g_{cd}|} {\hat g}^{a0} L_{a0} - {\pi}^{nn} \partial_0 \left( \frac {{\pi}^{n0}}{{\pi}^{nn}} \right)
  \, .
\ee

The last boundary term in \eq{boundtermcanc} cancels the corresponding term in the above formula when we add them together:
\begin{eqnarray}%\hspace*{-3cm}
 \lefteqn{
    \intdvol X^0 {\pi}^{\mu\nu} \delta  A^{n}_{\mu\nu} +X^0\delta \left(\pi^{n\mu}\Gamma^0_{\mu 0} -\pi^{0\mu}\Gamma^{n}_{\mu 0} \right) +  \underbrace{\partial_{\mu} X^0 \delta \pi^{n\mu}}_{ =(1)+\ \partial \mcV-\mbox{\rm \scriptsize term from    } (3)\,,\  \mbox{\rm \scriptsize cf.~\eqref{4IV15.12} and \eqref{4IV15.12+}}} = }
 \nonumber &  &
\\ \nonumber
 & & \hspace*{-1cm}   { \frac{(n-3)}{16 \pi} \intdvol X^0 \delta (\sqrt{|\det g_{cd}|}L)}
 - \frac 1{16 \pi} \intdvol X^0 g_{ab} \delta Q^{ab}   + \frac1{8\pi}\intdvol X^0 \delta \left[   \sqrt{|\det g_{cd}|} {\hat g}^{a0} L_{a0} \right] \\ \nonumber
 & & %\hspace*{-2cm}
 + \intdvol
  \partial_{\mu} X^0 \delta \pi^{n\mu}
 + \intdvol X^0 \partial_A \left[ {\pi}^{nn} \delta \left( \frac {{\pi}^{nA}}{{\pi}^{nn}} \right) \right]
 \nonumber
\\
 &&
  + {
\intdvol X^0\left[ \partial_0{\pi}^{nn} \delta \left( \frac {{\pi}^{n0}}{{\pi}^{nn}} \right)
 - \partial_0\left( \frac {{\pi}^{n0}}{{\pi}^{nn}} \right) \delta {\pi}^{nn} \right] }
\, .
\label{boundtermrazem}
\end{eqnarray}
%\mnote{JJ: w ten sposob dla pola $X=X^0\partial_0$ czlony brzegowe po prawej stronie \eq{tgtdL-Waldbis} mamy opanowane.}
%

Now, a rather lengthy calculation shows that
%%
%\begin{eqnarray}
% \nonumber
% &&   \frac1{8\pi}  \delta \intdvol
%    \sqrt{|\det g_{cd}|} ({\hat g}^{a0} L_{a0} { +\frac12(n-3)L})
%     \,
% - \frac 1{16 \pi} \intdvol  g_{ab} \delta Q^{ab}
%  \\
%   \nonumber
% & &
%  =  \frac 1{{ 16\pi}} \intdvol  (   2 \nthree\delta {\bf Q}
%- 2\nthree^A \delta {\bf Q}_A
%+ { \nu}{\bf Q}^{AB}
% \delta g_{AB})
%\, .
%   \label{26III15.1}
%\end{eqnarray}
%Uwaga: powyzszy wzor jest sluszny bez calkowania,
%tj.
\begin{eqnarray}
 \nonumber
 &&   \frac1{8\pi}  \delta \left[
    \sqrt{|\det g_{cd}|} ({\hat g}^{a0} L_{a0} { +\frac12(n-3)L}) \right]
     \,
 - \frac 1{16 \pi}   g_{ab} \delta Q^{ab}
  \\
   %\nonumber
 & &
  =  \frac 1{{ 16\pi}}  (   2 \nthree\delta {\bf Q}
- 2\nthree^A \delta {\bf Q}_A
+ { \nu}{\bf Q}^{AB}
 \delta g_{AB})
\, ,
   \label{Qnuidentity}
\end{eqnarray}
which enables one to rewrite the second line of \eq{boundtermrazem} as
\[ \frac 1{{ 16\pi}} \intdvol X^0 (   2 \nthree\delta {\bf Q}
- 2\nthree^A \delta {\bf Q}_A
+ { \nu}{\bf Q}^{AB}
 \delta g_{AB}) \, .\]
%
%Wzory \eq{X0hamform} and \eq{boundtermrazem} zebrane razem daja nastepujacy wynik dla pola $X=X^0\partial_0$:
Formulae \eq{X0hamform} and \eq{boundtermrazem}  give the following intermediate result:
\bean
 \lefteqn{
\intvol   X^0 \left(\partial_0 g_{kl} \delta P^{kl} - \partial_0 P^{kl}\delta g_{kl}\right)
+ 2 \intdvol  X^0 (\partial_0\alpha\delta\lambda - \partial_0\lambda \delta \alpha )
 }
 &&
\\
 &&
- \intdvol X^0 (   2 \nthree\delta {\bf Q}
- 2\nthree^A \delta {\bf Q}_A + { \nu}{\bf Q}^{AB}
 \delta g_{AB}) + \intvol (**) \times \partial_k X^0
  \nn
\\
 &
   =
   &
    -
  \intdvol \partial_A X^0  \left[ {\pi}^{nn} \delta \left( \frac {{\pi}^{nA}}{{\pi}^{nn}} \right) \right] + 16 \pi  \intvol X^0\left(
  -2\delta {\cal E}^0{}_{0}
   +  {\cal E}^ {\mu\nu}  \delta g_{\mu\nu}
    \right)
   \,,
    \phantom{xx}
   \label{intermdteres}
   \eea
where the expression (**) is given by
\begin{equation} \label{**bis}
\frac{(**)}{16\pi} =
 \underbrace{\pi^{\mu\nu}\delta A^{k}_{\mu\nu}}_{=:(a)/(16 \pi)} +\delta(\underbrace{\pi^{k\mu}\Gamma^{0}_{\mu 0} -\pi^{0\mu}\Gamma^{k}_{\mu 0}}_{:= \sqrt{|\det g_{ij}|}(b)/(16 \pi)}) +
 \underbrace{ \partial_0{\pi}^{00} \delta \left( \frac {{\pi}^{k0}}{{\pi}^{00}} \right)
 - \partial_0\left( \frac {{\pi}^{k0}}{{\pi}^{00}} \right) \delta {\pi}^{00}
 }_{=:(c)/(16 \pi)}
 \,.
\end{equation}
This is precisely the multiplicative factor of $\partial_k X^0$  in the sum of volume terms  in \eq{4IV15.12+} and \eq{X0hamform}.

\Eq{intermdteres} ends the proof of \eq{finalresultbis} when   $X^0=\const$, because then $\partial_k X^0$ vanishes, while the first term in the last line of \eq{intermdteres} integrates  to zero.

%% file: logikawzorow.tex
%logikawzorow.tex
For this, set
$$\displaystyle X=X^0\frac{\partial}{\partial y^0}+Y
 \,,
$$
%,
where $\displaystyle Y:=X^k \frac{\partial}{\partial y^k}$ is tangent to $\mcV$.

Let $f$ be some field on spacetime. We shall denote by $\mymcL_X f $ the usual Lie-derivative operator on a manifold, and by ${\cal L}_X f$ the restriction of $\mymcL_X f$ to $\mcV$. Note that $\mymcL_Y F$ coincides with ${\cal L}_Y F$ for vector fields $Y$ tangent to $\mcV$ and geometric fields $F$ on $\mcV$, and both notations will be used interchangeably in such cases.
See Appendix~\ref{A4VI15.1} for some explicit expressions.

We have the following:

\begin{Theorem}
  \label{T4VI15.1}
\begin{enumerate}
\item Suppose that $X=X^0 \partial_0$ on $\mcV$, then:
\bean
 \lefteqn{
  \intvol   \left({\cal L}_X g_{kl} \delta P^{kl} - {\cal L}_X P^{kl}\delta g_{kl}\right)
 + 2 \intdvol  ({\cal L}_X \alpha\delta\lambda - {\cal L}_X \lambda \delta \alpha )
 }
 &&
\\
  &&
    - \intdvol X^0 (   2 \nthree\delta {\bf Q}
    - 2\nthree^A \delta {\bf Q}_A + { \nu}{\bf Q}^{AB}\delta g_{AB})
 \nn
\\
 &=&       16 \pi  \intvol X^0\left(
  -2\delta {\cal E}^0{}_{0}
   +  {\cal E}^ {\mu\nu}  \delta g_{\mu\nu}
    \right)\,,
  \label{finalresultbis}
\eea
where
\begin{eqnarray}
{\cal L}_{X^0 \partial_0} g_{kl} &:=& %\nonumber
  X^0 \partial_0 g_{kl} + N_k \partial_l X^0 +
 N_l  \partial_k X^0
  \,,
\\
 \nonumber
  {\cal L}_{X^0 \partial_0} P^{kl} &:=& X^0 \partial_0 P^{kl}  + N \sqrt{{g}}\left( D^k D^l X^0 - {\tilde g}^{kl} \Delta X^0
   \right) \\ \nonumber
  & & + \left\{ \sqrt{g}\left(
  {\tilde g}^{lr} D^k N  + {\tilde g}^{kr}D^l N  - {2} {\tilde g}^{kl}
  D^r N
  \right) \right.\\
  & &  + \left. P^{kl} N^r -P^{kr}N^l -P^{lr}N^k
    \right\} \partial_r X^0 \, ,
\\
{\cal L}_{X^0 \partial_0} \lambda
 &  := &
 %X^0 \partial_0 \lambda + \left(N^A - \frac{{\tilde g}^{nA}}{{\tilde g}^{nn}}N^n\right) \lambda \partial_A X^0 =
 X^0 \partial_0 \lambda + \lambda\nu^A \partial_A X^0
  \,,
\\
{\cal L}_{X^0 \partial_0} \alpha &:=&
 %X^0 \partial_0 \alpha -\frac{{\tilde g}^{nk}\partial_k X^0}{\cosh\alpha\sqrt{|g^{00}g^{nn}|}} =
  X^0 \partial_0 \alpha { -\frac{N {\tilde g}^{nk}}{\sqrt{|{\tilde g}^{nn}|}}\partial_k X^0 }
  \,.
\end{eqnarray}
\item  Suppose that $X^0=0 \ \Leftrightarrow \ X=X^i\partial_i =: Y$ on $\mcV$,  then:
\begin{eqnarray}
 \nonumber
 \lefteqn{
     \intvol   \left({\cal L}_Y g_{kl} \delta  P^{kl} - {\cal L}_Y P^{kl}\delta  g_{kl}\right)
     }
      &&
\\
    &&=   \int_{\partial \mcV}
    \left(
  2 Y^k  \delta{P }^{n}{}_{  k} -  Y^n P ^{kl}  \delta g_{kl}
   \right)
    -32  \pi  \intvol
   Y^i\delta {\cal E}^0{}_{i}
     \, ,
      \label{26IV15.15bis}
\end{eqnarray}
where ${\cal L}_Y =\mymcL_Y$  is the usual Lie-derivative operator on $\mcV$.
\end{enumerate}
\end{Theorem}

\begin{Remarks}
 \label{R4VI15.1}
 {\rm
\begin{enumerate}
 \item Adding \eq{finalresultbis} and \eq{26IV15.15bis}, one obtains the variational formula for a general vector field $X=X^0\partial_0+Y$:
\bean
 \lefteqn{
  \intvol   \left({\cal L}_X g_{kl} \delta P^{kl} - {\cal L}_X P^{kl}\delta g_{kl}\right)
 + 2 \intdvol  ({\cal L}_{X^0\partial_0} \alpha\delta\lambda - {\cal L}_{X^0\partial_0} \lambda \delta \alpha )
 }
 &&
\\
  &&
    - \intdvol X^0 (   2 \nthree\delta {\bf Q}
    - 2\nthree^A \delta {\bf Q}_A + { \nu}{\bf Q}^{AB}\delta g_{AB})
 \nn
\\
    && -  \int_{\partial \mcV}
    \left(
  2 Y^k  \delta{P }^{n}{}_{  k} -  Y^n P ^{kl}  \delta g_{kl}
   \right)
 \nn
\\
 && \phantom{xxxxxxxxxxxxxx} =      16 \pi  \intvol
    X^0  {\cal E}^ {\mu\nu}  \delta g_{\mu\nu}
  -2X^\mu\delta {\cal E}^0{}_{\mu}
   \,,
  \label{finalresultbisbis}
\eea
with ${\cal L}_X$ for $g_{ij}$ and $P^{ij}$ obtained by adding ${\cal L}_{X^0\partial_0}$ and  ${\cal L}_Y$.
\item
For solutions of the field equations, \eq{finalresultbis} with $X^0=1$ clearly coincides with Kijowski's formula (\ref{homogeneous}).
 \item
We show in  Appendix \ref{JKforY} below that  (\ref{homogeneous}) with the additional condition that $Y$ is transversal to $\partial\mcV$ (equivalently, $\nu$ has no zeros) coincides with \eq{26IV15.15bis}.
\end{enumerate}
} % end of \rm
\end{Remarks}

\proof
Since the proofs are quite lengthy and  computationally intensive, we start with roadmaps.

The proof of formula (\ref{26IV15.15bis}) proceeds as follows:

\begin{itemize}
\item We integrate (\ref{dL-Waldbis}) over $\mcV$ and show that the left-hand side gives
(\ref{tgtboundtermpidpi}).
\item Using (\ref{pi-dA0}), (\ref{Pkl}) and (\ref{KomartoK}), the boundary terms on the right-hand side of (\ref{tgtdL-Waldbis}) lead to the formula (\ref{tgtgdP-Pdg}),
which is equivalent to  (\ref{26IV15.15bis}).
\end{itemize}

For \eq{finalresultbis} more work is needed:

\begin{itemize}
\item We integrate (\ref{dL-Waldbis}) over $\mcV$ and show that the left-hand side splits into
several terms as in (\ref{4VI15.11}). The  terms proportional to $X^0$ are given by (\ref{X0hamform}), while
the next term is a straightforward divergence.  The  terms  involving $\partial_k X^0$ are the most problematic ones, we return to them in the last three steps of the calculation.
\item The divergence terms from the right-hand side of (\ref{dL-Waldbis}) are collected as a   boundary integral in (\ref{boundtermrazem}). The second line of this formula simplifies to (\ref{Qnuidentity}).
\item An intermediate result is provided by (\ref{intermdteres}), where all volume terms containing
$\partial_k X^0$ are collected in \eq{**bis}. The proof is then complete when $X^0$ is space-independent, as all $\partial_k X^0$ terms vanish under this hypothesis.
\item To analyze  \eq{**bis} one  introduces an ADM-type parameterization of the fields. A    lengthy calculation gives \eq{volterm}.
\item Using the Lie derivatives worked-out in Appendix \ref{A4VI15.1},   one is led to \eq{summa}.
\item All  boundary terms containing $\partial_A X^0$ cancel out and we obtain our final formula \eq{finalresultbis}.

\end{itemize}

%% file: Christoffel.tex
Let us pass now to the details of the argument. We start by recalling the transport formula for Christoffel symbols. Letting $\Gamma$ denote the connection coeffiences in a coordinate system $y$, and $
\hat\Gamma$ those transported by a diffeomorphism $\psi$, it holds that
\bel{2IV15.1}
 {\displaystyle \Gamma^\alpha_{\beta\gamma}=
 \hat
\Gamma^\tau _{\sigma \rho} \frac{\partial
y^\alpha}{\partial{\psi^\tau}}\frac{\partial
\psi^\sigma}{\partial{y^\beta}}\frac{\partial \psi^\rho}{\partial{y^\gamma}}
 +
  \frac{\partial y^\alpha}{\partial{\psi^\tau}}
\frac{\partial^2 \psi^\tau}{\partial{y^\gamma}\partial{y^\beta}} }
 \,,
 \ee
where we have denoted by $\partial y /\partial \psi$ the derivatives of the map inverse to $\psi$.
Let us denote by $\delta_\psi {\displaystyle \Gamma^\alpha_{\beta\gamma}}$ those variations of ${\displaystyle \Gamma^\alpha_{\beta\gamma}}$ which arise from a family of maps $\psi(\lambda)$ such that $\psi(0)$ is the identity. It follows from
\eq{2IV15.1} that
\bean
 \delta_\psi
  \displaystyle \Gamma^\alpha_{\beta\gamma}
  & = &
  \delta X^\mu \pa_\mu \Gamma^\alpha_{\beta\gamma}
  -
 \Gamma^\sigma_{\beta\gamma} \frac{\partial \delta
 X ^\alpha} {\partial y^\sigma}
  +
 \Gamma^\alpha_{\sigma\gamma} \frac{\partial \delta
 X ^\sigma} {\partial y^\beta}
  +
 \Gamma^\alpha_{\beta\sigma} \frac{\partial \delta
 X ^\sigma} {\partial y^\gamma}
  \nonumber
\\
 \label{2IV15.4}
 &&
   +
\frac{\partial^2 \delta X^\alpha}{\partial{y^\gamma}\partial{y^\beta}}
 \,,
\\
 \nonumber
 \delta_\psi
  \displaystyle A^\alpha_{\beta\gamma}
  & = &
  \delta X^\mu \pa_\mu A^\alpha_{\beta\gamma}
  -
 A^\sigma_{\beta\gamma} \frac{\partial \delta
 X ^\alpha} {\partial y^\sigma}
  +
 A^\alpha_{\sigma\gamma} \frac{\partial \delta
 X ^\sigma} {\partial y^\beta}
  +
 A^\alpha_{\beta\sigma} \frac{\partial \delta
 X ^\sigma} {\partial y^\gamma}
\\
 \label{2IV15.2}
 &&
   +
 \frac{\partial^2 \delta X^\alpha}{\partial{y^\gamma}\partial{y^\beta}}
    -
     \delta^\alpha_{(\beta} \frac{\partial^2 \delta
    X^\sigma}{\partial{y^{\gamma)}}\partial{y^\sigma}}
 \,,
\\
 \nonumber
 \mymcL_X \displaystyle A^\alpha_{\beta\gamma}
  & = &
   X^\mu \pa_\mu A^\alpha_{\beta\gamma}
  -
 A^\sigma_{\beta\gamma} \frac{\partial
 X ^\alpha} {\partial y^\sigma}
  +
 A^\alpha_{\sigma\gamma} \frac{\partial
 X ^\sigma} {\partial y^\beta}
  +
 A^\alpha_{\beta\sigma} \frac{\partial
 X ^\sigma} {\partial y^\gamma}
\\
 \label{2IV15.3}
 &&
   +
 \frac{\partial^2  X^\alpha}{\partial{y^\gamma}\partial{y^\beta}}
    -
    \delta^\alpha_{(\beta} \frac{\partial^2
    X^\sigma}{\partial{y^{\gamma)}}\partial{y^\sigma}}
 \,.
  \phantom{xxx}
 \eea

Now, it should be kept in mind that in all formulae above part of the variations of the field arise from variations of the map $\psi$, which is allowed to vary. To make this clear,
let us  denote by $\delta_f$ those variations of the fields for which $\psi$ is the identity (here the index $f$ stands for ``fields'').
We thus have
\bel{26IV16.11}
\delta = \delta_f + \delta_\psi
\,.
\ee
The distinction between $\delta$ and $\delta_f$ is somewhat arbitrary for tensor fields defined in spacetime in a fixed coordinate system, since $\delta_\psi$ can always be absorbed in a redefinition of $\delta_f$ and vice-versa. The question arises, whether this remains true after pull-backs to the image of $\psi_0(t,\cdot)$ have been taken. In order to clarify this, let us denote by $\hat g_{ij}$ the  metrics induced by $g_{\mu\nu}(\lambda)$ on the images of $\psi_\lambda (t,\cdot)$
$$
 \hat g_{ij}(\lambda, t, \vec x) = g_{\mu\nu}(\lambda, \psi_\lambda (t, \vec x))
  \frac{\partial \psi^\mu _\lambda (t, \vec x)}{\partial x^i}
  \frac{\partial \psi^\nu _\lambda (t, \vec x)}{\partial x^i}
 \,.
$$
Then, in coordinates such that $y^0( \psi_0 (0,x^k))=0$, $y^i( \psi_0 (0,x^k))=x^i$:
\bean
 \delta_f \hat g_{ij}(  0, \vec x)
  & : = & \frac{d}{d\lambda}
   \left.
    \left(
    g_{\mu\nu}(\lambda, \psi_\lambda (0, \vec x))
  \frac{\partial \psi^\mu _0 (0, \vec x)}{\partial x^i}
  \frac{\partial \psi^\nu _0 (0, \vec x)}{\partial x^i})
   \right)
   \right|_{\lambda=0}
\\
 &   = &
   \left.\frac{d g_{ij}(\lambda, \vec x)}{d\lambda}
   \right|_{\lambda=0}
 \,,
\\
  \nonumber
 \delta_\psi \hat g_{ij}( 0, \vec x)
  & = & \frac{d}{d\lambda}
   \left.
    \left(
    g_{\mu\nu}(0, \psi_\lambda (0, \vec x))
  \frac{\partial \psi^\mu _\lambda (0, \vec x)}{\partial x^i}
  \frac{\partial \psi^\nu _\lambda (0, \vec x)}{\partial x^j}
   \right)
   \right|_{\lambda=0}
\\
 & = &
 {\cal L}_{\delta X} g_{ij}
\\
  \nonumber
 &&    \mbox{\rm (space-components of a spacetime Lie derivative),}
\\
 \delta  \hat g_{ij}( 0, \vec x)
  & = & \frac{d}{d\lambda}
   \left.
    \left(
    g_{\mu\nu}(\lambda, \psi_\lambda (0, \vec x))
  \frac{\partial \psi^\mu _\lambda (0, \vec x)}{\partial x^i}
  \frac{\partial \psi^\nu _\lambda (0, \vec x)}{\partial x^i})
   \right)
   \right|_{\lambda=0}
  \nonumber
\\
 & = &
 \delta_f  \hat  g_{ij}( 0, \vec x)
 +
 \delta_\psi \hat   g_{ij}( 0, \vec x)
 \,.
\eeal{4IV15.31}
Since our variations are arbitrary, for any $\delta_\psi \hat g_{ij}$ we can redefine $\delta_f  \hat  g_{ij}$ so that $\delta \hat g_{ij}$ takes any desired value, and vice-versa. This remains true even if the variations are constrained to satisfy the linearised field equations (which we do not assume in most of our calculations, but which might be convenient for some purposes), since $\delta_\psi$-variations do indeed satisfy the linearised field equations, respectively the linearized constraint equations, when the fields being varied satisfy the full equations, respectively the constraint equations.

 From now on we will not make a distinction between $\hat g_{ij}$ and $g_{ij}$.

A similar argument applies to $P^{ij}$.
Henceforth, from now on we will not make a distinction between $\delta_f$ and $\delta$ unless a significant ambiguity arises.

%% file: ExtensionIndependence.tex
We continue with a preliminary result which deserves highlighting:

\begin{Proposition}
  \label{P26IV15.1}
  The integral  \eq{hamformXApi}  depends upon the extension of $X \equiv \psi_* \partial_t$ off $\mcV$ only   through boundary terms arising at $\partial\mcV$.
\end{Proposition}

\proof
We will need the constraints implied by the identities $\nabla_k {\pi}^{0k} = 0$
and $\nabla_k {\pi}^{00} = 0$:
Indeed, expressing the left-hand sides
in terms of ${\pi}^{\mu\nu}$ and $A^{0}_{\mu\nu}$, we obtain
the following constraints:
\begin{eqnarray}
A^0_{00} & = & \frac 1{{\pi}^{00}}
 \left(
    \partial_k {\pi}^{0k} + A^0_{kl} {\pi}^{kl} \right)
     \, ,
      \label{A000}
\\
 A^0_{0k} & = & - \frac 1{2 {\pi}^{00}} \left(
 \partial_k {\pi}^{00} + 2 A^0_{kl} {\pi}^{0l} \right) \, .
 \label{A00k}
\end{eqnarray}

To continue, we insert \eq{2IV15.3} into \eq{hamformXApi}.
We start by noting that  all second-order derivatives $\partial_\mu \partial_\nu X^\sigma$  there with $\mu=\nu=0$ cancel out.
Next, all terms involving $\partial_0 X^0$ and  $\partial_k\partial_0 X^0$ can be collected into
\bean
(1)
 & :=
  &
  \intvol \partial_k (\partial_0 X^0 \delta \pi^{0k}) + \partial_0 X^0 \delta\big(
 \underbrace{
 2 A^0_{0\nu} \pi^{0\nu} - A^0_{\mu\nu} \pi^{\mu\nu} - \partial_k\pi^{0k}
  }_{=0 \ \mbox{\scriptsize \rm{by \eq{A000}}}}
   \big)
\\
  & =
  &
    \intdvol \partial_0 X^0 \delta \pi^{0n}
  \,.
\eeal{4IV15.12}
A similar calculation using \eq{A00k} gives the following contribution of terms involving $\partial_0 X^k$  in \eq{hamformXApi} after  inserting \eq{2IV15.3} there:
\bel{26IV15.12}
(2):=-\intvol\partial_k  \left( \partial_0 X^k \delta  \pi^{00}
 \right)
 =
 -\intdvol \partial_0 X^n \delta \pi^{00}
  \,,
\ee
which had to be established.
\qedskip

For further reference, we note that those terms in \eq{hamformXApi} which involve $\partial_i X^0$ and its space-derivatives take the form
% \ptcr{added and checked by me and Jacek 4 IV 15}
%
\bean
(3)
 & :=
  &
  \intvol \partial_k X^0 \left[-\partial_i \delta \pi^{ik}  +
 2  \delta (\pi^{k\nu} A^0_{0\nu}) - A^k_{\mu\nu}\delta\pi^{\mu\nu} \right]
\\
 &&  +
    \intdvol \partial_i X^0 \delta \pi^{in}
  \,.
\eeal{4IV15.12+}

We continue with the terms which do \emph{not} involve $X^0$ and $\partial_0 X^\mu$. Recall that $\mcV$ is given by the equation $\{y^0=0\}$. To avoid ambiguities, we will  use the notation
\bel{3IV15.11}
 Y:= X^k \partial_k
 \,,
 \
 \mbox{ \rm thus $X = X^0 \partial_0 +Y$, with $dy^0(Y)=0$.}
\ee

\input{tangentXsubset}
We note that for any field $f$, with $\delta f := \partial_\lambda f|_{\lambda=0} $, it holds that
\bean
 \partial_t \delta f = X^\alpha \partial _\alpha \partial_\lambda f |_{\lambda=0}
 & = & \partial_\lambda \left (X^\alpha \partial _\alpha f\right) |_{\lambda=0} - \partial_\lambda X^\alpha |_{\lambda=0} \left ( \partial _\alpha f\right)
\\
  &  = &
  \delta (\partial_t f) - \delta X^\alpha \partial_\alpha f
  \,.
\eeal{25VIII14.2}
While this can be used to analyze the commutator terms  in \eq{boundtermpidpi+k+}, it is calculationally advantageous to proceed as follows:
Using \eq{2IV15.3} and \eq{pochodnaLiegoPiwApendiksie}, Appendix~\ref{A4VI15.1}, and taking an extension of $X^k$ off $\mcV$ such that $\partial_0 X^\mu=0$ we have
 %\ptcr{dodac wzor na pochodna pi do appendiksu }
%
\begin{eqnarray}
 \nonumber
 \lefteqn{ ({\mycal L}_Y A^{0}_{\mu\nu}) \delta {\pi}^{\mu\nu} - ({\mycal L}_Y{\pi}^{\mu\nu}) \delta  A^{0}_{\mu\nu}
   =     }
   &&
\\ & & = Y^i
\underbrace{\left( \partial_i A^{0}_{\mu\nu}\delta{\pi}^{\mu\nu} -\partial_i{\pi}^{\mu\nu} \delta A^{0}_{\mu\nu}\right)}_{\ast} -\partial_k Y^k \pi^{\mu\nu}\delta A^{0}_{\mu\nu}+
\underbrace{2\partial_\mu Y^k
\delta \left( {\pi}^{\mu\nu}  A^{0}_{k\nu} \right)}_{=   2\partial_i Y^k
\delta \left( {\pi}^{i\nu}  A^{0}_{k\nu} \right) } \nonumber \\
& & { - \partial_i\partial_k Y^k \delta\pi^{0i} }
 \,.
 \end{eqnarray}
For the term $\ast$, we can use \eq{boundtermpidpi+k+} with $\partial_t=\partial_{y^i}$ and with vanishing commutators:
\begin{eqnarray}
 \partial_i A^{0}_{\mu\nu}\delta{\pi}^{\mu\nu} -\partial_i{\pi}^{\mu\nu} \delta A^{0}_{\mu\nu}
 & = &
  \frac 1{16 \pi}    \left(\partial_i g_{kl} \delta P^{kl} - \partial_i P^{kl}\delta g_{kl}
    \right)
     \nn
  \\
\label{boundtermpidpi+k+i}
 &  &
  -
  \partial_k \left[ \partial_i{\pi}^{00} \delta \left( \frac {{\pi}^{k0}}{{\pi}^{00}} \right)
 - \partial_i\left( \frac {{\pi}^{k0}}{{\pi}^{00}} \right) \delta {\pi}^{00}
\right]
 \, .
  \phantom{xxx}
\end{eqnarray}
The middle term $\pi^{\mu\nu}\delta A^{0}_{\mu\nu}$ is given by \eq{pi-dA0}.
The term involving
${\pi}^{i\nu}  A^{0}_{k\nu}$ can be rewritten as
\bean
 {\pi}^{i\nu}  A^{0}_{k\nu}
 & = & {\pi}^{il}  A^{0}_{kl}+{\pi}^{i0}  A^{0}_{k0}=
    \frac{N\sqrt{\det g_{mn}}}{16\pi}\left( g^{il}\Gamma^0{_{kl}}+g^{i0}A^{0}_{k0} \right)
\\
 &=&
    -\frac{\sqrt{\det g_{mn}}}{16\pi}{\tilde g}^{il}K{_{kl}}
        -\frac{\pi^{i0}}{2 {\pi}^{00}}\partial_k {\pi}^{00}
 \, ,
\eeal{7VI15.1}
where we have used the second line of \eq{5VI15.1}, Appendix~\ref{A4VI15.1}:
\begin{eqnarray*}
A^0_{0k} & = &
 - \frac 12 \frac{\partial_k   \pi^{00}  }{ \pi^{00} } - \Gamma^0_{kl} \frac{g^{0l}}{g^{00}}
 \, .
\end{eqnarray*}
In the first line of \eq{7VI15.1}, and elsewhere, we use the notation
\bel{3IV15.15}
 N^k:=- g^{0k}/g^{00}
 \,,
  \quad
  N:=(-g^{00})^{- 1/2}
 \,.
\ee
Now, we add all terms depending on $P^{kl}$:
\bean
 \lefteqn{\hspace*{-3cm}
  \frac1{16\pi}\left\{   Y^i \left(\partial_i g_{kl} \delta P^{kl} - \partial_i P^{kl}\delta g_{kl}
    \right) -\partial_k Y^k \left(  \delta  P -  g_{kl} \delta  P^{kl}\right)
    -2\partial_i Y^k \delta P^i{_{l}} \right\}
    }
    & &
\\
   & &
     =\frac1{16\pi} \left({\cal L}_Y g_{kl} \delta_f P^{kl} - {\cal L}_Y P^{kl}\delta_f g_{kl}\right)
      \,.
\eeal{5VI15.2}
The remaining terms   gather to a full divergence,
%\mnote{JJ: Juz sie zgadza! Zgubiony zostal czlon z drugimi pochodnymi (na czerwono).}
%
\begin{eqnarray} \nn -
  Y^i\partial_k \left[ \partial_i{\pi}^{00} \delta \left( \frac {{\pi}^{k0}}{{\pi}^{00}} \right)
 - \partial_i\left( \frac {{\pi}^{k0}}{{\pi}^{00}} \right) \delta {\pi}^{00} \right]
 -\partial_i Y^i    \partial_k \left[
{\pi}^{00} \delta  \left( \frac {{\pi}^{0k}}{{\pi}^{00}}
\right) \right] \\ \nn
    - \partial_i Y^k \delta \left(\frac{\pi^{i0}}{ {\pi}^{00}}\partial_k {\pi}^{00}\right)
     +{ \partial_i\partial_k Y^k \delta (\pi^{00} N^i)} \\
 = \partial_k \bigg[ \underbrace{\partial_i (Y^i\pi^{00})}_{{\cal L}_Y\pi^{00}} \delta N^k
  - \underbrace{(Y^i \partial_i N^k   - \partial_i Y^k  N^i)}_{{\cal L}_Y N^k} \delta \pi^{00} \bigg]
  \, ,
\end{eqnarray}
leading to
\begin{eqnarray}
  (4) := \lefteqn{
   \frac 1{16 \pi}  \intvol   \left({\cal L}_Y g_{kl} \delta_f P^{kl} - {\cal L}_Y P^{kl}\delta_f g_{kl}\right)
   }
   &&
    \nonumber
  \\
\label{tgtboundtermpidpi(4)}
 &  &
  - \frac1{16 \pi}
  \intdvol \left( {\cal L}_Y \frac{\sqrt{\det g_{mn}}}{N} \delta_f N^n
 - {\cal L}_Y N^n \delta_f \frac{\sqrt{\det g_{mn}}}{N}
    \right)
     \, .
\end{eqnarray}

A case of interest in its own is that of vector fields $X$ which are tangent  to $\mcV$. An alternative, standard and rather more straightforward treatment of this case is presented in Section~\ref{ss26IV15.1} below. Here we show how the relevant identity follows from the calculations above:

\begin{Proposition}
  \label{P26IV15.2}
  Suppose that the map $\psi$ leaves $\mcV$ invariant:
\bel{26IV15.13}
 \psi(t, \mcV) \subset \mcV
 \,,
\ee
and let $Y$ denote the generator of the flow  $t\mapsto \psi(t,\cdot)$ on $\mcV$. Then
\begin{eqnarray}
 \nonumber
 \lefteqn{
     \intvol   \left({\cal L}_Y g_{kl} \delta  P^{kl} - {\cal L}_Y P^{kl}\delta  g_{kl}\right)
     }
      &&
\\
    &&=   \int_{\partial \mcV}
    \left(
  2 Y^k  \delta{P }^{n}{}_{  k} -  Y^n P ^{kl}  \delta g_{kl}
   \right)
    -32  \pi  \intvol
   X^i\delta {\cal E}^0{}_{i}
     \, .
      \label{26IV15.15}
\end{eqnarray}
\end{Proposition}

\proof
We return to \eq{dL-Waldbis} with $\lambda=0$, which we repeat here for the convenience of the reader
\begin{eqnarray}
 \nonumber
 \lefteqn{ ({\mycal L}_X A^{\lambda}_{\mu\nu}) \delta {\pi}^{\mu\nu} - ({\mycal L}_X{\pi}^{\mu\nu}) \delta  A^{\lambda}_{\mu\nu}
   =  -2  X^\sigma \delta {\cal E}^\lambda{}_{\sigma}
   + X^\lambda {\cal E}^ {\mu\nu}  \delta g_{\mu\nu}
   }
   &&
\\
 \nonumber
 &&  +\frac {1}{16 \pi} \partial_\mu \bigg\{
 \delta\big[ \sqrt{|g|} (\nabla^\mu X^\lambda - \nabla^\lambda
X^\mu ) \big]
\bigg\}
% \\ & &
+ \partial_\sigma \left[ X^\lambda {\pi}^{\mu\nu} \delta A^{\sigma}_{\mu\nu}
 -X^\sigma {\pi}^{\mu\nu} \delta A^{\lambda}_{\mu\nu} \right]
   \\
    & & {
   -\frac {1}{16 \pi}  %\bigg\{
\partial_\mu \left[ \sqrt{|g|} (\nabla^\mu \delta X^\lambda - \nabla^\lambda
\delta X^\mu ) \right]
%\bigg\}
}
\, .
   \label{tgtdL-Waldbis}
\end{eqnarray}
Integrating over $\mcV$,  the last term above leads to the following boundary integrand:
\[ \frac 1{16 \pi} \left[\sqrt{|g|} (\nabla^{n} X^0 - \nabla^0 X^{n} )\right]
 = \frac 1{16 \pi} \left[\sqrt{|g|} (g^{n\mu} g^{0\nu} - g^{n\nu} g^{0\mu} )\partial_\mu(g_{\nu\lambda}X^\lambda)\right]
 \] \begin{equation} \label{KomartoK}
 = - \frac {\sqrt{|\det g_{kl}|}}{16 \pi N} \left( 2NX^i K^n{_i}-\partial_0 X^n +N^iD_i X^n -X^iD_i N^n
 \right)
% =  \left(\pi^{n\mu}\Gamma^0_{\mu 0} -\pi^{0\mu}\Gamma^{n}_{\mu 0} \right)
 \, .
\end{equation}
Here, and elsewhere, $D$ denotes the covariant derivative operator of the space-metric $g_{ij}$.
Now, we have seen that
\bea
    \intvol  \left[({\cal L}_X A^{\lambda}_{\mu\nu}) \delta {\pi}^{\mu\nu} - ({\cal L}_X{\pi}^{\mu\nu}) \delta  A^{\lambda}_{\mu\nu}\right]\rd\Sigma_\lambda = (1)+(2)+(3)+(4)
  \, , \label{26IV15.16}
\eea
where $(1)$ has been defined in \eqref{4IV15.12}, $(2)$ in \eqref{26IV15.12}, $(3)$ in \eqref{4IV15.12+} and $(4)$ in \eqref{tgtboundtermpidpi(4)}.
Let us  take  an extension  $X$ of the vector field $Y$ (which is tangent to $\mcV$ and defined so far only on $\mcV$) such that  $X^0\equiv   0$.
Then $(1)$ and $(3)$  vanish  by \eq{4IV15.12} and \eq{4IV15.12+}. From \eq{26IV15.12} and \eq{tgtboundtermpidpi(4)} one finds
% \ptcr{stopping here 26 IV 15 evening}
%
\begin{eqnarray}
 \nonumber
  \lefteqn{
\intvol
({\cal L}_X A^{0}_{\mu\nu}) \delta_f {\pi}^{\mu\nu} - ({\cal L}_X{\pi}^{\mu\nu}) \delta_f  A^{0}_{\mu\nu}
 }
  &&
\\
  \nonumber &=& { (2)} +
\frac 1{16 \pi}  \intvol   \left({\cal L}_X g_{kl} \delta_f P^{kl} - {\cal L}_X P^{kl}\delta_f g_{kl}\right)
   \\
\label{tgtboundtermpidpi}
 &  &
  - \frac1{16 \pi}
  \intdvol \left[ {\cal L}_X \frac{\sqrt{\det g_{mn}}}{N} \delta N^n
 - {\cal L}_X N^n \delta \frac{\sqrt{\det g_{mn}}}{N}
\right] \, .
\end{eqnarray}
Using  \eq{pi-dA0} and \eq{KomartoK} to rewrite the right-hand side of  \eq{tgtdL-Waldbis}, with some work \eq{tgtboundtermpidpi} can be rewritten as
\begin{eqnarray} \nonumber
    \lefteqn{ 16\pi \underbrace{{ \intdvol
-\partial_0 X^n \delta \pi^{00}} }_{=(2)} + \intvol
\left({\cal L}_X g_{kl} \delta P^{kl} - {\cal L}_X P^{kl}\delta
g_{kl}\right) =} \nn && \\ & & \intdvol \left\{ 2\delta(P^n{_i}X^i)-X^nP^{kl}\delta
g_{kl} \right\}
\nonumber
 %\intdvol \left\{ 2\delta(P^n{_i}X^i)-X^nP^{kl}\delta g_{kl} \right\}
+ \underbrace{\intdvol
\partial_i \left[\frac{\sqrt{\det g_{kl}}}{N}
\left( X^i\delta N^n - X^n\delta N^i\right)\right]}_{=0}
\\
  \label{tgtgdP-Pdg}
& & + \intdvol \left[ { \partial_0
X^n\delta\frac{\sqrt{\det g_{kl}}}{N}} -2P^n{_i}\delta X^i \right] -32  \pi  \intvol
   X^i\delta {\cal E}^0{}_{i}
    \,.
\end{eqnarray}
The first term in the first line cancels out the first term in the last line, providing the required result.
\qedskip

%% file: tangentXsubset.tex
It has been shown in~\cite{KijowskiGRG}, for variations such that the image of $V$ in $\mcM$ remains fixed (equivalently, $\delta \psi^0=0$), that we have
\begin{eqnarray} %& &\hspace*{-1cm}
{\pi}^{\mu\nu} \delta   A^{0}_{\mu\nu}|_{\delta \psi^0=0}
  &= & {\frac{(n-3)}{16 \pi} \delta  (\sqrt{|\det g_{mn}|}K)} - \frac 1{16 \pi} g_{kl} \delta  P^{kl}
 \nonumber \\ & & +
\partial_k \left[
{\pi}^{00} \delta  \left( \frac {{\pi}^{0k}}{{\pi}^{00}}
\right) \right]   \, ;
 \phantom{xxxxx}
 \label{pi-dA0}
\end{eqnarray}
recall that we allow $\det g_{mn}$ to have either sign.
Here we use the usual definitions
\begin{eqnarray}
 K_{kl} & := & - \frac 1{\sqrt{|g^{00}|}} {\Gamma}^0_{kl} =
  - \frac 1{\sqrt{|g^{00}|}} A^0_{kl}
 \, ,
\\
  \label{Pkl}
 P^{kl} & := & \sqrt{|\det g_{mn}|} \ (K {\tilde g}^{kl} - K^{kl} )
 \, ,
\end{eqnarray}
where ${\tilde g}^{kl}$ is the $n$-dimensional inverse
of the metric  $g_{kl}$  induced by $g_{\mu\nu}$ on $\mcV$. (More precisely, the calculation in~\cite{KijowskiGRG} has been done in dimension $3+1$, but the same calculation in higher dimensions leads to the formula above.) To avoid the overburdening of notation, we will not make a distinction between the fields on $\mcV$ and their pull-backs to $V$.

We use \eq{pi-dA0} as follows: Consider any differentiable family of fields $t\mapsto (\pi(t),A(t))$. We first note the identity
\begin{eqnarray}
 \partialstot  A^{0}_{\mu\nu}\delta{\pi}^{\mu\nu} -\partialstot {\pi}^{\mu\nu} \delta A^{0}_{\mu\nu}
 & = &
  -\partialstot ({\pi}^{\mu\nu}\delta A^{0}_{\mu\nu}) + \delta({\pi}^{\mu\nu}\partialstot  A^{0}_{\mu\nu})
   \nonumber
\\
 &&
    + \pi^{\mu\nu} [\partialstot ,\delta] A^0_{\mu\nu}
  \, .
   \label{8IV15.1}
\end{eqnarray}
\Eq{pi-dA0} implies
\begin{eqnarray}
\lefteqn{ \partialstot  ({\pi}^{\mu\nu} \delta  A^{0}_{\mu\nu})=}
 & & \nonumber \\ & & \partialstot \left(
 {\frac{(n-3)}{16 \pi} \delta (\sqrt{|\det g_{mn}|}K)}
  - \frac 1{16 \pi} g_{kl} \delta P^{kl} +
\partial_k \left[
{\pi}^{00} \delta \left( \frac {{\pi}^{0k}}{{\pi}^{00}}
\right) \right] \right)  \, ,
 \phantom{xxxx}
\\
\lefteqn{ \delta
 ({\pi}^{\mu\nu}\partialstot  A^{0}_{\mu\nu}) = }
 & &  \nn \\
  && \delta \left({\frac{(n-3)}{16 \pi} \partialstot  (\sqrt{|\det g_{mn}|}K)} - \frac 1{16 \pi} g_{kl} \partialstot  P^{kl} +
\partial_k \left[
{\pi}^{00} \partialstot  \left( \frac {{\pi}^{0k}}{{\pi}^{00}}
\right) \right] \right)  \, .
 \phantom{xxxxx}
 \label{8IV15.2}
\end{eqnarray}
Inserting this into the right-hand side of \eq{8IV15.1}, we find
\begin{eqnarray}
\lefteqn{
 \partialstot  A^{0}_{\mu\nu}\delta{\pi}^{\mu\nu} -\partialstot {\pi}^{\mu\nu} \delta A^{0}_{\mu\nu}
 =
  \frac 1{16 \pi}    \left(\partialstot  g_{kl} \delta P^{kl} - \partialstot  P^{kl}\delta g_{kl}
    \right)
    }
    &&
     \nn
  \\
 &  &
 -{\frac{(n-3)}{16 \pi} [\partialstot ,\delta] (\sqrt{|\det g_{mn}|}K)}
  + \frac 1{16 \pi} g_{kl} [\partialstot ,\delta] P^{kl} +
 [\delta, \partial_k] \left[
{\pi}^{00} \partialstot  \left( \frac {{\pi}^{0k}}{{\pi}^{00}}
\right) \right]
     \nn
  \\
 &  & -
 [\partialstot , \partial_k] \left[
{\pi}^{00} \delta \left( \frac {{\pi}^{0k}}{{\pi}^{00}}
\right) \right]
 +
 \partial_k  \left[
{\pi}^{00} [\partialstot ,  \delta] \left( \frac {{\pi}^{0k}}{{\pi}^{00}}
\right) \right]
     \nn
  \\
 &  &
  -
  \partial_k \left[ \partialstot {\pi}^{00} \delta \left( \frac {{\pi}^{k0}}{{\pi}^{00}} \right)
 - \partialstot \left( \frac {{\pi}^{k0}}{{\pi}^{00}} \right) \delta {\pi}^{00}
\right]
 \, .
  \phantom{xxx}
 \label{boundtermpidpi+k+}
\end{eqnarray}
When $\partialstot=\partial_{y^0}\equiv \partial_0$ and all commutators vanish,
one obtains
\begin{eqnarray}
 \partial_0 A^{0}_{\mu\nu}\delta{\pi}^{\mu\nu} -\partial_0{\pi}^{\mu\nu} \delta A^{0}_{\mu\nu}
 & = &
  \frac 1{16 \pi}    \left(\partial_0 g_{kl} \delta P^{kl} - \partial_0 P^{kl}\delta g_{kl}
    \right)
     \nn
  \\
\label{boundtermpidpi+k}
 &  &
  -
  \partial_k \left[ \partial_0{\pi}^{00} \delta \left( \frac {{\pi}^{k0}}{{\pi}^{00}} \right)
 - \partial_0\left( \frac {{\pi}^{k0}}{{\pi}^{00}} \right) \delta {\pi}^{00}
\right]
 \, .
  \phantom{xxx}
\end{eqnarray}

%% file: tangentX.tex
\input{wzoryADM}

Adding all the results above, and using the formula for ${\cal L}_X \lambda$ (see \eq{4VI15.6}, Appendix~\ref{A4VI15.1})) we end up with the following identity when $X=X^0\partial_0$:

\bean
 \lefteqn{
  \intvol   \left({\cal L}_X g_{kl} \delta P^{kl} - {\cal L}_X P^{kl}\delta g_{kl}\right)
 + 2 \intdvol  ({\cal L}_X \alpha\delta\lambda - {\cal L}_X \lambda \delta \alpha
 +
 { \lambda\nu^A \partial_A X^0 \delta\alpha})
 }
 &&
\\
 \nn
 &= &
  \intdvol  \partial_A X^{0} \left[ N \lambda
 { \sqrt{{\tilde g}^{nn}} \delta
 \left(\frac{{\tilde g}^{nA}}{{\tilde g}^{nn}}\right)}
 + \sqrt{g}\left( { \frac{N^A\delta N^n}{N}
 -  \frac{N^n\delta N^A}{N} } \right) \right]
\\
 \nn
  &&
+ \intdvol X^0 (   2 \nthree\delta {\bf Q}
- 2\nthree^A \delta {\bf Q}_A + { \nu}{\bf Q}^{AB}
 \delta g_{AB}) + 16\pi\intdvol X^0 \partial_A \left[ {\pi}^{nn} \delta \left( \frac {{\pi}^{nA}}{{\pi}^{nn}} \right) \right]
\\
  &&
  + 16 \pi  \intvol X^0\left(
  -2\delta {\cal E}^0{}_{0}
   +  {\cal E}^ {\mu\nu}  \delta g_{\mu\nu}
    \right)
 \;.
  \label{summa}
 \eea
Using the identity
\[ N^A=\nu^A + N^n \left(\frac{{\tilde g}^{nA}}{{\tilde g}^{nn}}\right) \, ,
\]
one finds
 \[
 \intdvol  \partial_A X^{0} \left[ N \lambda
 { \sqrt{{\tilde g}^{nn}} \delta
 \left(\frac{{\tilde g}^{nA}}{{\tilde g}^{nn}}\right)}
 + \sqrt{g}\left( { \frac{N^A\delta N^n}{N}
 -  \frac{N^n\delta N^A}{N} } \right)  - 2  { \lambda\nu^A  \delta\alpha} \right]
\]
\[ = \intdvol \partial_A X^{0} N \sqrt{g} g^{nn}\delta \left(\frac{g^{nA}}{g^{nn}}\right)
= - 16\pi \intdvol  X^{0} \partial_A \left[\pi^{nn} \delta \left(\frac{\pi^{nA}}{\pi^{nn}}\right)\right]
 \,.
\]
This, together with some obvious cancelations of boundary terms in \eq{summa} ends the proof of \eq{finalresultbis}.
\qed

%% file: wzoryADM.tex
%wzory ADM

To calculate $(**)$, we will need the usual ADM formulae \eq{4VI15.21} for an $(n+1)$ splitting, together with
$$\Gamma^0_{k0}=\partial_k\log N -\frac{N^l}N K_{lk} \, ,\quad
\Gamma^0_{00}= \partial_0\log N +N^k \partial_k\log N -\frac{N^lN^k}N K_{lk}
%\, ,
$$
(recall that
$K_{kl}=-N\Gamma^0_{kl}=\frac1{2N}(D_l N_{k} +D_k N_{l}-\partial_0 g_{kl})$). Furthermore,
\[  \Gamma^k_{ij}= {\tilde\Gamma}^k_{ij} { +}\frac{N^k}N K_{ij}
 \,,
 \quad
 \Gamma^k_{0j}= D_j N^k -NK^k{_j}+ \frac{N^k}N \left( N^lK_{lj}- D_j N \right)
 \,.
\]
Using the above, we find
\begin{eqnarray}
 \nonumber
(b) & = &  N g^{k\mu}\Gamma^{0}_{\mu 0} - Ng^{0\mu}\Gamma^{k}_{\mu 0} = 2D^k N +
\frac1N \left( {\tilde g}^{kl}{\dot N}_l - D_l N^k N^l-N_l D^k N^{l} \right) \\
& =& 2D^k N + \frac1N {\dot N}^k - 2 N_l K^{lk}
 \,,
\\
(c) & = &
%=16\pi \left[ \partial_0{\pi}^{00} \delta \left( \frac {{\pi}^{k0}}{{\pi}^{00}} \right)
% - \partial_0\left( \frac {{\pi}^{k0}}{{\pi}^{00}} \right) \delta {\pi}^{00}
%\right]
%\\
%  & = &
    \partial_0 (\frac{\sqrt{\det g_{mn}}}{N})\delta N^k -\dot N^k\delta (\frac{\sqrt{\det g_{mn}}}{N})
     \,.
\eea
To calculate $(a)$, the following intermediate results are useful:
\begin{equation} A^k_{00}=\Gamma^k_{00}=
\underbrace{{\tilde g}^{kl}{\dot N}_l -N_l D^k N^{l}}_{{\dot N}^k +N_l D^lN^{k} - 2N N_l K^{lk}}
+ND^k N+
\frac{N^k}{N}\left( N^iN^jK_{ij}-\dot N - N^l D_l N \right) \end{equation}

\[ A^k_{0j}=\Gamma^k_{0j}-\frac12\delta^k_j \Gamma^\lambda_{0\lambda}=
\Gamma^k_{0j} -\frac12\delta^k_j \Gamma^l_{0l} -\frac12\delta^k_j \Gamma^0_{00}
\]
\begin{equation} A^k_{0j} = D_j N^k -N K^k{_j} +\frac{N^k}N (N^lK_{lj}-D_j N)-\frac12 \delta^k_j \left(
\partial_0\log N +D_l N^l -N K\right) \end{equation}

\bean
 A^k_{ij}
  &= &
   \Gamma^k_{ij}-\delta^k_{(j} \Gamma^\lambda_{i)\lambda}= \Gamma^k_{ij}-\delta^k_{(j} \Gamma^l_{i)l} -\delta^k_{(j} \Gamma^0_{i)0}
\\
    & = &
     {\tilde\Gamma}^k_{ij} -\delta^k_{(j}{\tilde\Gamma}^l_{i)l} -
\delta^k_{(j}\partial_{i)}\log N { +} \frac{N^k}N K_{ij}
\eea
%
% \ptcr{nastepny chyba niepotrzebny}
%\[ 16\pi\pi^{\mu\nu}A^k_{\mu\nu}= \sqrt{g}\left\{ 2N^k  K + N  \left( {\tilde g}^{ij}{\tilde\Gamma}^k_{ij}
%-{\tilde g}^{ik}{\tilde\Gamma}^l_{il}\right) -2 D^k N +
%\frac{N^k}{N}\partial_j N^j - \frac{N^j}{N}\partial_j N^k -\frac1N
%\partial_0 N^k \right\}\]
%{ Trzeba zebrac to do kupy}

We pass now to
\[
 \partial_k X^0 \frac{(a)}{16 \pi} =
 \partial_k X^0 \pi^{\mu\nu}\delta A^{k}_{\mu\nu} = \partial_k X^0 \left(\pi^{00}\delta A^{k}_{00} + 2\pi^{0j}\delta A^{k}_{0j}
+ \pi^{ij}\delta A^{k}_{ij} \right)
 \,.
  \]%
%To continue our analysis we have to analyze our most difficult volume term:
%\begin{equation} \label{**}
%\frac{(**)}{16\pi} = \pi^{\mu\nu}\delta A^{k}_{\mu\nu} +\delta(\pi^{k\mu}\Gamma^{0}_{\mu 0} -\pi^{0\mu}\Gamma^{k}_{\mu 0}) +
%\partial_0{\pi}^{00} \delta \left( \frac {{\pi}^{k0}}{{\pi}^{00}} \right)
% - \partial_0\left( \frac {{\pi}^{k0}}{{\pi}^{00}} \right) \delta {\pi}^{00}
%\end{equation}
Using the above, one can first notice that $(**)$ does not contain $\partial_0 N$ or $\partial_0 N^k$.
After some further work we obtain:
\begin{eqnarray}\nonumber
(**) &=& N \sqrt{g} \left( {\tilde g}^{ij} -\frac{N^iN^j}{N^2} \right)
\delta \left(\tilde\Gamma^{k}_{ij} - \delta^k_{(i}\tilde\Gamma^{l}_{j)l} \right)
+2D^k N \delta\sqrt{g}+\sqrt{g}D_i N \delta {\tilde g}^{ki} + \\ & & \nonumber
 + \frac{\sqrt{g}}{N^2} D_i N (N^k\delta N^i -N^i\delta N^k) - { \sqrt{g}K \delta N^k} \\ & & \nonumber
+\frac{\sqrt{g}}N \left[ N^i \delta D_i N^k - D_i N^k\delta N^i -N^k \delta (D_i N^i) + { D_i N^i\delta N^k}
\right]
\\ \label{volterm}
& & + \sqrt{g} N^k \delta K  - 2N^j\delta (\sqrt{g} K_j^k ) + \sqrt{g}{\tilde g}^{ji}\delta (N^k K_{ij})
 \,.
\end{eqnarray}
%\mnote{czlony na czerwono pochodza z $\partial_0{\pi}^{00} \delta \left( \frac {{\pi}^{k0}}{{\pi}^{00}} \right)$
%(bez $\dot N$) i tozsamosci $\partial_0\sqrt{g}=\partial_l(\sqrt{g}N^l) -N\sqrt{g}K$}
Those terms in  \eq{volterm} which depend on the shift vector $N^k$ but do not involve $K_{ij}$ may be simplified as follows:
\[ -\sqrt{g}  \frac{N^iN^j}{N}
\delta \left(\tilde\Gamma^{k}_{ij} - \delta^k_{(i}\tilde\Gamma^{l}_{j)l} \right)
 + \frac{\sqrt{g}}{N^2} D_i N (N^k\delta N^i -N^i\delta N^k) +\] \[
+\frac{\sqrt{g}}N \left[ N^i \delta D_i N^k - D_i N^k\delta N^i -N^k \delta (D_i N^i) + D_i N^i\delta N^k
\right] = \]
\[ = \partial_i\left( \sqrt{g} \frac{N^i\delta N^k}{N} - \sqrt{g} \frac{N^k\delta N^i}{N} \right)
 \,.
\]
After multiplying by $\partial_k X^0$ one obtains a full divergence:
\begin{eqnarray} \nonumber
\lefteqn{
 \partial_k X^0 \partial_i\left( \sqrt{g} \frac{N^i\delta N^k}{N} - \sqrt{g} \frac{N^k\delta N^i}{N} \right)
  }
   &&
 \\ \label{bdterm}
 &&
   =
 \partial_k \left[ X^0\partial_i\left( {\sqrt{g} \frac{N^i\delta N^k}{N}
 - \sqrt{g} \frac{N^k\delta N^i}{N} } \right)\right]
  \,.
\end{eqnarray}
We continue with all terms in  \eq{volterm} that involve $K_{ij}$:
\begin{equation} \label{KwV}
 {\sqrt{g} N^k \delta K } -{\sqrt{g} K \delta N^k} - 2N^j\delta (\sqrt{g} K_j^k ) + \sqrt{g}{\tilde g}^{ji}\delta (N^k K_{ij}) =
  2N^j\delta P_j{}^k   - N^k{P}^{ij}\delta g_{ij} \, .\end{equation}
Now, using \eq{4VI15.1} and \eq{4VI15.3}, Appendix~\ref{A4VI15.1},
one finds
\begin{eqnarray} \nonumber
 \lefteqn{
 {\cal L}_X g_{kl} \delta P^{kl}-{\cal L}_X P^{kl}\delta g_{kl}
}
&&
\\ & =&
    X^0[ \partial_0 g_{kl} \delta P^{kl}-\partial_0 P^{kl}\delta g_{kl} ] \nn
\\
&& +
   \underbrace{ D_k X^0 \left[ 2N_l\delta P^{kl} - (N^k{P}^{ij} - N^j{P}^{ik}-N^i{P}^{jk})\delta g_{ij}\right]}_{(*)}
  \nonumber
\\
 & &
   - D_r X^0  \left[ \sqrt{g}\left(
  {\tilde g}^{lr} D^ k N  + {\tilde g}^{kr}D^ l N  -  2 {\tilde g}^{kl}
  D^ r N
  \right)\right]\delta g_{kl}
  \nonumber \\ & &
- N \sqrt{g}\left( \nabla^k \nabla^l X^0 - {\tilde g}^{kl} \Delta X^0  \right)\delta g_{kl}
 \,.
  \label{4VI15.4}
\end{eqnarray}
We thus see that the right-hand side of  \eq{KwV} corresponds precisely to the $(*)$-part of
${\cal L}_X g_{kl} \delta P^{kl}-{\cal L}_X P^{kl}\delta g_{kl}$:
\[ \partial_k X^0\left[ 2N^j\delta  P_j{}^k   - N^k{P}^{ij}\delta g_{ij} \right] =
(*)
 \,.
\]
Using
\[ \partial_k X^0 \sqrt{\det g_{mn}}{\tilde g}^{ij} \delta \left(\tilde\Gamma^{k}_{ij} - \delta^k_{(i}\tilde\Gamma^{l}_{j)l} \right) = D^l X^0 \sqrt{\det g_{mn}}{\tilde g}^{ij}
(D_j\delta g_{li}-D_l\delta g_{ij})=
\]
 \[
 -\partial_l\left[ \sqrt{\det g_{mn}}\left(D^l X^{0 }{\tilde g}^{ij}\delta{ g}_{ij}  +D_k  X^0  \delta{\tilde g}^{kl}\right)\right] +
\sqrt{\det g_{mn}}\left({\tilde g}^{kl}D^iD_iX^0 - D_l D_k X^{0 }\right)\delta g_{kl}
\,, \]
all the  terms in $(**)  D_k X^0$ which involve $DN$ and $DD X^0$, and which we denote by $(***)$, take the form
\begin{eqnarray} \nonumber
(***)  & = & D_k X^0  \left[ N \sqrt{g} {\tilde g}^{ij}
\delta \left( \tilde\Gamma^{k}_{ij} - \delta^k_{(i}\tilde\Gamma^{l}_{j)l} \right)
+2D^k N \delta\sqrt{g}+\sqrt{g}D_i N \delta {\tilde g}^{ki} \right] \\
& = & \nonumber
 -\partial_l\left[ N\sqrt{g}\left(D^l X^0 {\tilde g}^{ij}\delta{ g}_{ij}
 + D_k X^0  \delta{\tilde g}^{kl}\right)\right]
\\
 \nn
 &&
 + 2 D_k X^0
 (2D^k N \delta\sqrt{g}+\sqrt{g}D_i N \delta {\tilde g}^{ki}) \\ & &
+ N\sqrt{g}\left({\tilde g}^{kl}D^iD_iX^0 - D^k D^l X^0 \right)\delta g_{kl} \, .
\end{eqnarray}
Comparing with the last two lines of \eq{4VI15.4}, we see that $(***)$ differs from the desired expression by a  divergence term. Further, all the volume terms in \eq{finalresultbis} have at this stage been accounted for. Now, this divergence term  leads to one more boundary integral, with integrand equal to
\begin{eqnarray} \nonumber
 \lefteqn{
 - N\sqrt{g}\left(D^n X^0 {\tilde g}^{ij}\delta{ g}_{ij}
 + D_k X^0  \delta{\tilde g}^{kn}\right)
 }
  &&
\\
 \nn
   &=& - N\sqrt{g} D_k X^0
 \left({\tilde g}^{ij}{\tilde g}^{nk} - {\tilde g}^{ik}{\tilde g}^{jn}\right) \delta{ g}_{ij} \\
 & = &
 \nonumber    {
 -2N \left({\sqrt{{\tilde g}^{nn}}} D_n X^0 +
  D_C X^0 \frac{{\tilde g}^{nC}}{\sqrt{{\tilde g}^{nn}}} \right) } \delta\lambda
 \\ & &
 + N\lambda \ttg^{AC}\partial_C X^0
 \underbrace{\left(\frac{{\tilde g}^{nB}}{\sqrt{{\tilde g}^{nn}}}\delta g_{AB}
 +\sqrt{{\tilde g}^{nn}}\delta g_{nA}\right)}_{ -\sqrt{{\tilde g}^{nn}}g_{AB} \delta
 \left(\frac{{\tilde g}^{nB}}{{\tilde g}^{nn}}\right)}
  \,.
 \label{dodczlonb}
 \end{eqnarray}
The before-last line above will be part of ${\cal L}_X \alpha$.

%% file: ShortAreaHamiltonian.tex
\subsection{$\kappa_*^{-1} X_*$}
 \label{ss28III15.4}

We have seen so far two sets of variational identities of thermodynamical type which are satisfied in Kerr-de Sitter spacetimes, and one can clearly produce an infinite collection of such identities by considering all possible Hamiltonian dynamical systems associated with all Hamiltonian asymptotic Killing vectors of Theorem~\ref{T18VII14.1}. It is thus natural to raise the question, whether any such identities  are singled-out by the geometry.
It turns out that this is indeed the case, and can be seen as follows:

Recall that the equation $\Delta_r=0$ can have up to four distinct roots, and that there are always at least two distinct zeros when no naked singularities occur. Further, there is always a negative simple root. Consider, then, a Kerr-de Sitter spacetime with a $a^2 \Lambda <3$, and choose a \emph{non-degenerate} Killing horizon $\mcH_*=\{r=r_*\}$.
There exists precisely one preferred Killing vector associated with this horizon, namely the unique Killing vector which is tangent to $\mcH_*$ and has surface gravity equal to one:
\bel{28III15.21}
 \tilde X_* :=  \kappa_*^{-1} X_* =
 \frac{2(r^2_*+a^2) \Xi }  {\partial_r \Delta_r \big|_{r=r_*} } \left(\partial_t + \frac{a}{a^2+r_*^2} \partial_\varphi\right)
  \,,
\ee
where $X_*$ has been given in \eq{3X13.2}, and we have used \eq{5X13.2}.

It turns out that $\tilde X_*$ is Hamiltonian. To see this, we rewrite \eq{28III15.4+} as
\bea
 \delta \left(\frac{A_*}{8\pi} \right)
  &  = &
   \frac 1 {\kappa_*}\left(\delta M_H -\Omega_* \delta J \right)
   \,.
\eeal{nullham}
This means that the Hamiltonian $ {A_*}/{8\pi}$ generates the vector field
$$
  \frac 1 {\kappa_*}\left(X_{\mathrm{KS}} + \Omega_* \partial_\varphi \right) =  {\kappa_*^{-1} }X_*
\,.
$$
Equivalently, the flow of  $  {\kappa_*^{-1} }X_*$ is generated by
\bel{28III15.22}
 \displaystyle \tilde M_H := \frac{A_*}{8\pi}=\frac{r^2_*+a^2}{2\Xi}
 \,.
\ee

Consider, then, two Killing horizons with areas $A_a$ and generating Killing vectors $X_a$.
Suppose that $A_1$ is non-degenerate and normalise $X_1$ to unit surface gravity. We will  denote by $\Omega_{21}$ the relative angular velocity of the second horizon with respect to the first one:
\bean
 X_1
  & = &
 \frac{2\Xi }  {\partial_r \Delta_r \big|_{r=r_1} } ((r^2_1+a^2) \partial_t + a\partial_\varphi)
  \,,
\\
 X_2
  &  = &
 \frac{2 \Xi }  {\partial_r \Delta_r \big|_{r=r_2} } ((r^2_2+a^2)\partial_t + a \partial_\varphi)
  \,,
   \nonumber
\\
  X_1
  & = &
 \underbrace{\frac{(r^2_1+a^2)\partial_r \Delta_r \big|_{r=r_2}}  {(r^2_2+a^2)\partial_r \Delta_r \big|_{r=r_1} }}_{=:\kappa_{21}}  X_2  + \underbrace{\frac{ 2 a \Xi ( r_2^2-r_1^2)}  { (r^2_2+a^2)\partial_r \Delta_r \big|_{r=r_1} }}_{-\Omega_{21}} \partial_\varphi
  \,.
\eeal{28III15.31}
We then have the variational identity
\bean
 \delta \left(\frac{A_1}{8\pi} \right)
  & = &
   \frac{\kappa_{21}}{8\pi}  \delta  {A_2}  +\Omega_{21} \delta J
\\
  & = &
    \frac{\kappa_{2 }}{8\pi\kappa_1}   \delta  {A_2} + \frac{\Omega_{2 }-\Omega_1}{\kappa_1} \delta J
   \,,
\eeal{29III15.1}
 with $\kappa_i$ given by \eq{5X13.2}, and $\Omega_i$ given by the last line of \eq{3X13.2}.

For Schwarzschild-de Sitter spacetime, the areas $A_*$ are determined by $r^2_*$, and we note the following relation between the roots of  the polynomial $\Delta_r|_{a=0} $
\bel{29III15.2}
 r_1^2 + r_1  r_2 +  r_2^2 =\frac 3 \Lambda =: \ell^2
%\,.
\ee
which allows one to express explicitly $A_1$ as a function of $A_2$.

In the general case $a\ne 0$, \eq{3X13.4} can be rewritten as
\bel{30III15.2}
 A_*=      4 \pi \ell^2  \frac{M^2 r^2_*+ {J^2} } {M^2 \ell^2 +  {J^2} }\,;
 \
 \mbox{\rm recall that $M=\Xi^{-2}m$ and $J=Ma$, $\ell^2= 3 \Lambda^{-1}$.}
%       \, .
\ee
This can be solved for $r_*$ as a function of $A_*$, $M$ and $J$:
\bel{7VII15.1}
 r_*^2 = \frac{(\ell ^2 M^2+J^2)A_* -4 \pi  \ell ^2 J^2}{4 \pi  \ell ^2 M^2}
 \,.
\ee
Rewriting the equation $\Delta_r=0$ as
\bel{7VII15.2}
 (r^2+a^2)^2\left(1-\frac{\Lambda}3 r^2\right)^2-4 m^2 r^2 =0
\ee
one obtains, dropping stars:
\bel{30III15.1}
  z^2(1 -z)^2 -   4z \mu^2  - 4 j^2(z-1) =0\,,
%       \, .
\ee
where
\bel{30III15.1+}
 \mbox{\rm  $z:=A/(4\pi\ell^2) $, $\mu:=   \ell^{-1}M   $, and $j = \ell^{-2} J$.}
%       \, .
\ee
Let $z_a$, $a=1,2$, be two distinct roots of \eq{30III15.1}. Eliminating $\mu$ between the resulting equations, one finds the relation
\bel{1IV15.1}
 {z_{1} z_{2}
   \left(1-2z_1 -2
   z_{2}+z_{1}z_2+z_{1}^2+z_{2}^2\right)} =  4 j^2
    \,.
\ee
This can be explicitly solved for $z_2(z_1)$, but the resulting formulae are not very illuminating.
 %\ptcr{the following material went to OriginalAreaHamiltonian, to be returned to later}
A detailed analysis of the resulting equations will be presented elsewhere.

%% file: Notations.tex
\section{Notations}
 \label{A4VI15.2}

We summarize some of our notations, which largely follow~\cite{KijowskiGRG} with, however, some exceptions:

The coordinates $y^\mu$ are always local coordinates on the spacetime $\mcM$. In Section~\ref{s16IV13.1} the coordinates $(t,\vec x)$ are local coordinates on $\R\times \Sigma$. However, $t$ denotes a Boyer-Lindquist time coordinate when discussing Kerr-de Sitter and asymptotically Kerr-de Sitter metrics.

The coordinate $y^0$ is constant on $\mcV:= \psi_0(V)$.

The symbol $\tilde g^{ij}$, with $i,j\in \{1,\ldots,n\}$, denotes the $n$-dimensional inverse of the metric $g_{ij}$ induced on the level sets of $\psi_t(V)$. Strictly speaking, we have a one-parameter family of metrics $g_{ij}(t)$, but we will not use this notation. We use the symbol $D$ to denote the covariant derivative operator of $g_{ij}$.

The coordinate $y^n$ is constant on $\partial \mcV$.

The symbol  $\hat g^{ab}$, with $a,b\in \{0,1,\ldots,n-1\}$, denotes the $n$-dimensional inverse of the metric $ g_{ab}$ induced on the level sets of $y^n$.

The symbol $\ttg^{AB}$, with $A,B\in \{ 1,\ldots,n-1\}$, denotes the $(n-1)$-dimensional inverse of the metric $g_{AB}$ induced on $\partial \mcV$.

%% file: Liederivatives.tex
\section{Lie derivatives of geometric fields on $\mcV$}
 \label{A4VI15.1}

The aim of this appendix is to derive the derivative operators which are obtained by restricting the Lie derivative of spacetime objects to a non-characteristic hypersurface. For composite objects such that $\alpha$ or $\lambda$, the method is to use systematically the chain-rule together with the spacetime expression for the Lie derivative of the relevant components of the metric. We use the ADM decomposition of the metric:
\bel{4VI15.21}
    \mbox{
$g^{kl}={\tilde g}^{kl}-\frac{N^lN^k}{N^2}$,
$g_{0k}=N_k$, $g^{0k} =\frac{N^k}{N^2}$, $N^2=-\frac1{g^{00}}$, $g_{00}=N^kN_k-N^2$.}
\ee

Here, as elsewhere, we let $X=X^0\partial_0+Y$, with $Y = X^i \partial_i$. We will use the symbol ${\cal L}_X$ to denote the restriction to the hypersurface $\{y^0=\const\}$ of the spacetime Lie derivative operator, while $\mymcL_Y$ is the usual Lie derivative operator on the level sets of the function $y^0$, viewed as an $n$-dimensional manifold of its own.

\subsection{The induced metric}

We start with the following straightforward formulae:
\begin{eqnarray}
{\cal L}_X g_{kl} &=&
  X^0 \partial_0 g_{kl} + N_k \partial_l X^0 +
 N_l  \partial_k X^0
  + {\cal L}_Y g_{kl}
  \label{4VI15.1}
   \,,
\\
{\cal L}_X {\tilde g}^{kl} &=& %\nonumber
 - {\tilde g}^{ki}{\tilde g}^{lj}{\cal L}_X g_{ij} =
 X^0 \partial_0 {\tilde g}^{kl} - N^k D^l X^0 - N^l  D^k X^0
  + {\cal L}_Y {\tilde g}^{kl}
  \label{4VI15.2}
    . \phantom{XX}
%\\
% \nonumber
%  {\cal L}_X P^{kl} &=& X^0 \partial_0 P^{kl}  + N \sqrt{{g}}\left( \nabla^k \nabla^l X^0 - {\tilde g}^{kl} \Delta X^0
%   \right) \\ \nonumber
%  & & + \left\{ \sqrt{g}\left(
%  {\tilde g}^{lr} D^k N  + {\tilde g}^{kr}D^l N  - {2} {\tilde g}^{kl}
%  D^r N
%  \right) \right.\\
%  & &  + \left. P^{kl} N^r -P^{kr}N^l -P^{lr}N^k
%    \right\} \partial_r X^0
%  \label{4VI15.3}
\end{eqnarray}

\subsection{The volume density $\lambda$}
We pass now to
\[ \lambda^2=\det g_{AB}=\det g_{\mu\nu} \left(g^{00}g^{nn}-g^{0n}g^{0n}\right) \, , \]
which can equivalently be written as
\[ \left(-g^{00}g^{nn}+g^{0n}g^{0n}\right)^{1/2}=\frac{\lambda}{N\sqrt{g}}=\frac{\sqrt{|{\tilde g}^{nn}|}}{N}
 \,.
\]
The definition of the Lie derivative
\[ {\mycal L}_X g^{\mu\nu}= X^\lambda \partial_{\lambda} g^{\mu\nu}
-\partial_\alpha X^\mu g^{\alpha\nu}-\partial_\alpha X^\nu g^{\mu\alpha}
 %\,,
\]
together with
\begin{equation}\label{Liesqrg}
     {\mycal L}_X {\sqrt{|\det g_{\alpha\beta}|}} = \partial_\mu\left(X^\mu {\sqrt{|\det g_{\alpha\beta}|}} \right)
%     \,.
\end{equation}
gives the following for $X$ of the form $X=X^0\partial_0$:
\begin{equation}
{\cal L}_X \lambda =
 X^0 \partial_0 \lambda + \left(N^A - \frac{{\tilde g}^{nA}}{{\tilde g}^{nn}}N^n\right) \lambda \partial_A X^0 =
 X^0 \partial_0 \lambda + \lambda\nu^A \partial_A X^0
  \,.
   \label{4VI15.6}
\end{equation}
For further reference we have
\begin{equation}
\label{pochodnaLiegoPiwApendiksie}
{\mycal L}_X \pi^{\mu\nu}= \partial_{\lambda}\left( X^\lambda  \pi^{\mu\nu}\right)
-\partial_\alpha X^\mu \pi^{\alpha\nu}-\partial_\alpha X^\nu \pi^{\mu\alpha}
 \, .
\end{equation}

\subsection{The angle $\alpha$}

Using the definition \eq{defq} of $q$ and   \eq{4VI15.2}, we get
\[ \frac2q {\cal L}_X q= \frac2{g^{0n}} {\cal L}_X g^{0n}
-\frac1{g^{00}} {\cal L}_X g^{00} -\frac1{g^{nn}} {\cal L}_X g^{nn}
 \,,
\]
which gives, again for $X=X^0\partial_0$,
\begin{eqnarray}
{\cal L}_X q &=&
 X^0 \partial_0 q -\frac{{\tilde g}^{nk}}{\sqrt{|g^{00}g^{nn}|}}\partial_k X^0
  \,,
\end{eqnarray}
and finally for $q=\sinh\alpha$ we have the following:
\begin{eqnarray}
{\cal L}_X \alpha &=&
 X^0 \partial_0 \alpha -\frac{{\tilde g}^{nk}\partial_k X^0}{\cosh\alpha\sqrt{|g^{00}g^{nn}|}} =
  X^0 \partial_0 \alpha {  -\frac{N {\tilde g}^{nk}}{\sqrt{|{\tilde g}^{nn}|}}\partial_k X^0 }
   \,.
\end{eqnarray}

\subsection{The ADM momentum}

The definition \eq{Pkl} of $P^{kl}$ can be rewritten as follows:
\begin{eqnarray} \label{Pkl+}
 P^{kl} & := &  \sqrt{\det g_{mn}} \ (K {\tilde g}^{kl} - K^{kl} )
 \nonumber \\
 &=&
  \frac 1{\sqrt{|g^{00}|}} \sqrt{\det g_{mn}} \ {\Gamma}^0_{mn}
 \left( {\tilde g}^{mk}{\tilde g}^{nl}-{\tilde g}^{mn}{\tilde g}^{kl}\right)
 = \sqrt{|\det g_{\alpha\beta}|} \ {\Gamma}^0_{mn}\left( {\tilde g}^{mk}{\tilde g}^{nl}-{\tilde g}^{mn}{\tilde g}^{kl}\right)
 \nonumber \\
 &=&
 {\cal K}_{mn} \left( {\tilde g}^{mk}{\tilde g}^{nl}-{\tilde g}^{mn}{\tilde g}^{kl}\right)
 %\ ,
\end{eqnarray}
(summation over spatial indices $m,n$ only),
where we have defined the purely spatial tensor density
\[
    {\cal K}_{mn} := \sqrt{|\det g_{\alpha\beta}|} \ {\Gamma}^0_{mn}  \equiv - \sqrt{\det g_{mn}} K_{mn}  \ .
\]

Now, when $X$ has no zeros,
${\dot P}^{kl}$ can be defined as a derivative with respect to time, as calculated in a coordinate system in which $X=\partial_t$. This definition implies  the Leibniz rule. Applying it to \eq{Pkl+}, we express ${\dot P}^{kl}$ in terms of ``dots'' acting on $\Gamma^{\lambda}_{\mu\nu}$ and $g_{\mu\nu}$, i.e. in terms of Lie derivatives of the metric and of the connection. To obtain the general formula, valid for any vector field, we use  \eq{2IV15.5} to obtain
\begin{eqnarray}
 {\mycal L}_X \Gamma^{\lambda}_{\mu\nu} &=&
 \nabla_\mu \nabla_\nu X^\lambda - X^\sigma R^\lambda{_{\nu\mu\sigma}} \nonumber\\
 &=& \partial_\mu \partial_\nu X^\lambda + \Gamma^\lambda_{\mu\sigma}\partial_\nu X^\sigma
 + \Gamma^\lambda_{\nu\sigma}\partial_\mu X^\sigma - \Gamma^\sigma_{\mu\nu}\partial_\sigma X^\lambda + X^\sigma \partial_\sigma \Gamma^\lambda_{\mu\nu} \, ,\nonumber
 \\
 {\cal L}_X \Gamma^{0}_{kl}
 &=&
 \partial_k \partial_l X^0 + \Gamma^0_{kn}\partial_l X^n +
 \Gamma^0_{k0}\partial_l X^0
 + \Gamma^0_{ln}\partial_k X^n + \Gamma^0_{l0}\partial_k X^0 \nonumber\\
 & &
 - \Gamma^0_{kl}\partial_0 X^0 - \Gamma^n_{kl}\partial_n X^0
 +
 X^\sigma \partial_\sigma \Gamma^0_{kl}
 %\ ,
 \label{LieK}
% \\
% {\cal L}_X  g_{kl} &=&
% X^\lambda \partial_\lambda g_{kl} +g_{k\mu} \partial_l X^\mu
% + g_{l\mu} \partial_k X^\mu \nonumber \\
% &=& X^0\partial_0 g_{kl}+
% {\cal L}_{Y}g_{kl} + N_k \partial_l X^0 + N_l \partial_k X^0  \,. \label{Liegdown} \\
% \\
%  {\cal L}_X {\tilde g}^{rs} &=& - {\tilde g}^{rk} \left({\cal L}_X g_{kl}
%  \right) {\tilde g}^{ls} \nonumber \\
%  &=& X^0 \partial_0 {\tilde g}^{rs} +
%  {\cal L}_{Y}{\tilde g}^{rs} - \left( N^r D^s X^0 +
%  N^s D^r X^0 \right)
 \,.
% \label{Liegup}
\end{eqnarray}
Recall that $A^0_{k0} = \Gamma^0_{k0}  - \frac 12 \Gamma^\lambda_{k\lambda}$.
Using the constraints \eq{A000}
\begin{eqnarray}
 \nn
    A^0_{0k} & = & - \frac 1{2 {\pi}^{00}} \left(
    \partial_k {\pi}^{00} + 2 A^0_{kl} {\pi}^{0l} \right)
\\
 \nn
 &=&
  - \frac 12 \frac
    {\partial_k \left( g^{00} \sqrt{|\det g_{\alpha\beta}|} \right)}{g^{00} \sqrt{|\det g_{\alpha\beta}|}} - \Gamma^0_{kl} \frac{g^{0l}}{g^{00}}
\\
 &=&
  -\frac 12 \frac {\partial_k{\sqrt{|\det g_{\alpha\beta}|}}}{{\sqrt{|\det g_{\alpha\beta}|}}} - \frac 12 \frac {\partial_k g^{00}}{g^{00}} - \Gamma^0_{kl} \frac{g^{0l}}{g^{00}}
 \, ,
  \label{5VI15.1}
\end{eqnarray}
together with the formula
\begin{equation}\label{gammak}
       \Gamma^\lambda_{k\lambda} = \frac {\partial_k{\sqrt{|\det g_{\alpha\beta}|}}}{{\sqrt{|\det g_{\alpha\beta}|}}}
%        \,,
\end{equation}
we find
\[
    \Gamma^0_{k0} =  - \frac 12 \frac {\partial_k g^{00}}{g^{00}} - \Gamma^0_{kl} \frac{g^{0l}}{g^{00}} \,.
\]
The identity
\[
    g_{m0}\Gamma^{0}_{kl} + g_{mn}\Gamma^{n}_{kl} = \Gamma_{nkl} =
    g_{mn}\tilde{\Gamma}^{n}_{kl}
%    \,,
\]
implies
\begin{equation}\label{Gammankl}
    \Gamma^n_{kl} = \tilde{\Gamma}^{n}_{kl} - \tilde{g}^{mn} g_{m0} \Gamma^{0}_{kl} =
    \tilde{\Gamma}^{n}_{kl} + \frac {g^{n0}}{g^{00}} \Gamma^{0}_{kl} \,,
\end{equation}
where $\tilde{\Gamma}^{n}_{kl}$ describes the spatial connection on hypersurfaces $\{t = {\const}\}$.
Inserting \eq{Gammankl} and \eq{gammak} into \eq{LieK} we obtain
\begin{eqnarray*}
  {\cal L}_X \Gamma^{0}_{mn} &=& D_m D_n X^0 + X^\sigma \partial_\sigma \Gamma^{0}_{mn} +\Gamma^{0}_{mk}\partial_n X^k +\Gamma^{0}_{nk}\partial_m X^k
  - \Gamma^{0}_{mn} \partial_0 X^0
  \\
    & & - \frac {g^{0k}}{g^{00}} \left(
  \Gamma^{0}_{mn}\partial_k X^0 + \Gamma^{0}_{mk}\partial_n X^0
  + \Gamma^{0}_{nk}\partial_m X^0\right) \\
  & & - \frac {1}{2g^{00}} \left(
  \partial_m g^{00} \partial_n X^0 + \partial_n g^{00} \partial_m X^0\right) \ ,
\end{eqnarray*}
where $D$ denotes the covariant derivative in each hypersurface $\{ y^0 = {\rm const.}\}$ separately. Using now \eq{Liesqrg}, we are led to
\begin{eqnarray*}
 {\cal L}_X {\cal K}_{mn} &=& {\cal L}_X \left(\sqrt{|\det g_{\alpha\beta}|} \ {\Gamma}^0_{mn}\right)\\
  &=& {\Gamma}^0_{mn}X^\sigma \partial_\sigma\left(  \sqrt{|\det g_{\alpha\beta}|} \right) + \sqrt{|\det g_{\alpha\beta}|}\  {\cal L}_X {\Gamma}^0_{mn} \\
  &=& X^0 \partial_0 {\cal K}_{mn} + \sqrt{|\det g_{\alpha\beta}|}\ D_m D_n X^0 \\
  & & + \partial_k \left( X^k {\cal K}_{mn} \right) + {\cal K}_{mk} \partial_n X^k
  + {\cal K}_{nk} \partial_m X^k \\
   & & - \frac {g^{0k}}{g^{00}} \left(
  {\cal K}_{mn}\partial_k X^0 + {\cal K}_{mk}\partial_n X^0
  + {\cal K}_{nk}\partial_m X^0\right) \\
  & & - \frac {\sqrt{|\det g_{\alpha\beta}|}}{2g^{00}} \left(
  \partial_m g^{00} \partial_n X^0 + \partial_n g^{00} \partial_m X^0\right)
  \,.
\end{eqnarray*}
Using ADM notation,  we have $\sqrt{|\det g_{\alpha\beta}|} = N \sqrt{|\det {g_{mn}|} {}}$ and
\begin{eqnarray*}
  {\cal L}_X {\cal K}_{mn} &=& X^0 \partial_0 {\cal K}_{mn} + N \sqrt{|\det {g_{mn}|} {}}\ D_m D_n X^0 + {\cal L}_{\vec{X}} {\cal K}_{mn}\\
    & &    +N^k \left(
  {\cal K}_{mn}\partial_k X^0 + {\cal K}_{mk}\partial_n X^0
  + {\cal K}_{nk}\partial_m X^0\right) \\
  & & + \sqrt{|\det {g_{mn}|} {}}\left(
  \partial_m N \partial_n X^0 + \partial_n N \partial_m X^0\right)
   \,.
\end{eqnarray*}
On the other hand,
\[
  {\cal L}_X \left( {\tilde g}^{mk}{\tilde g}^{nl} - {\tilde g}^{mn}{\tilde g}^{kl}\right) = {\tilde g}^{nl} {\cal L}_X  {\tilde g}^{mk} +  {\tilde g}^{mk} {\cal L}_X {\tilde g}^{nl}
  -{\tilde g}^{kl} {\cal L}_X {\tilde g}^{mn} - {\tilde g}^{mn} {\cal L}_X {\tilde g}^{kl}\ .
\]
Consequently
\begin{eqnarray*}
{\cal L}_X \left( {\tilde g}^{mk}{\tilde g}^{nl} - {\tilde g}^{mn}{\tilde g}^{kl}\right)  =
  X^0 \partial_0 \left( {\tilde g}^{mk}{\tilde g}^{nl} - {\tilde g}^{mn}{\tilde g}^{kl}\right)
  + {\cal L}_{\vec{X}} \left( {\tilde g}^{mk}{\tilde g}^{nl} - {\tilde g}^{mn}{\tilde g}^{kl}\right) \\
       - \left\{ {\tilde g}^{mk}
    \left( N^n {\tilde g}^{lr} + N^l {\tilde g}^{nr} \right) + {\tilde g}^{nl}
   \left( N^m {\tilde g}^{kr} + N^k {\tilde g}^{mr} \right) \right.\\
    \left.
   - {\tilde g}^{mn}
   \left( N^k {\tilde g}^{lr} + N^l {\tilde g}^{kr} \right)
   - {\tilde g}^{kl} \left( N^m {\tilde g}^{nr} + N^n {\tilde g}^{mr} \right)
    \right\} \partial_r X^0 \, .
\end{eqnarray*}
Finally, we obtain:
\begin{eqnarray*}
  {\cal L}_X P^{kl} &=& {\cal L}_X \left\{{\cal K}_{mn} \left( {\tilde g}^{mk}{\tilde g}^{nl}-{\tilde g}^{mn}{\tilde g}^{kl}\right) \right\}\\
   &=&  X^0 \partial_0 P^{kl} + {\cal L}_{\vec{X}} P^{kl} + N \sqrt{|\det {g_{mn}|} {}}\left( D^k D^l X^0 - {\tilde g}^{kl} \Delta X^0
   \right)\\
    & &    +N^r \left(
  P^{kl}\partial_r X^0 + {\cal K}^k_{\ r}D^l X^0
  + {\cal K}^l_{\ r}D^k X^0- 2{\tilde g}^{kl} {\cal K}^m_{\ r}\partial_m X^0\right) \\
  & & + \sqrt{|\det {g_{mn}|} {}}\left(
  D^k N D^l X^0 + D^l N D^k X^0 - 2{\tilde g}^{kl}
  D^m N \partial_m X^0
  \right) \\
  & & - \left\{ {\cal K}^k_{\ n}
    \left( N^n {\tilde g}^{lr} + N^l {\tilde g}^{nr} \right)+ {\cal K}^l_{\ m}
   \left( N^m {\tilde g}^{kr} + N^k {\tilde g}^{mr} \right) \right.\\
   & & \left.
   -{\cal K}^n_{\ n}
   \left( N^k {\tilde g}^{lr} + N^l {\tilde g}^{kr} \right)
   - {\tilde g}^{kl}
   \left( 2 N^m {\cal K}^r_{\ m}   \right)
    \right\} \partial_r X^0 \ .
\end{eqnarray*}
After obvious cancelations we obtain the desired formula:
\begin{eqnarray}
  {\cal L}_X P^{kl} &=& X^0 \partial_0 P^{kl} + {\cal L}_{\vec{X}} P^{kl} + N \sqrt{|\det {g_{mn}|} {}}\left( D^k D^l X^0 - {\tilde g}^{kl} \Delta X^0
   \right) \nn
   \\
  & & + \left\{ \sqrt{|\det {g_{mn}|} {}}\left(
  {\tilde g}^{lr} D^k N  + {\tilde g}^{kr}D^l N  - 2 {\tilde g}^{kl}
  D^r N
  \right) \right. \nn
  \\ \label{4VI15.3}
  & &  + \left. P^{kl} N^r -P^{kr}N^l -P^{lr}N^k
    \right\} \partial_r X^0 \ .
\end{eqnarray}

%% file: invarianceJK.tex
\section{Stationarity with respect to variations of the field $X$}
 \label{A5VI15.1}

One of the consequences of Theorem~\ref{T4VI15.1} is, that  variations of $X$ enter formula \eq{homogeneous} with vanishing coefficients. In this appendix we give an alternative proof of this, under the supplementary assumptions that $X$ is a space-time vector field, which is everywhere transverse to $\mcV$, with {$\nu$ without zeros, and assuming variations of the map $\psi$ which leave $\mcV$ invariant}.

Given a field $X$ on $\mcM$ which is transversal
with respect to a hypersurface $\mcV$, and give a variation field $\delta X$, we choose an arbitrary 1-parameter family $X(\lambda)$ of (transversal) vector fields such that $X(0)=X$, fulfilling:
\bel{27VI15.1}
 \delta X = \frac{ \partial}{ \partial \lambda }  \bigg|_{\lambda=0} X(\lambda)
 \,.
\ee
A possible contribution of the variation of $X$ to the right-hand side of formula \eq{homogeneous} is linear with respect to $\delta X$. We denote it by
${\cal A}(\delta X)$.

Let ${\cal G}^{X(\lambda)}_t$ be the local flow generated by $X(\lambda)$. We set
\bel{3VII15.1}
 \R \times \mcV\ni (t,x) \mapsto \phi_\lambda(t,x) ={\cal G}^{X(\lambda)}_t(x)
 \,.
\ee
Next, given a field configuration $g$ on $\mcM$, we define a 1-parameter family $g(\lambda)$ of metric tensors as
\bel{27VI15.2}
 g(\lambda):= (\phi_\lambda^*)^{-1} g
 \quad
 \Longleftrightarrow
 \quad
 \phi_\lambda^* g(\lambda) = \phi_0^* g(0) = g
 \,.
\ee
This means that in local coordinates  $(t,x^i)$  in a neighborhood of $\mcV$ as in \eq{3VII15.1}, the  metric coefficients $g_{\mu\nu}(\lambda)$ of $g(\lambda)$ do not depend upon $\lambda$:
\[
    g_{\mu\nu}(\lambda) = g_{\mu\nu} \,.
\]
So, from the point of view of the manifold
$$
 \R\times \Sigma \supset \mathbb{R}\times V \approx \R \times \mcV
$$
and the above coordinate system on it, the resulting variations satisfy
\[
    \delta g_{\mu\nu} \equiv \delta_\psi g_{\mu\nu} = \frac{ \partial}{ \partial \lambda }  \bigg|_{\lambda=0} g_{\mu\nu}(\lambda) \equiv 0 \,,
\]
and, therefore, all the terms in the right-hand side of formula \eq{homogeneous} vanish.

Now, it has been  shown  in \cite{KijowskiGRG} that \eq{homogeneous} is coordinate invariant. This implies that the right-hand side of \eq{homogeneous} vanishes when calculated in any coordinates $(x^\alpha)$ on $\mcM$, not necessarily the ones adapted to the flow of $X(\lambda)$ as above. Observe that by \eq{27VI15.2} we have
\[
    \delta g = \frac{ \partial}{ \partial \lambda }  \bigg|_{\lambda=0} g(\lambda) =
    \frac{ \partial}{ \partial \lambda }  \bigg|_{\lambda=0} \left(\phi_\lambda^{-1}\right)^*g =
    -\mcL _{Z} g
     \,,
\]
where the Lie derivative  $\mcL_Z g$ of the metric $g$ is calculated with respect to the field
\[
    Z(t,x):= \frac{ \partial}{ \partial \lambda }  \bigg|_{\lambda=0} \phi_\lambda (t,x) \,.
\]
Hence, the sum of all contributions of $\delta_\psi g_{\mu\nu}$ to the right-hand side of \eq{homogeneous} is canceled by the contribution ${\cal A}(\delta X)$ of $\delta X$. In other words: ${\cal A}(\delta X)$ is equal to the contribution of $\delta g = \mcL _{Z} g$.

In fact, this last contribution vanishes identically. To show this, observe that $Z$ vanishes identically on $\mcV$ because $\phi_\lambda (0,x) =  x$ for every $x\in \mcV $. This implies that variations of all the Cauchy data, i.e. $\delta g_{kl}$, $\delta P^{kl}$ and,  therefore, also $\delta \lambda$ and $\delta g_{AB}$, together with Cauchy data of matter fields: $\delta \varphi$ and $\delta p$, vanish identically. The only non-vanishing contribution could, therefore, come from $\delta {\bf Q}$, $\delta {\bf Q}_A$ and $\delta \alpha$. Hence, the total contribution of these fields, according to \eq{homogeneous}, is equal to:
\begin{equation}
 {\cal A}(\delta X)= \frac 1{{ \gamma}} \intdvol  \left(   {\dot \lambda} \delta \alpha+ \nthree\delta {\bf Q}
- \nthree^A \delta {\bf Q}_A \right)
 \,.  \label{contribution}
\end{equation}
These quantities obey, however, the following identities:
\begin{eqnarray}
\delta {\bf Q} & = &
- \frac 1\nu \left( \dot{\lambda} - \partial_A (\lambda \nu^A) \right)\delta \alpha \,, \label{bfirst}\\
{\bf Q}_A + P^n_{\ A} & = & - \lambda \partial_A \alpha \,, \\
\delta {\bf Q}_A &=& - \lambda \partial_A \delta \alpha
 %\,.
\label{last}
\end{eqnarray}
(see Equations~(7.11)-(7.13) in \cite{KijowskiGRG}). Inserting (\ref{bfirst}) and (\ref{last}) into \eq{contribution} and integrating by parts  they cancel each other, so that ${\cal A}(\delta X)=0$.

For further reference,
we note the following formulae for the variation of
the remaining fields involved:
\begin{eqnarray*}
    \delta g_{AB}&=& 0 \,,
\\
  \delta g_{00}  &=& \delta (X|X) = 2 (X |\delta X)
  = - 2 \nu\delta\nu + 2\nu_A\delta \nu^A
   \,,
\\
  \delta g_{0A} &=& \delta (X | \partial _A) = g_{AB} \delta \nu^B
  \,.
\end{eqnarray*}
Moreover, using the invariance of ${\bf Q}= \nu Q^{00}$ and ${\bf Q}_A = Q^0_{\ A}$, we have:
\begin{eqnarray*}
  0 &=& \delta  \left( \nu Q^{00} \right)= Q^{00} \delta \nu + \nu \delta Q^{00} \,,\\
  0 &=& \delta Q^0_{\ A} = \delta \left( Q^{00} g_{0A} + Q^{0B} g_{BA} \right) \,,
\end{eqnarray*}
whence,
\begin{eqnarray*}
  \delta Q^{00}  &=& - \frac 1\nu Q^{00} \delta \nu \,, \\
  \delta Q^{0B} &=& - {\tilde{\tilde{g}}}^{BA}\left( g_{0A} \delta Q^{00} + Q^{00} \delta g_{0A} \right) = Q^{00} \left( \frac 1\nu \nu^B \delta \nu - \delta \nu^B \right) \,.
\end{eqnarray*}

%% file: SpacelikeVectorsJK.tex
\section{Spacelike vectors in adapted coordinates}
 \label{JKforY}

In this appendix we show that the variational formula \eq{homogeneous} for vector fields $X$ tangent to the initial data surface coincides with \eq{26IV15.15bis}. For simplicity, and consistency with~\cite{KijowskiGRG}  we assume  variations  that satisfy the linearized constraint equations.

Using \eq{Qnuidentity}, one
can
define a Hamiltonian by
performing a Legendre transformation which leads to the so-called
 ``purely metric'' formula (9.1) in Kijowski's paper~\cite{KijowskiGRG}:
 %provides the following version:
\begin{eqnarray}
- \delta \overline{\overline{\cal H}} & = &
\frac 1{16 \pi} \int_V  \left( {\dot P}^{kl}  \delta g_{kl} -
{\dot g}_{kl} \delta P^{kl} \right) +
\frac 1{8 \pi} \int_{\partial V} ( {\dot \lambda} \delta \alpha  -
{\dot \alpha} \delta \lambda ) \nonumber
\\
 & + & \frac 1{16 \pi}
\int_{\partial V}  Q^{ab}  \delta g_{ab}
 \label{dbarbarH-grav}
\,,
\end{eqnarray}
where
\begin{equation}
\overline{\overline{\cal H}}
%= - \frac 2{16 \pi}  \int_{\partial V}  \left( Q^{00} \, g_{00} + Q^{0A} \, g_{0A} \right) %- E_0
= - \frac 1{8 \pi}  \int_{\partial V}
 Q^{0}_{\ 0} %- E_0
 = - \frac1{8\pi}  \int_{\partial V} \left[
    \sqrt{|\det g_{cd}|} ({\hat g}^{a0} L_{a0} { +\frac12(n-3)L}) \right]
\,. \label{barH-grav-metric}
\end{equation}

For the purpose of the calculation here, it is convenient to change the notation so far as follows: Instead of $P^{kl}$ we write $P_{V}^{kl}$, and instead of $Q^{ab}$ we write $P_{\cal T}^{ab}$. This notation emphasises the fact that those fields describe  the ``ADM  momentum'' of the surface $V$, respectively the ``world tube'' ${\cal T}$  obtained  by flowing $\partial V$ along the vector field $X$:
\[
    {\cal T}:= \{ {\cal G}^X_t(x) | x \in \partial V\} \,.
\]
In the case of current interest, where $X$ is tangent to $V$ both surfaces coincide, hence so do the corresponding ADM momenta.

In this notation, \eq{dbarbarH-grav} takes the form
\begin{eqnarray}
 \frac 1{8 \pi}  \delta \int_{\partial V}
 {P_{\cal T}}^{0}_{\ 0} & = &
\frac 1{16 \pi} \int_V  \left( {\dot P}_{V}^{kl}  \delta g_{kl} -
{\dot g}_{kl} \delta P_{V}^{kl} \right) +
\frac 1{8 \pi} \int_{\partial V} ( {\dot \lambda} \delta \alpha  -
{\dot \alpha} \delta \lambda ) \nonumber \\
 & + & \frac 1{16 \pi}
\int_{\partial V}  P_{\cal T}^{ab}  \delta g_{ab}  \label{dbarbarH-grav-2}
\,.
\end{eqnarray}
So far the variation  $\delta X$ of $X$ was assumed to be zero, and
we used adapted coordinates so that $X = \partial_0$. The reader is warned that $\partial_0$ has nothing to do with a time coordinate in space-time, but is related to a parameter along the flow of $X$. Now, we want to rewrite the formula in a way which allows variations of $X$. For this purpose observe that the upper index $0$ in \eq{dbarbarH-grav-2} describes the transversal, or ``normal'' (with respect to $\partial V$) direction, which we denote by $n$, whereas the lower index $0$ describes the direction of the field $X$. Hence:
\[
    {P_{\cal T}}^{0}_{\ 0}= {P_{\cal T}}^{n}_{\ k} X^k \,.
\]
This leads to an associated rewriting of \eq{dbarbarH-grav-2}:
\begin{eqnarray}
 \frac 1{8 \pi}   \int_{\partial V}
  X^k\delta {P_{\cal T}}^{n}_{\ k} & = &
\frac 1{16 \pi} \int_V  \left( {\dot P}_{V}^{kl}  \delta g_{kl} -
{\dot g}_{kl} \delta P_{V}^{kl} \right) +
\frac 1{8 \pi} \int_{\partial V} ( {\dot \lambda} \delta \alpha  -
{\dot \alpha} \delta \lambda ) \nonumber \\
 & + & \frac 1{16 \pi}
\int_{\partial V} X^n P_{\cal T}^{kl}  \delta g_{kl}  \label{dbarbarH-grav-2+}
\,.
\end{eqnarray}
Observe, now, that for $X$ tangent to $V$ we have  $\alpha \equiv 0$ (see formula (\ref{alfa})). Using further $P_{\cal T}^{kl}=P_{V}^{kl}$, we are led to
\begin{eqnarray}
 \frac 1{16 \pi}   \int_{\partial V}
  2 X^k  \delta{P_{V}}^{n}_{\ k} -  X^n P_{V}^{kl}  \delta g_{kl} & = &
\frac 1{16 \pi} \int_V  \left( {\dot P}_{V}^{kl}  \delta g_{kl} -
{\dot g}_{kl} \delta P_{V}^{kl} \right)
\,,
 \phantom{xxx}
   \label{dbarbarH-grav-3}
\end{eqnarray}
where a ``dot'' denotes the Lie derivative with respect to $X$. This is a particular case, with $V={\cal T}$, of formula \eq{homogeneous}, and coincides with \eq{26IV15.15bis} within the collection of solutions of field equations, as desired.

For further reference, we note that in fact we also have the pointwise identity:
\begin{eqnarray}
    2 \nthree\delta {\bf Q} - 2\nthree^A \delta {\bf Q}_A
    + { \nu}{\bf Q}^{AB}
    \delta g_{AB}
    & = &
    2 X^k  \delta{P_{V}}^{n}_{\ k} -  X^n P_{V}^{kl}  \delta g_{kl}
     \,.
     \label{homogeneous-tangent}
\end{eqnarray}
which is equivalent to \eq{Qnuidentity} when $X$ is tangent to $\mcV$.

%% file: NegativeLambda.tex
\section{Negative $\Lambda$}
 \label{A5VI15.11}

There exists a clear prescription how to calculate the Hamiltonian mass of a family of metrics asymptotic to a fixed background metric~\cite{ChAIHP}. This is the case for Kerr-anti de Sitter metrics with negative cosmological constant, which are all asymptotic
to the anti de Sitter metric~\cite{CJL,HT,CMT,ChruscielSimon}. We emphasise that this the key difference between $\Lambda <0$ and $\Lambda >0$, as considered in this work: there is no single metric to which the Kerr de Sitter metrics converge as one recedes to infinity along asymptotically periodic ends of initial data sets.

More precisely, we use the standard form of the background anti de Sitter metric,
\bel{3IV15.1}
 b = - \left(1+ \frac{R^2}{\ell^2}\right) dT^2
  + \left(1+ \frac{R^2}{\ell^2}\right)^{-1} dR^2
  +R^2 \left(
   d\Theta^2 + \sin^2 \Theta \, d\Phi^2
    \right)
%\,,
\ee
where, as usual, $\ell^2 = - 3/\Lambda$. We consider the space of initial data sets for the vacuum Einstein equations which along $\hyp:=\{T=0\}$ approach the initial data for $b$ as $R$ tends to infinity at a rate made precise in \eq{30IX13.3} below. We wish to check whether the Kerr-de Sitter metrics can be put in the relevant form, and calculate their Hamiltonian mass.

In this appendix we apply this prescription to the Kerr-anti de Sitter metrics.  As such,
for such metrics the mass is not a global invariant anymore, but the component of a  linear functional on the set of KIDs for the anti-de Sitter metrics~\cite{ChNagy,ChHerzlich}, which transforms as a Lorentz covector under asymptotic isometries of the anti de Siter background. We will ignore the remaining components of the functional and consider only the ``energy component'', since the transformation properties of the associated object are well understood.

When $\Lambda<0$ and $\Xi > 0$ (as needed for non-singular rotating black holes with negative cosmological constant~\cite{CMT,HT}),
to calculate the Hamiltonian mass we need to find the leading order behaviour of the metric and compare it to anti de Sitter. For this one needs first to transform the Boyer-Lindquist form of the metric to a new coordinate system defined by (see, e.g., \cite{AMatzner})
\begin{eqnarray}
  \nonumber
    T &=& \frac{t}{\Xi} \,, \\
  \nonumber
    R^2 &=& \frac1{\Xi} \left(r^2 \Delta_{\theta}+a^2 \sin^2(\theta)\right) \,, \\
  \nonumber
    R \cos(\Theta) &=& r \cos({\theta}) \,, \\
  \label{AkcayMatzner}
  \Phi &=& \varphi - a \frac{\Lambda}{3 \Xi} t \,,
\end{eqnarray}
Under \eq{AkcayMatzner} we have, with expansions for large $R$,
% \ptcr{wrong sign in the square root? actually not, because $\Xi>0$ guarantees that the
%   square roots are real}
%
\bean
 g_{TT} &= & \Xi^2 g_{tt}+ \frac{2 \Xi a \Lambda}3g_{t\varphi} + \frac{a^2\Lambda^2}9 g_{\varphi\varphi}
\\
 \label{29IX13.11}
  & = & \frac{\Lambda  R^2}{3}-1+\frac{72
   \sqrt{6} m}{R \left(a^2 \Lambda
   -a^2 \Lambda  \cos (2
   \Theta)+6\right)^{5/2}}+O(R^{-2})
  \, ,
   \phantom{xxx}
\\
 g_{T\Phi} &= & \Xi g_{t\varphi} + \frac{a \Lambda }3 g_{\varphi\varphi}
  \nonumber
\\
 & = &-\frac{72 \left(\sqrt{6} a m \sin
   ^2(\Theta)\right)}{R
   \left(a^2 \Lambda -a^2 \Lambda
   \cos (2
   \Theta)+6\right)^{5/2}}+O(R^{-2})
    \, ,
\\
 g_{RR} & = & -\frac{3  }{\Lambda
   R^2}-\frac{9
    }{\Lambda ^2
   R^4}
 \nonumber
\\
 &&    +\frac{108 \sqrt{6}
    m}{\Lambda ^2 R^5
   \left(a^2 \Lambda -a^2 \Lambda
   \cos (2
   \Theta)+6\right)^{3/2}}+O(R^{-6})
   \, ,
\\
 g_{R\Theta}  & = &\frac{216 \sqrt{6} a^2
    m \sin (2
   \Theta)}{\Lambda  R^4
   \left(a^2 \Lambda -a^2 \Lambda
   \cos (2
   \Theta)+6\right)^{5/2}}+O(R^{-5})
   \, ,
\\
 g_{\Theta\Theta}    & = &
    R^2+\frac{108
   \sqrt{6} a^4   m
   \sin ^2(2 \Theta)}{R^3
   \left(a^2 \Lambda -a^2 \Lambda
   \cos (2
   \Theta)+6\right)^{7/2}}+O(R^{-4})
   \, ,
\\
 g_{\Phi\Phi}  & = &  R^2 \sin ^2(\Theta)
\nonumber
\\
 &&
  +\frac{72
   \sqrt{6} a^2 m \sin
   ^4(\Theta)}{R \left(a^2
   \Lambda -a^2 \Lambda  \cos (2
   \Theta)+6\right)^{5/2}}+O(R^{-2})
 \, .
\eeal{29IX13.12}

To calculate the Hamiltonian mass of the Kerr-anti de Sitter metrics we can use the results of~\cite{ChruscielSimon}: For this, let $b=b_{\mu\nu}dx^\mu dx^\nu$ denote the anti de Sitter metric in the coordinate system \eq{ads.static}. Let $e_{\hat a}$, $\hat a \in \{0,1,2,3\}$ be the following ON frame for $b$,
\bel{30IX13.1}
 e_{\hat 0} = \frac 1 {\sqrt{1+\frac{R^2}{\ell^2}}} \partial_T
  \, ,
  \
 e_{\hat 1} = \sqrt{1+\frac{R^2}{\ell^2}} \partial_R
  \, ,
  \
 e_{\hat 2} =  \frac 1 R \partial_\Theta
  \, ,
  \
 e_{\hat 3} =  \frac 1 {R\sin \Theta
 } \partial_\Phi
  \, .
\ee
Define
\bel{30IX13.2}
  e^{\mu\nu}  :=  g^{\mu\nu} - b^{\mu\nu}
 \, ,
\ee
and let $e^{\hat a \hat b}$ denote the components of $e^{\mu\nu}$ in the coframe dual to  $\{e_{\hat a}\}$.
If there exists $\epsilon>0$ such that
\bel{30IX13.3}
  e^{\hat a \hat b} =O(R^{-3/2-\epsilon})
 \, ,
 \quad
 e_{\hat c}(
  e^{\hat a \hat b}) =O(R^{-3/2-\epsilon})
  \, ,
\ee
then the Hamiltonian mass $M_H$ of a hypersurface $\{t=\tau\}$ equals~\cite[Equation~(5.22)]{ChruscielSimon}
\bean
 M_H &= & \lim_{R\to\infty} \frac{R^3}{16 \pi \ell^2} \int_{S_{\tau,R}}\big( R\partial_R e^{\hat A \hat A}  - 2  e^{\hat1 \hat 1}) d^2S
\\
 &= & \lim_{R\to\infty} \frac{R^3}{16 \pi \ell^2} \int_{S_{\tau,R}}\big(- R\partial_R e_{\hat A \hat A}  + 2  e_{\hat1 \hat 1}) d^2S
%  \, ,
\eeal{30IX13.4}
(summation over $\hat A$).

One finds
\bean
 e_{\hat 1 \hat 1} & = & \big(1 + \frac {R^2}{\ell^2}\big) (g_{RR}-b_{RR})
\\
 & = &-\frac{36 \left(\sqrt{6} m\right)}{R^3 \left(\Lambda
   \left(a^2 \Lambda -a^2 \Lambda  \cos (2
   \Theta)+6\right)^{3/2}\right)}
   +O(R^{-4})
 \label{30IX13.5}
 \, ,
\\
 e_{\hat 1 \hat 2} & = &  O(R^{-4})
 \label{30IX13.5+}
 \, ,
\\
 e_{\hat 2 \hat 2} & = &   \frac 1 {R^2}  (g_{\Theta\Theta}-b_{\Theta\Theta})
  \nonumber
\\
 & = &
 \frac{108 \sqrt{6} a^4 m \sin ^2(2 \Theta)}{R^5
   \left(a^2 \Lambda -a^2 \Lambda  \cos (2
   \Theta)+6\right)^{7/2}}
   +O(R^{-6})
 \label{30IX13.6}
 \, ,
\\
 e_{\hat 3 \hat 3} & = &   \frac 1 {R^2\sin^2(\Theta)}  (g_{\Phi\Phi}-b_{\Phi\Phi})
  \nonumber
\\
 & = &
 \frac{72 \sqrt{6} a^2 m \sin ^2(\Theta)}{R^3
   \left(a^2 \Lambda -a^2 \Lambda  \cos (2
   \Theta)+6\right)^{5/2}}
   +O(R^{-4})
 \, ,
\\
 e_{\hat 0 \hat 0} & = &
   -\frac{216 \left(\sqrt{6} m\right)}{R^3 \left(\Lambda
   \left(a^2 \Lambda -a^2 \Lambda  \cos (2
   \Theta)+6\right)^{5/2}\right)}+O(R^{-4})
 \, ,
\\
 e_{\hat 0 \hat 3} & = & -\frac{216 \left(\sqrt{2} a m \sin
   ^3(\Theta)\right)}{R \left(\sqrt{-\Lambda }
   \left(a^2 \Lambda -a^2 \Lambda  \cos (2
   \Theta)+6\right)^{5/2}\right)}+O(R^{-2})
 \, ,
\eeal{30IX13.8}
with vanishing remaining components.
The fall-off requirements are therefore satisfied, and we obtain a mass integrand
$$
-\frac{36 \left(\sqrt{6} m \left(a^2 (-\Lambda )+a^2
   \Lambda  \cos (2 \Theta)+12\right)\right)}{R^3
   \left(\Lambda  \left(a^2 \Lambda -a^2 \Lambda  \cos (2
   \Theta)+6\right)^{5/2}\right)}+O(R^{-4})
$$
which integrates over $S^2$ to
\bel{30IX13.9}
 M_H=\frac{9 m}{\left(a^2 \Lambda +3\right)^2}
  =\frac{m}{\Xi^2}
 \, .
\ee

%% file: CYK.tex
\section{Conformal Yano-Killing tensors}
 \label{A4VII15.1}

In four-dimensional spacetimes admitting non-trivial conformal Yano-Killing (CYK) tensors $Y_{\alpha\beta}$, or asymptotic CYK tensors, global invariants can be defined by integrating $Y_{\alpha\beta}C^{\alpha\beta\gamma\delta}$ over two-dimensional submanifolds (cf., e.g., \cite{JJCYKADS,JJCYKCQG}%,houri,krtous}
and references therein).
In Kerr-de Sitter spacetime~\cite{KubicekKrtous} a solution of the CYK equations is given by
% \ptc{Zostalo wielokrotnie sprawdzone, ze ponizszy CYK tensor jest nim (bez zmian) dla KdS.}
%
% { JJ: Zostalo wielokrotnie sprawdzone, ze ponizszy CYK tensor jest nim (bez zmian) dla KdS.
% Ponadto po zwezeniu z tensorem Weyla daje odpowiednia zamknieta dwuforme, ktorej calka tez bez zmian daje $m$
% tzn. nie ma wspolczynnika zaleznego od $\Xi$.}
\begin{equation}\label{Kerr_Q}
 Y:=  %Q_{{\rm Kerr}}=
 r\sin\theta \rd\theta \wedge \left[ \left( r^2+a^2\right)\rd\phi - a\rd t
 \right]+ a\cos\theta \rd r\wedge\left(\rd t - a\sin^2\theta \rd\phi \right) \,.
\end{equation}
%Theorem~\ref{dual_th}
The Hodge dual of $Y$ is also a CYK tensor:
\begin{equation}\label{Kerr_Qstar}
    \ast Y=a\cos\theta\sin\theta \rd\theta \wedge \left[ \left(r^2+a^2
    \right) \rd\phi - a\rd t \right]+ r\rd r \wedge \left(a\sin^2\theta \rd\phi -
    \rd t \right)
     \,.
\end{equation}
% \ptcr{ale tu jest Kerr de Sitter a nie Kerr?}
According to~\cite[p.~17]{Aksteiner}, these two tensors form a basis of the set of solutions.
As discussed extensively above, there is no unique choice of a preferred Killing vector $X$ which can be used to define mass.
Similarly, when we define the mass via CYK tensor, we have a freedom to multiply $Y$ by any function of $m$ and $a$, and there does not seem to be a preferred choice for asymptotically KdS metrics.

The Weyl tensor for KdS depends on the cosmological constant $\Lambda$ in a non-trivial way. Surprisingly enough, the CYK--Weyl contractions $F(C,Y):=Y^{\lambda \kappa} C_{\mu \nu \lambda \kappa }$ and $F(C,\ast Y):= \ast Y^{\lambda \kappa} C_{\mu \nu \lambda \kappa }$ do not depend on $\Lambda$, which results in the following closed two-forms:
\begin{eqnarray} F(C,Y)
%Y^{\lambda \kappa} C_{\mu \nu \lambda \kappa } \mathrm{d} x^{\mu} \wedge \mathrm{d} x^{\nu}
    &=&\frac{4 m}{\rho^4} \left\{[r^{2}-a^{2} \cos^{2} \theta] \sin \theta \mathrm{d} \theta \wedge [(r^{2}+a^{2}) \mathrm{d} \varphi- a \mathrm{d} t]
    \right. \nonumber
\\
& &
 +\left. 2 a r \cos \theta \mathrm{d} r \wedge [a \sin^{2} \theta \mathrm{d} \theta - \mathrm{d} t] \right\} \label{YC_contraction}
 \,,
\\
    F(C,\ast Y)
%\ast Y^{\lambda \kappa} C_{\mu \nu \lambda \kappa }  \mathrm{d} x^{\mu} \wedge \mathrm{d} x^{\nu}
    &=&
     \frac{4 m}{\rho^{4}} \left\{2 a r \sin \theta \cos \theta \mathrm{d} \theta \wedge [a \mathrm{d} t- (r^{2}+a^{2}) \mathrm{d} \varphi] \right. \nonumber
\\
 & &
  +\left. [r^{2}-a^{2} \cos^{2} \theta] \mathrm{d} r \wedge (a \sin^{2} \theta \mathrm{d} \varphi -\mathrm{d} t) \right\}
   \,,
   \label{YC_contraction*}
\end{eqnarray}
where%
\footnote{The symbolic software {\sc Waterloo Maple} has been used to check the above.}
$\rho^{2}=r^{2}+a^{2} \cos^{2} \theta$.
Integration of \eq{YC_contraction} over a sphere $r=\textrm{const.}$ and~\mbox{$t=\textrm{const.}$}
gives
\be\label{mKerr}
\frac{1}{16\pi}\int_{S} F(C,Y) = m \, ,
\ee
while a similar integral for \eq{YC_contraction*} vanishes.